\documentclass[11pt, a4paper]{article}
\pdfoutput=1
\usepackage{mathtools}
\usepackage{amsfonts}
\usepackage{amssymb}
\usepackage{graphicx, rotating}
\usepackage{epstopdf}
\usepackage{epsfig}
\usepackage{latexsym}
\usepackage{color}
\usepackage{cite}
\usepackage{slashed}
\usepackage[hyperfootnotes=true]{hyperref}
\definecolor{bluscuro}{rgb}{0.15, 0.2, .85}
\definecolor{ndc}{rgb}{0.7, 0.3, .3}
\hypersetup{colorlinks, citecolor=black, linkcolor=black, urlcolor=bluscuro}
\usepackage[font={small,it}]{caption}
\usepackage{geometry}
\usepackage{subcaption}
\usepackage{tikz}
\usepackage{ulem}
\definecolor{GRed}{rgb}{0.9,0.18,0.2}
\definecolor{CBlue}{rgb}{0.1,0.12,0.9}

\definecolor{amber}{rgb}{1.0, 0.49, 0.0}
\definecolor{auburn}{rgb}{0.43, 0.21, 0.1}

\newcommand{\0}{^{(0)}}
\newcommand{\1}{^{(1)}}
\newcommand{\2}{^{(2)}}

\newcommand{\eq}[1]{(\ref{#1})}

\newcommand{\nn}{\nonumber}

\newcommand{\be}{\begin{equation}}
\newcommand{\ee}{\end{equation}}
\newcommand{\bea}{\begin{eqnarray}}
\newcommand{\eea}{\end{eqnarray}}

\newcommand{\gsim}{\lower.7ex\hbox{$\;\stackrel{\textstyle>}{\sim}\;$}}
\newcommand{\lsim}{\lower.7ex\hbox{$\;\stackrel{\textstyle<}{\sim}\;$}}

\renewcommand\S{Section}

\numberwithin{equation}{section}

\begin{document}
\thispagestyle{empty}
\enlargethispage{3cm}
\vspace*{-2.5cm}
\begin{minipage}{.45\linewidth}
\begin{flushleft}                           
{SACLAY-t15/176\\CERN-PH-TH-2015-247\\ IFT-UAM/CSIC-15-111}
\end{flushleft} 
\end{minipage}
\hfill
\begin{minipage}{.45\linewidth}
\begin{flushright}    
DCPT/15/116\\                                 
IPPP/15/58
\end{flushright} 
\end{minipage}

{\flushright 
\vspace{1.5cm}}

\begin{center}
{\Huge \textbf{Radiative plateau inflation}}\\
\bigskip
\vspace{0.7cm}
{ \Large \bf \sc Guillermo Ballesteros}$^{1,2,3}$ {\large and} {\Large \bf \sc Carlos Tamarit}$^{4}$\\[7mm]
{$^1$ \it Institut de Physique Th\'eorique, Universit\'e Paris Saclay, CEA, CNRS}\\
{\it 91191 Gif-sur-Yvette, France}\\[2.4mm]
{$^2$ \it CERN, Theory Division, 1211 Geneva, Switzerland}\\[2.4mm]
{$^3$ \it Instituto de F\'{\i}sica Te\'orica IFT-UAM/CSIC}\\
{ \it C/ Nicol\'as Cabrera 13-15, Cantoblanco, 28049 Madrid, Spain}\\[2.4mm]
{ \it $^4$ Institute for Particle Physics Phenomenology}\\
{\it Durham University, South Road, DH1 3LE, United Kingdom}\\
[17mm]
{\bf Abstract}
\end{center}\vspace{0.3cm}
\noindent We describe how monomial chaotic inflation becomes compatible with the latest CMB data thanks to radiative corrections producing a plateau. The interactions of the inflaton with other fields, required for reheating, can flatten the potential and moderate the production of primordial gravitational waves, keeping these below the current upper bound. We show that the appearance of a plateau requires that the inflaton couples to fermions and to another scalar or a gauge group. We give concrete examples of minimal particle physics models leading to plateaus for quadratic and quartic chaotic inflation. We also provide a three-parameter model-independent description of radiatively corrected inflation that is amenable to CMB analyses. 

\begin{center} 

\vspace*{1.5cm} 
October 2015
\\
\vfill\flushleft
\noindent\rule{6cm}{0.4pt}\\
{\small  E-mail addresses: \tt guillermo.ballesteros@cea.fr, carlos.tamarit@durham.ac.uk}

\end{center}
\bigskip

\newpage

\tableofcontents

\section{Introduction}

Since its conception \cite{Starobinsky:1979ty,Starobinsky:1980te,Guth:1980zm,Kazanas:1980tx} and subsequent early development \cite{Mukhanov:1981xt,Linde:1981mu,Albrecht:1982wi,Mukhanov:1982nu,Hawking:1982cz,Starobinsky:1982ee,Guth:1982ec,Bardeen:1983qw}, primordial inflation  remains the best proposal for understanding the homogeneity and flatness of the universe at very large scales, as well as  the initial conditions leading to the inferred spectrum of fluctuations on this background. In particular, single-field slow-roll chaotic inflation \cite{Linde:1981mu,Linde:1983gd} provides the simplest and most successful framework so far to approach these key questions in cosmology. Multiple variants of this idea have been proposed during the years, while the precision of CMB observations was steadily increasing. Thanks to the latest Planck CMB data, several of those proposals are now discarded \cite{Planck:2015xua,Ade:2015lrj}. Remarkably, the most economical models of chaotic inflation, based on monomial potentials with a positive and even exponent, $p$, are either strongly $(p>2)$ or moderately ($p=2$) disfavoured by the data  \cite{Ade:2015lrj}, since they predict primordial gravitational waves with an amplitude in excess of the present upper bound \cite{Planck:2015xua,Ade:2015lrj,Ade:2015tva}. 

Indeed, the tensor-to-scalar ratio, $r$, is today one of the strongest constraints on the landscape of inflationary models. The latest analysis of Planck CMB data, including some external data as well, has established the upper bound $r<0.11$ at the scale $k=0.002$ Mpc$^{-1}$ and 95\% confidence level \cite{Planck:2015xua}, whereas the inclusion of B-mode polarization data from the BICEP2/Keck collaboration gives $r<0.12$ at $k=0.05$ Mpc$^{-1}$ and the same statistical significance \cite{Ade:2015tva}.  It is expected that this barrier will be surpassed in the near future thanks to ongoing observations\footnote{See the note added just before the Appendices. The current upper bound on $r$ has been significantly lowered to $r<0.07$ at $k=0.05$ Mpc$^{-1}$, which motivates even more our study.} and planned probes; see e.g.\ \cite{Creminelli:2015oda} for a discussion concerning specifically B-modes. Projected CMB observations may reach a sensitivity of $\Delta r\simeq 10^{-3}$ \cite{Matsumura:2013aja} or even $\Delta r\simeq 5\times10^{-4}$ \cite{Andre:2013afa}; and it has been claimed that lensing measurements of the 21-cm line of atomic Hydrogen might potentially get down to the astonishing level of $\Delta r\sim 10^{-9}$ \cite{Book:2011dz}.

Various theoretical constructions which modify monomial chaotic inflation have been engineered in order to satisfy the current observational constraints on the primordial power spectrum \cite{Nakayama:2013jka,Kallosh:2013tua,Ashoorioon:2013eia,Nakayama:2014wpa,Harigaya:2014fca,Li:2015mwa,Achucarro:2015rfa,
Harigaya:2015pea,Pallis:2015mga,Kannike:2015apa,Boubekeur:2015xza,Buchmuller:2015oma}. Many of these proposals, though not all, involve the ad hoc addition of extra scalar degrees of freedom or non-canonical couplings to gravity, which alter the appealing minimal structure of the original idea. However, whereas simplicity should undoubtedly be an important guiding principle, there are rather compelling reasons which indicate that a monomial potential, standing alone, cannot be considered a completely satisfactory model on its own.  As we will now argue, these reasons come from the requirement of reheating the universe after inflation and from basic considerations of quantum field theory. 

A complete inflationary model has to provide a mechanism to reheat the universe after inflation. This implies that a coupling (direct or not) between the inflaton and the Standard Model of particle physics  (SM) must be contained in the model.\footnote{We do not consider for this argument the possibility of gravitational reheating. In relation to this, it has to be mentioned that, for simplicity, we will assume through the paper that direct non-minimal couplings to the metric are suppressed, although it must be noted that these are generically created through loop corrections.}
  Let us entertain the simple possibility that the inflaton field, $\phi$, decays into a fermion, through a Yukawa interaction, when inflation ends. It is then straightforward to show that all possible renormalizable powers of $\phi$ are generated radiatively. A similar  conclusion holds if the inflaton decays (or anihilates) to another scalar field.\footnote{However, discrete symmetries which may involve fermions and scalars can forbid certain terms, e.g.  those involving odd powers of the inflaton.} Therefore, there is clearly an implicit tuning in all monomial chaotic models of inflation and a natural assumption would be to consider simultaneously all renormalizable powers of $\phi$. This implies that the potential always grows as $\phi^4$ (with logarithmic corrections) at sufficiently large field values, unless some symmetry prevents it.\footnote{For simplicity, we assume in this work that non-renormalizable operators can be neglected, assuming that the range of validity of the model is well beyond the region where inflation takes place. These operators could be included for generality and they would typically introduce small corrections to our results, suppressed by some high energy scale.}

Coupling monomial inflation  to a fermion can also be motivated as a way of lowering the prediction for the amplitude of primordial gravitational waves. At tree-level, the inflationary dynamics remains unchanged by such a coupling. However, radiative corrections with fermionic loops tend to flatten the potential, possibly reducing $r$ to a value compatible with the data if the corrections are sufficiently large. The effect of fermionic radiative corrections to quadratic and quartic chaotic inflation has been studied previously \cite{NeferSenoguz:2008nn,Enqvist:2013eua,Okada:2014lxa,Ahmed:2014cma}, leading to effective potentials that can fit the data and become negative at large field values. 

A qualitatively less radical modification of the overall shape of the potential is possible if the destabilizing effect of fermionic loops is compensated by radiative corrections produced by other fields, which may be scalars (including the inflaton itself) and gauge bosons.\footnote{Quantum corrections of bosons and fermions have also been considered e.g.\ in \cite{Croon:2015fza}, where the inflaton is a pseudo-Goldstone boson.  Other examples of models in which the interplay between bosonic and fermionic loops is essential are Higgs-plateau inflation (see e.g.\cite{Isidori:2007vm,Fairbairn:2014nxa,Ballesteros:2015iua} and Section \ref{subsubgf} of the present work) and the Higgs false-vacuum model \cite{Masina:2012yd,Notari:2014noa,Fairbairn:2014nxa,Ballesteros:2015iua}. None of these two models can actually fit the CMB the data. Bosonic loops also intervene in other models, e.g.\ \cite{Kannike:2014mia,Kannike:2015kda}.} These extra loops may lead to an inflationary plateau, ideally suited for slow-roll, and keep the potential positive definite at all field values. Such a plateau is characterized by an inflection point, which in general will be just approximate. Intuitively, a plateau helps to fit the CMB by reducing $r$ (which is proportional to the first derivative of the potential) and providing a flat region where the inflaton can exhibit a prolonged period of slow-roll, enhancing the number of e-folds of expansion. Indeed, inflation in the neighbourhood of an inflection point has been studied in a number of examples of diverse origins \cite{Ballesteros:2005eg,Ballesteros:2007te,BuenoSanchez:2006xk,Allahverdi:2006we,Baumann:2007np,Baumann:2007ah,Itzhaki:2007nk,Linde:2007jn,Cicoli:2008gp,Badziak:2008gv,Chen:2009nk,Enqvist:2010vd,Hotchkiss:2011gz,Mazumdar:2011ih,Cerezo:2012ub,Ballesteros:2014yva} and plateau potentials are known to be among the best candidates to describe the current CMB constraints \cite{Martin:2013nzq,Ijjas:2013vea,Guth:2013sya,Ade:2015lrj,Martin:2015dha}.

Given this, we may ask which is the minimal model, well motivated from the point of view of particle physics, that is needed to reduce the amplitude of primordial gravitational waves of monomial inflation through the appearance of a plateau. In this work we show that a successful inflationary plateau with an inflection point can generically appear in monomial chaotic inflation at the two-loop level, provided that the inflaton couples to an appropriate small set of extra fields. 

In Section \ref{plateaugeneral} we discuss the generic conditions that the potential has to satisfy to have a plateau, focusing on quartic and quadratic chaotic models. We express these conditions as constraints on an effective quartic coupling or effective mass, respectively. Approximate (i.e.\ deformed) plateaus can be described in broad generality with just three parameters; see Appendix \ref{reppot}. Then, in Section \ref{phenodes} we study the predictions for inflation and show that these models can easily accommodate the current CMB constraints, effectively lowering $r$, as required. Finally, in section 4 we study in detail radiative corrections to monomial inflation due to the couplings to extra fermions, scalars, gauge bosons and SM fields, in particular the Higgs. We show that plateaus generically require that the inflaton couples to fermions and another scalar or a gauge group. In that section we present minimal particle physics models that are capable of realizing a successful inflationary plateau. As a by-product, we show that the SM field content corresponds to an augmented version of one of these minimal cases. Appendix \ref{2loop} contains the relevant beta functions and Appendix \ref{precision} provides extra details on the solutions of the plateau equations.

\section{\label{plateaugeneral}Radiative plateaus}

Radiative corrections to a renormalizable tree-level potential can induce an inflection point with vanishing first derivative along some direction. This can favour slow-roll inflation in the approximately flat region of the potential around that point. Considering the potential on the direction $\phi$ relevant for inflation, such a plateau will be characterized by the existence of a field value $\phi_0$ which solves the two equations
\begin{align} \label{conds1}
\frac{dV}{d\phi}=0\,,\quad \frac{d^2V}{d\phi^2}=0\,.
\end{align}

In the limit of large field values, only the quartic part of the potential matters and the radiative corrections can be accounted for through the running of the quartic coupling. The potential can be well approximated in this situation by
\begin{align}
\label{lambdaeff}
 V(\phi) \simeq \frac{\lambda(\phi) }{4!}\phi^4\,,
\end{align}
where $\lambda(\phi)$ is an effective field-dependent coupling.  The simplest way of understanding the field dependence of the potential \eq{lambdaeff} is using the Coleman-Weinberg form \cite{Coleman:1973jx}:
\begin{align} \label{CW}
V(\phi)=\Omega(\mu)+V_{0}(\phi)+\frac{1}{64\pi^2}\sum_i M_i^4(\phi)\left(\log\frac{M_i^2(\phi)}{\mu^2}-C_i\right)+\cdots\,,
\end{align}
recalling that the effective squared masses $M_i^2$ of the particles running in the loops are proportional to $\phi^2$ for large values of the field.\footnote{We assume that either $\phi$ is the only scalar of the model or, in a more general situation, that the field-dependent masses $M_i$ are dominated by the $\phi$ contribution.} In this expression, $V_0(\phi)$ is the tree-level potential and $\Omega(\mu)$ is the field-independent loop contribution to the cosmological constant. The third term on the right hand side is due to one-loop diagrams and the ellipsis stands for higher order loops.  At large values of $\phi$, the effective quartic coupling will be of the form 
\begin{align} \label{expNLO}
\lambda(\phi)=\tilde\lambda(\mu)+\frac{1}{2}\tilde c_1(\mu)\log\frac{\phi^2}{\mu^2}+\frac{1}{8}\tilde c_2(\mu)\left(\log\frac{\phi^2}{\mu^2}\right)^2+\cdots\,,
\end{align}
where the logarithm squared comes from two-loop and higher order terms in the Coleman-Weinberg expansion \eq{CW}. Without loss of generality, we absorb in $\tilde\lambda(\mu)$ all non-logarithmic terms that come from the constants $C_i$. The field independent part $\Omega(\mu)$ is irrelevant in the large field limit. 

It is convenient to choose the renormalization scale $\mu$ to be proportional to the location of the plateau, i.e.\ $\mu=\mu_0\equiv\varepsilon\,\phi_0$, where generically $\varepsilon \ll 1$. The (positive) constant $\varepsilon$ comes from $M_i^2(\phi)\propto\phi^2$ (which is valid in the large field limit) and parametrizes the smallness of the couplings implicit in these proportionality relations. One can choose $\varepsilon$ to correspond, for instance, to the smallest of the couplings appearing for the masses $M_i^2(\phi)$. A different choice does not make any practical difference in the computations if all the couplings are approximately of the same order of magnitude.\footnote{If the model contained widely dissimilar energy thresholds or particle masses, a treatment such as those of \cite{Bando:1992wy} and \cite{Casas:1998cf} could be implemented. For simplicity, we assume throughout that this is not required.}

With this choice of renormalization scale, a straightforward redefinition of the coefficients of \eq{expNLO} allows to write this expression as follows: 
\begin{align} \label{expNLO0}
\lambda(\phi)=\lambda(\phi_0)+\frac{1}{2}c_1(\phi_0)\log\frac{\phi^2}{\phi_0^2}+\frac{1}{8}c_2(\phi_0)\left(\log\frac{\phi^2}{\phi_0^2}\right)^2+\cdots\,,
\end{align}
where the coefficients $\lambda_0(\phi_0)$, $c_1(\phi_0)$, \ldots, are functions of the original $\tilde\lambda(\mu)$, $\tilde c_1(\mu)$, etc.\ evaluated at $\mu_0$. 
Thus, the equations \eqref{conds1} are solved at $\phi=\phi_0$ if 
\begin{align}
\label{conds2} c_2(\phi_0)=-4\,c_1(\phi_0)=16\lambda(\phi_0)\,.
\end{align}
The reason for the relative signs in \eq{conds2} can be easily understood. Since the potential has to be positive at the plateau, i.e.\ around the inflection point $\phi_0$, the effective quartic coupling $\lambda_0(\phi_0)$ has to be positive. A negative first derivative of $\lambda(\phi)$ with respect to $\log \phi$ at the plateau, i.e.\ $c_1(\phi_0)$, tends to drive the effective quartic coupling to negative values, flattening the potential. In order to avoid the appearance of an instability, $c_2(\phi_0)$ has to be positive. 

Using the conditions \eq{conds2} into the expression \eq{expNLO0}, the effective potential becomes
\begin{align}
\label{potanalytic}
  V(\phi)\simeq\frac{\lambda(\phi_0)}{4!}\left(1-2\log\frac{\phi^2}{\phi_0^2}+2\left(\log\frac{\phi^2}{\phi_0^2}\right)^2+\cdots\right)\phi^4\,,
\end{align}
where the ellipsis stand for higher powers of the logarithm, coming from higher order loops, which we are neglecting. Therefore, we see that a plateau may arise from the interplay of the one- and two-loop corrections to the effective potential. Actually, the expression \eq{potanalytic} shows that a plateau can arise already at the two-loop leading log level, as we will discuss next. 

Another way of understanding the potential is the following. Starting anew with the Coleman-Weinberg expansion \eq{CW}, we choose the renormalization scale to be $\mu=\varepsilon\,\phi$ and keep only the terms
containing the fourth power of $\phi$. With this choice, the logarithms are effectively resummed into an effective quartic coupling $\lambda(\phi)$, which multiplies $\phi^4$, as we anticipated in \eq{lambdaeff}. This effective coupling $\lambda(\phi)$ includes the quartic terms at all orders, arising from the the Coleman-Weinberg potential in the large field limit. Then, we expand $\lambda(\phi)$ around the location of the plateau, $\phi_0$, obtaining an expression analogous to $\eq{expNLO0}$, i.e.
\begin{align} \label{expNLO2}
\lambda(\phi)=\lambda(\phi_0)+\frac{1}{2}\beta_{\lambda}(\phi_0)\log\frac{\phi^2}{\phi_0^2}+\frac{1}{8}\beta'_{\lambda}(\phi_0)\left(\log\frac{\phi^2}{\phi_0^2}\right)^2+\cdots\,.
\end{align}
This shows explicitly that the coefficients $c_1$, $c_2$, etc.\ of \eq{expNLO0} are related to the beta function of the effective quartic coupling,
\begin{align}
\beta_\lambda=\frac{\partial\lambda}{\partial\log \mu}\,,
\end{align} 
and its logarithmic derivatives, indicated with primes in \eq{expNLO2}. Therefore, the conditions \eq{conds2} can be interpreted in terms of the variation of the beta function of the effective quartic coupling at the plateau. The need of including two loops to describe a plateau also becomes automatically apparent in this way, since $\beta'_{\lambda}$ is of order two in the loop expansion. By construction, evaluating the effective quartic coupling $\lambda$ at $\phi_0$ corresponds to evaluating the original couplings of the Lagrangian (and their beta functions) at $\varepsilon\,\phi_0$ since we have chosen our renormalization scale to be $\mu_0=\varepsilon\, \phi_0$. It is also important to stress that $\beta_\lambda$, the beta function of the effective quartic coupling $\lambda(\phi)$, is not the same as the one of its tree-level counterpart, because $\lambda(\phi)$ absorbs the loop corrections from the Coleman-Weinberg expansion.\footnote{In concrete examples, it is nonetheless straightforward to obtain $\beta_\lambda$ and $\beta'_{\lambda}$ from the expression of $\lambda(\phi)$ in terms of the rest of the couplings and their beta functions.} In the large field limit, a logarithmic expansion of the tree-level quartic coupling around $\phi_0$ would obviously lead to an expression for the potential with the same functional form as \eq{expNLO}, but failing to reproduce the loop effects appropriately.

Coming back to the choice of renormalization scale, $\mu$, it should be
pointed out that the Callan-Symanzik equation guarantees that the the
effective potential is independent of it, when computed to
all loop orders. The truncation of the loop expansion (needed for
practical computations) introduces a marginal scale dependence which is
never worse than the precision of the truncation. In other words, upon
truncation, the effective potential remains scale-invariant up to subleading
 terms. However, as we mentioned before, the numerical effect
of these subleading terms at large field values can actually be better accounted for with the
choice $\mu\propto\phi$,  because it can suppress the potentially large logarithms of the form  $(\log M_i^2(\phi)/\mu^2)$ \cite{Bando:1992np}. 
 It is  worth
remarking that the energies of the virtual fluctuations inside loops
are related to their inflaton-dependent masses, which may remain well below the Planck mass even when the field takes Planckian values (as it is generically the case in chaotic large-field inflation) if the couplings are sufficiently small.  This will be ensured in our concrete examples, for which successful inflation demands a very flat potential, which implies that $\eta\equiv M_P^2 V''/V\ll
1$, where $M_P$ denotes the reduced Planck mass, $M_P=m_P/\sqrt{8\pi}$. Concerning couplings along orthogonal directions, the masses
sourced by the inflaton will be under the Planck scale as long as the
mixed scalar couplings between the inflaton and the additional scalars
are suppressed, as it will be the case in the examples that we study later.\footnote{This is consistent with our
assumption that the effective masses are dominated by $\phi$ at large
field values.}
Similarly, the inflaton field can source fermionic masses through Yukawa
couplings, which again will be suppressed if the latter are small. In
the examples of Section \ref{pp}, it is indeed possible to have inflaton
excursions of the order of $10 M_P$  with inflaton-sourced masses well below
$m_P$. Moreover, the overall height of the potential (which
is a most relevant scale for inflation) remains below $m_P^4$.

A plateau supporting slow-roll inflation may also appear in a regime in which the potential can be approximated by an effective quadratic term with positive mass squared:
\begin{align} \label{Pq}
 V(\phi)\simeq\frac{1}{2}m^2(\phi)\phi^2.
\end{align}
In terms of the beta function of the effective mass squared, the conditions for the existence of the plateau in this case are:
\begin{align}
\label{condsquad}  \beta'_{m^2}(\phi_0)=-2\beta_{ m^2}(\phi_0)=4 m^2(\phi_0)\,.
\end{align}
Expanding the effective mass around $\phi_0$, the potential can be written as
\begin{align}
\label{potanalyticquad}
  V(\phi)\simeq\frac{m^2(\phi_0)}{2}\left(1-\log\frac{\phi^2}{\phi_0^2}+\frac{1}{2}\left(\log\frac{\phi^2}{\phi_0^2}\right)^2+\cdots\right)\phi^2\,.
\end{align}
A plateau of this type is only viable if the cubic and quartic corrections to the potential remain suppressed for the field values around $\phi_0$. The potential \eq{potanalyticquad} assumes that those interactions vanish at the scale $\phi_0$ of the plateau. Clearly, a similar assumption is needed for a standard $\phi^2$ chaotic model of inflation. In concrete implementations of the quadratic plateau, it has to be checked that the cubic and quartic corrections induced through loops are suppressed for the range of fields relevant for inflation.

Plateaus may also occur in a region of a potential that is dominated by a linear or cubic term. However, whereas the radiatively corrected $\phi^2$ and $\phi^4$ potentials have a vanishing absolute  minimum $V(0)\simeq 0$ (under the assumption that the odd terms are negligible at all scales), the situation is more complicated for odd monomials. For linear and cubic plateaus the minimum can only appear once the quadratic or quartic terms start to be relevant, which happens away from $\phi=0$ and necessarily breaks the monomial approximation. Unless the potential has a tuned field-independent piece, the minimum will then be negative. We recall that the stages of the universe following inflation require a long-lived minimum with a small and positive cosmological constant. As it is usually done, we will assume that this is achieved by some mechanism for which inflation has no bearing. For simplicity,  we will focus our attention on $\phi^2$ and $\phi^4$ plateaus, avoiding in this work the the complications of odd monomials.

\section{Inflation}\label{phenodes}

Before searching for concrete particle physics models giving rise to successful inflationary plateaus, we are going to study the generic properties of the potentials \eqref{potanalytic} and \eq{potanalyticquad}.  As discussed in the introduction, the appearance of a plateau from radiative corrections is a possibility for completing and rendering viable the classical quadratic and quartic chaotic models of inflation, which are now either under strong tension (in the case of $\phi^2$) or completely ruled out ($\phi^4$), mostly due to Planck CMB data \cite{Planck:2015xua,Ade:2015lrj}. In this section we will show how radiative corrections to standard monomial chaotic inflation make these models compatible with the data.

\subsection{Quartic plateau}

The potential \eq{potanalytic} can be generalized to allow small deformations of the plateau, alleviating the tuning implied by the conditions \eq{conds2}.  This may also serve the purpose of parametrizing the effect of higher order radiative corrections altering the shape of the plateau. We will then consider the potential 
\begin{align}
\label{potanalyticdef}
  V(\phi)=\frac{\lambda}{4!}\left(1-2\big(1-b_1\big)\log\frac{\phi^2}{\phi_0^2}+2\big(1+b_2\big)\left(\log\frac{\phi^2}{\phi_0^2}\right)^2\right)\phi^4\,,
\end{align}
where the absolute values of $b_1$ and $b_2$ are assumed to be smaller than 1. 

Notice that the running of the couplings is constrained, in such a way that the truncated potential is approximately scale invariant. Formally, the potential \eq{potanalyticdef} has just three parameters, which are combinations of $\lambda$, $b_1$, $b_2$ and the logarithm of $\phi_0/M_P$. In Appendix \ref{reppot} we provide the explicit expressions for the three independent parameters, which can be used for direct fits to CMB data.

We consider the field rolling from larger to smaller values and compute the primordial inflationary parameters at some field value $\phi_i$, such that the total amount of inflation from that point until the end should be approximately $50$--$60$ e-folds \cite{Liddle:2003as}. For very small deformations (and in particular if $b_1=b_2=0$), a much larger amount of inflation is actually possible due to the flatness of the plateau, but those extra e-folds are irrelevant for the solution of the horizon problem. Enough e-folds can be attained generically with an initial field value $\phi_i<\phi_0$, which is the case we consider in this work. However,  it is worth pointing out that if the plateau is tilted considerably, the rolling of the field will be faster and enough inflation may require $\phi_i > \phi_0$. 

Inflation is of chaotic type for the radiatively corrected potential \eq{potanalyticdef}, like in the standard monomial $\phi^4$ model. If the parameters $b_1$ and $b_2$ are such that the potential increases monotonically, the inflaton classically rolls down the potential  until it reaches the attractor solution and its velocity is determined by the slope. The attractor is guaranteed to be attained by the flatness of the potential. Then, we can think of the inflaton field as taking random values initially in different patches of the primordial universe. All the regions where the inflaton is larger than $\phi_i$ lead to the required final state. If the potential has a local minimum,\footnote{A necessary condition to avoid the formation of a local minimum is $b_1>0$.} the random inflaton distribution ensures that successful inflation can take place, provided that $\phi_i$ is smaller than the location of the minimum. The same argument holds if the potential is unstable, if the value of $\phi_i$ leading to $N_e\sim 50$ is smaller than the maximum of the potential. 

The primordial spectra of scalar and tensor perturbations in the slow-roll approximation are:
\begin{align} \label{Ps}
\log P_s(k) & =  \log A_s+\left(n_s-1+\frac{\alpha}{2}\log\frac{k}{k_*}+\cdots\right)\log\left(\frac{k}{k_*}\right)\,,\\
\log P_t(k) & =  \log A_t+\left(n_t+\cdots\right)\log\left(\frac{k}{k_*}\right)\,, 
\end{align}
where the ellipses stand for higher order terms in the slow-roll expansion. The primordial parameters $A_s$, $n_s$, etc.\ are understood to be evaluated at the fiducial scale $k_*$, which is often chosen to be $k_*=0.05$ Mpc$^{-1}$. Since the largest observable scale is $k_i\simeq 2\times 10^{-4} $Mpc$^{-1}$, the scale $k_*$ sits at $\log(0.05/10^{-4})\sim 5$ e-folds away from the beginning of observable inflation. Given that the CMB data implies that the change of the primordial parameters is very small between $k_i$ and $k_*$ and given the uncertainty on the number of e-folds of observable inflation, we can simply evaluate the primordial parameters at $\phi_i$ (corresponding to $k_i$). This approximation is particularly good for very flat potentials, as it is the case here. 

If $b_1$, $b_2$ and $\chi=1-\phi_i/\phi_0$ are sufficiently small, it is straightforward to write explicit formulas for the primordial parameters by Taylor expanding their standard slow-roll  expressions in these variables. The value of $\chi$ for the cases that fit the data tends to be too large to obtain expansions that are both accurate and concise, so we do not write them here. Instead, we compute numerically the primordial parameters and the number of e-folds of inflation. 

In what follows we will denote by $N_e$ the number of e-folds of inflation that take place between $\phi_i$ and the end of inflation. Notice that if we fix the values of $b_1$, $b_2$ and $\phi_0$, the value of $\phi_i$ is set solely by $n_s$, and the value of $\lambda$ by $A_s$. In order to determine the number of e-folds of inflation from $\phi_i$, we solve the equation of motion as a function of e-folds itself, see \cite{Ballesteros:2014yva}.  This allows a precise determination of $N_e$ at the end of inflation and does not rely on the slow-roll approximation.  

\renewcommand\textfraction{.0}
\begin{figure}[t]
\centering
\includegraphics[scale=0.71]{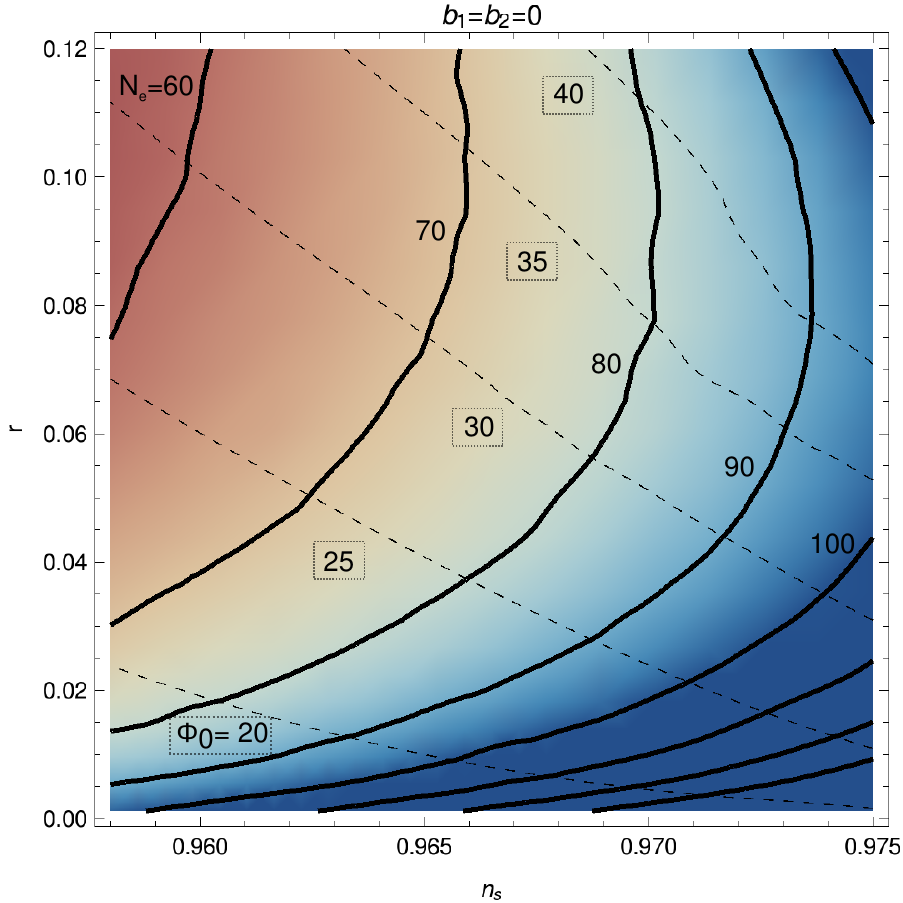}
\includegraphics[scale=0.71]{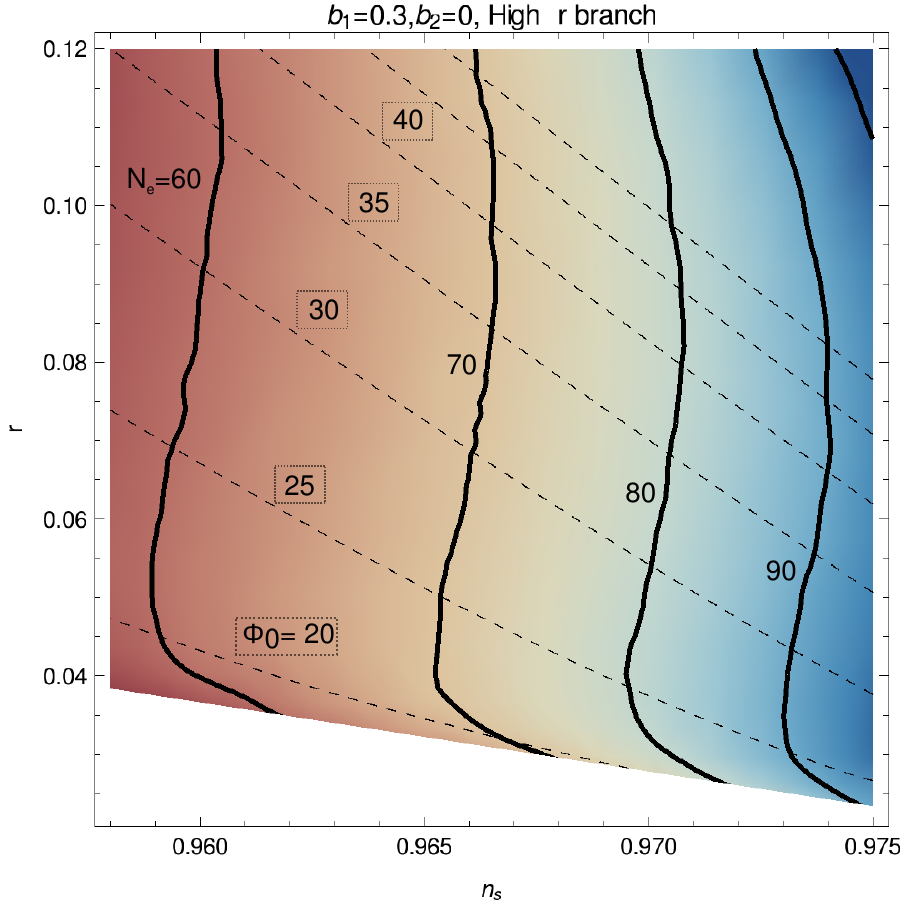}
\par
\vspace{0.4cm}
\includegraphics[scale=0.71]{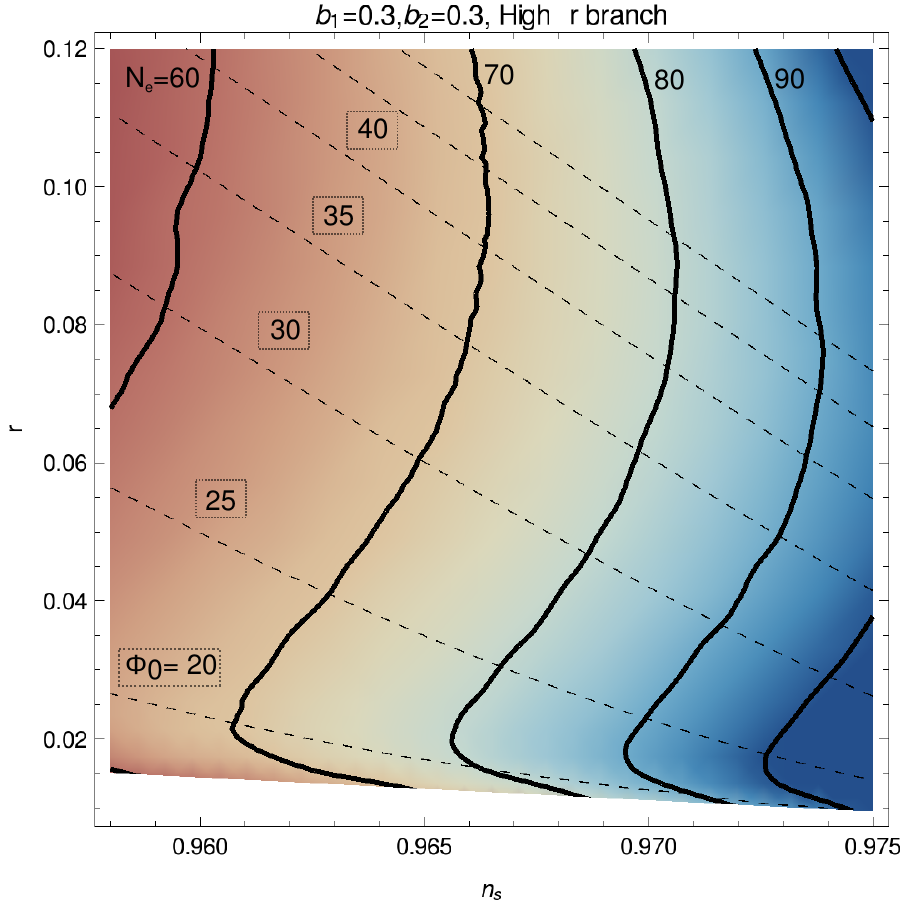}
\includegraphics[scale=0.71]{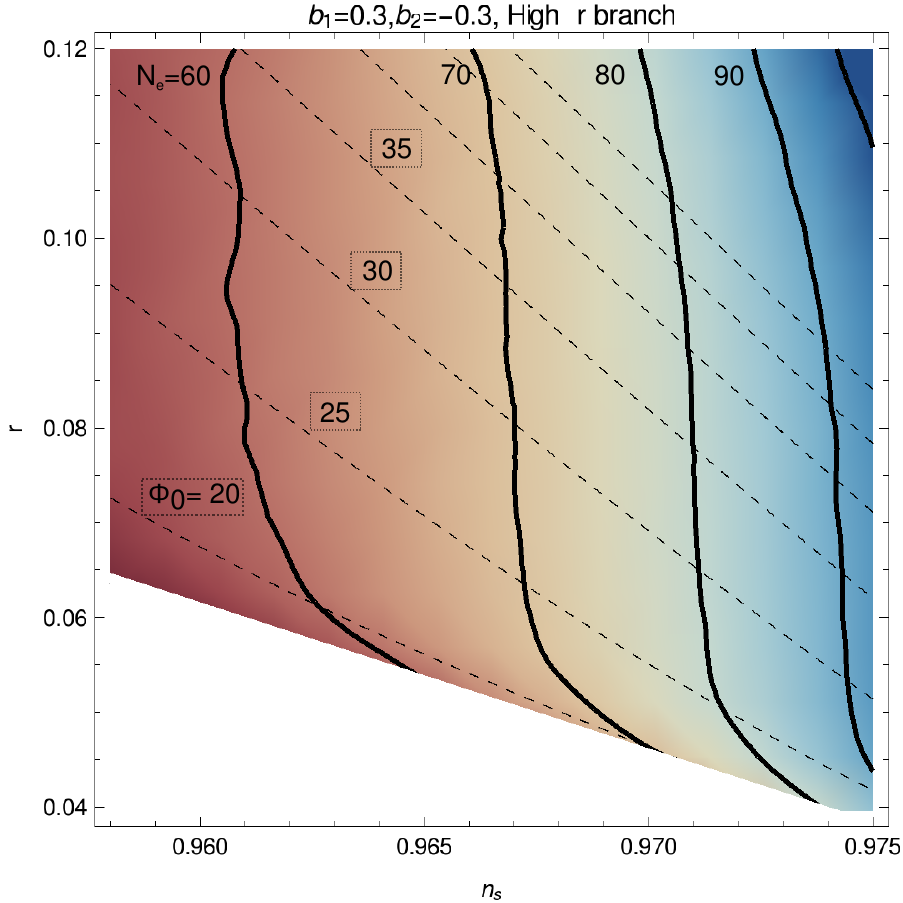}
\caption{\label{phi4rns} Curves of constant $N_e$ (solid) and constant $\phi_0$ (dashed, with boxed labels) in quartic plateau models \eq{potanalyticdef}, in terms of $r$ and $n_s$, for different choices of the deformation parameters $b_1,b_2$, and choosing the branch of solutions
giving smaller $\phi_i/\phi_0<1$.
The value of $A_s$ has been fixed to $2.142\times10^{-9}$. The white regions are excluded for these choices of $b_1$ and $b_2$.}
  \end{figure}
  
 \begin{figure}[t!]
\centering
\includegraphics[scale=0.8]{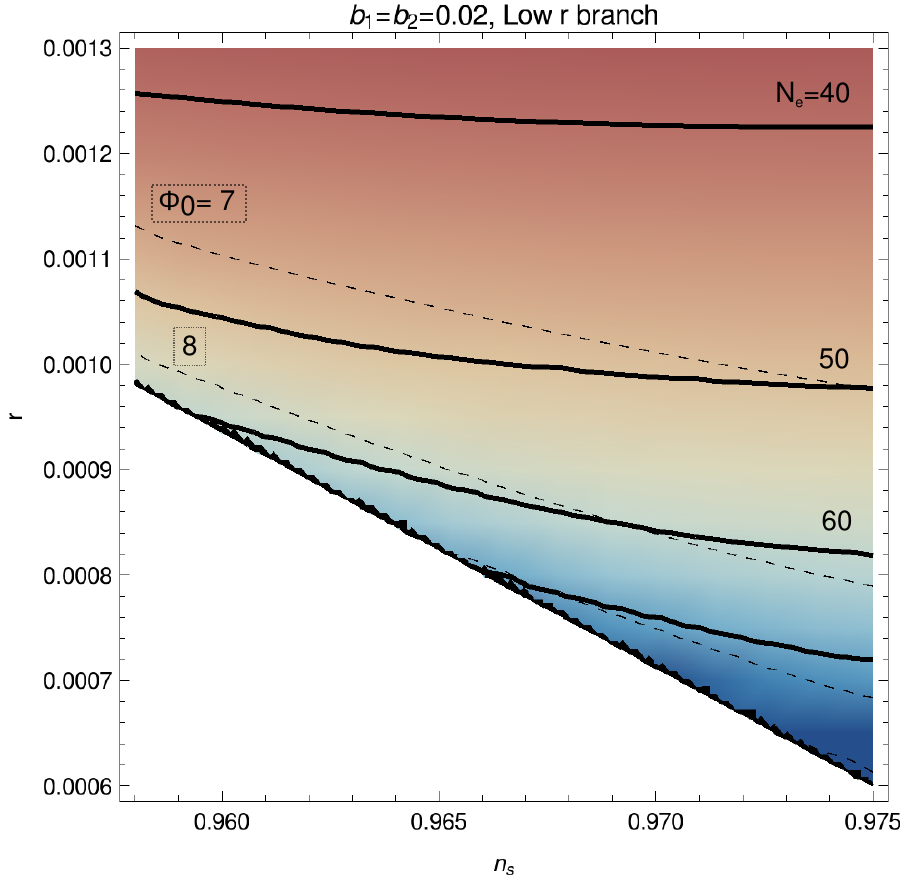}
\includegraphics[scale=0.8]{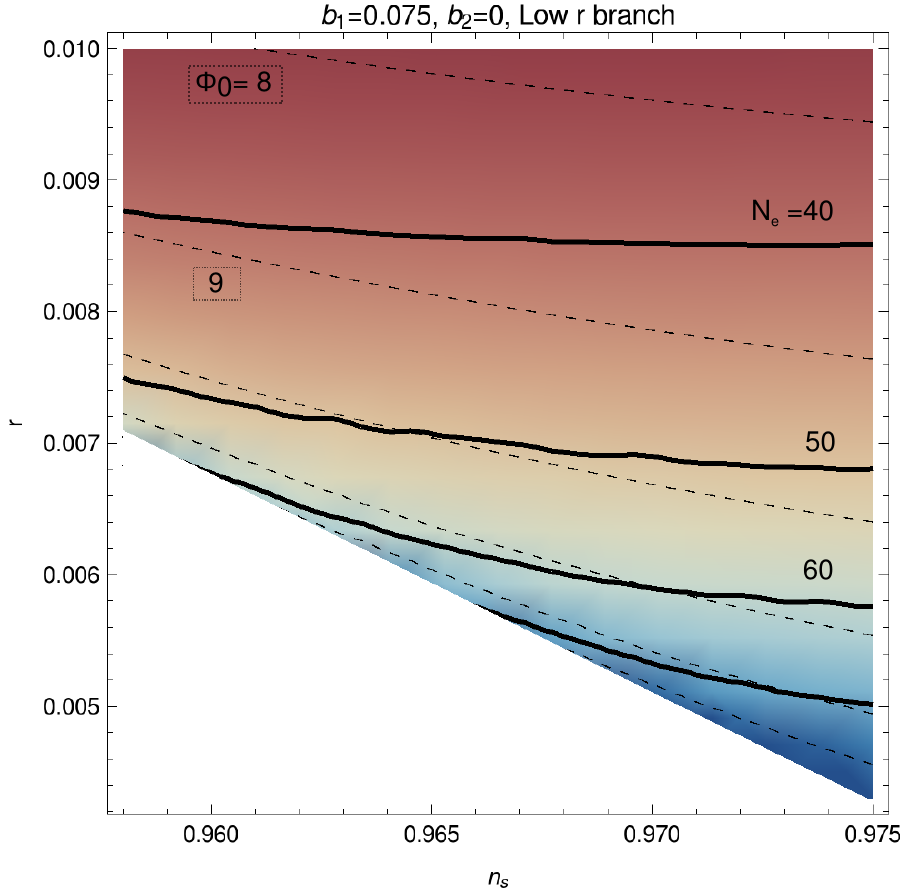}
\par
\vspace{0.4cm}
\includegraphics[scale=0.8]{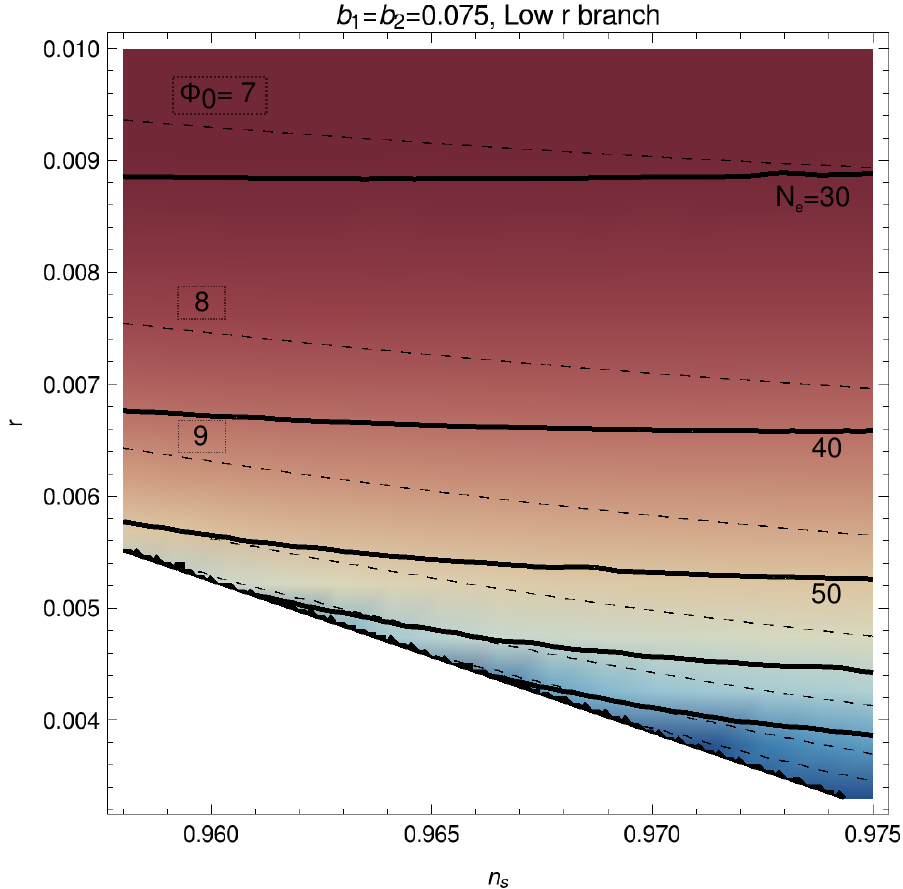}
\includegraphics[scale=0.8]{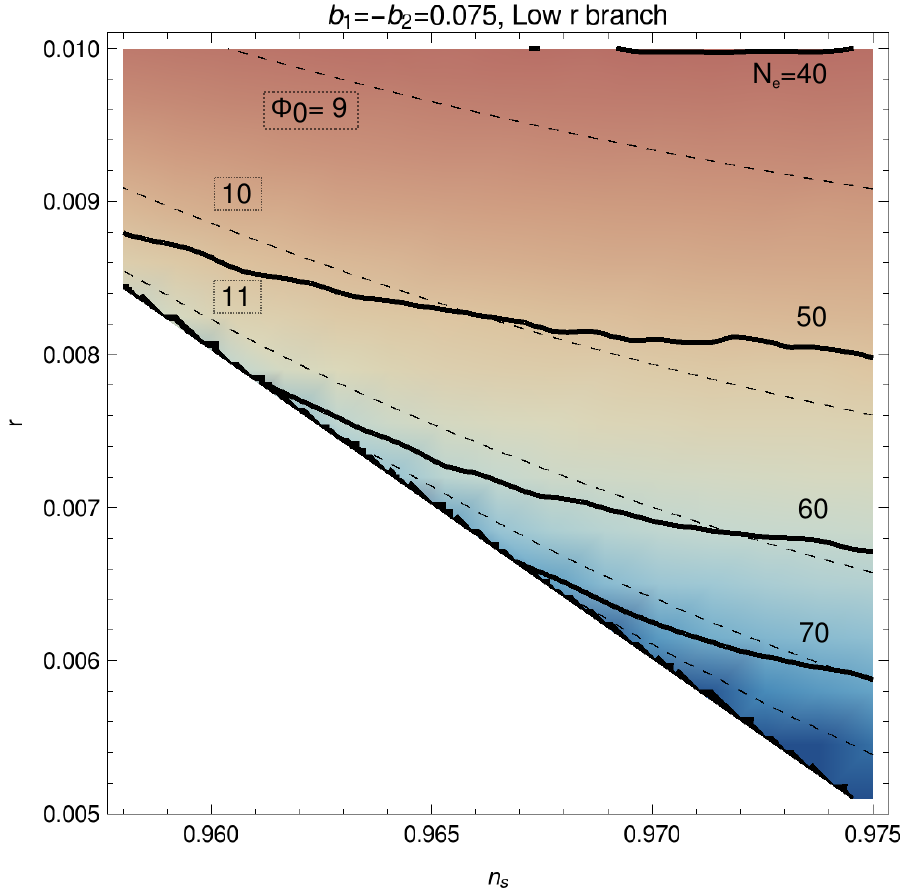}
\caption{\label{phi4rns2} Curves of constant $N_e$ (solid) and constant $\phi_0$ (dashed, with boxed labels) in quartic plateau models, in terms of $r$ and $n_s$, for different choices of the deformation parameters $b_1,b_2$, and choosing the branch of solutions
giving larger $\phi_i/\phi_0<1$.
The value of $A_s$ has been fixed to $2.142\times10^{-9}$. The white regions are forbidden}
  \end{figure}

Given $b_1\neq 0$, $b_2\neq 0$, $A_s$ and $n_s$, there are two branches of solutions for $\phi_i$ and $\phi_0$. Each branch corresponds to a different number of e-folds and different values of the tensor-to-scalar ratio $r=A_t/A_s$. One of them, which produces the largest values of $r$, connects with the exact plateau case ($b_1=b_2=0$), and is illustrated in Figure \ref{phi4rns}, which shows contour curves in the plane $r$--$n_s$, for four different choices of $b_1$ and $b_2$. Planck CMB data alone imply $n_s = 0.9666 \pm 0.0062$ at $k_*=0.05$ Mpc$^{-1}$ and $r < 0.103$ at $k_*=0.002$ Mpc$^{-1}$, both at $95\%$ c.l., and assuming negligible running of the scalar spectral index \cite{Ade:2015lrj}.  In the examples of Figure \ref{phi4rns} the running, $\alpha$, is of the order of $-10^{-4}$, or $-10^{-3}$ at most. Whereas these values of $\alpha$ are small, they may be probed with future data, specifically with measurements of the 21cm line in galaxy surveys \cite{Adshead:2010mc,Shimabukuro:2014ava} and CMB polarization \cite{Martin:2014rqa}. The solid black lines in Figure \ref{phi4rns} correspond to points with equal number of e-folds from $\phi_i$ to the end of inflation, while the dashed lines represent points with equal values of $\phi_0$. The values of the tensor spectral index, $n_t$, can be read directly from the figures, using that $n_t\simeq -r/8$. Figure \ref{phi4rns} shows that a plateau with $b_1=b_2=0$ and less than than $\sim 60$ e-folds is in very strong tension with the current bounds on $r$ and $n_s$. Assuming a higher number of e-folds, $r$ can be decreased (or $n_s$ increased) sufficiently to make such a  plateau compatible with current constraints. However, the possibility of $N_e \gg 60$ is unlikely \cite{Liddle:2003as}. A precise determination of $N_e$ requires a detailed knowledge of the post-inflationary physics and hence a specific particle physics model for the plateau. Increasing $b_1$ and allowing for a negative $b_2$ decreases the number of e-folds and eases the tension with the data, making a quartic plateau compatible, as Figure \ref{phi4rns} also illustrates. For example, for $b_1\gtrsim 0.3$, the spectral index can be $n_s=0.96$ and the number of e-folds $N_e\lesssim60$. This is because $b_1>0$ makes the potential steeper near the plateau, so the field can roll faster, reducing $N_e$ with respect to the case $b_1=0$. 

The second branch of solutions for $b_1>0$ and $b_2\neq 0$ is characterized by having larger values of the ratio $\phi_i/\phi_0$. In this branch $r$ is significantly lower, yielding a plateau in perfect agreement with the data. The reason for the existence of two distinct branches can be understood intuitively as follows. A very flat plateau implies a very slow rolling and therefore favours a large number of e-folds, and thus  $N_e$ can be reduced if $\phi_i$ is sufficiently smaller than $\phi_0$. However, this implies that the potential will be steeper at $\phi_i$, which translates into a larger $r\propto (V'/V)^2$. On the other hand, a positive $b_1$ implies $V'\neq 0$ at $\phi_0$ and a steeper potential. Then, $\phi_i$ can be closer to $\phi_0$, where $r$ remains small. The present CMB data suggest that the inflationary potential could indeed be a very flat plateau, but not excessively flat. In other words: the inflaton must have rolled slowly, {\it ma non troppo}.  The second branch of solutions yielding potentials in agreement with data is illustrated in Figure \ref{phi4rns2}.

\begin{figure}[t!]\centering
 \includegraphics[scale=0.43]{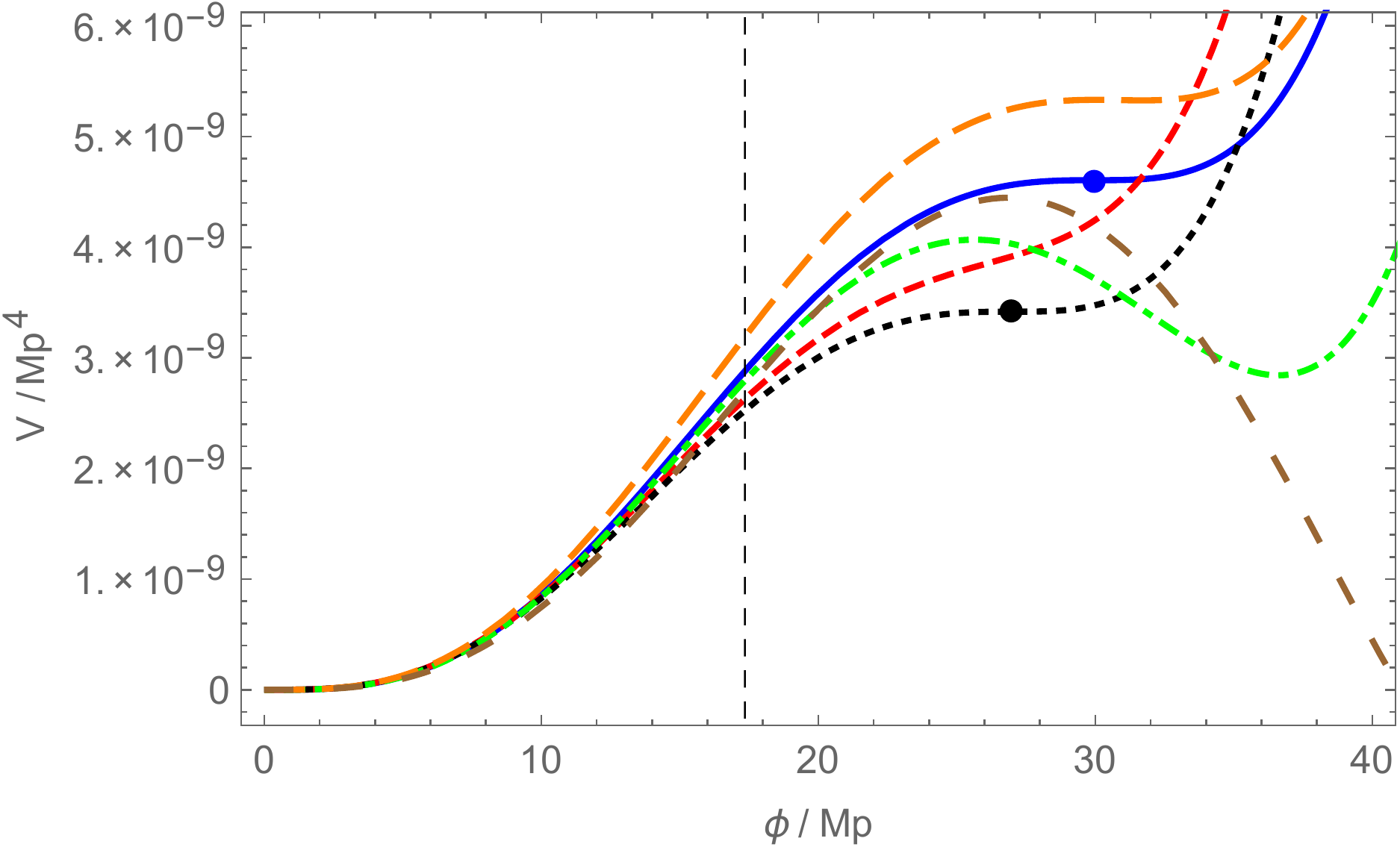}
\includegraphics[scale=0.43]{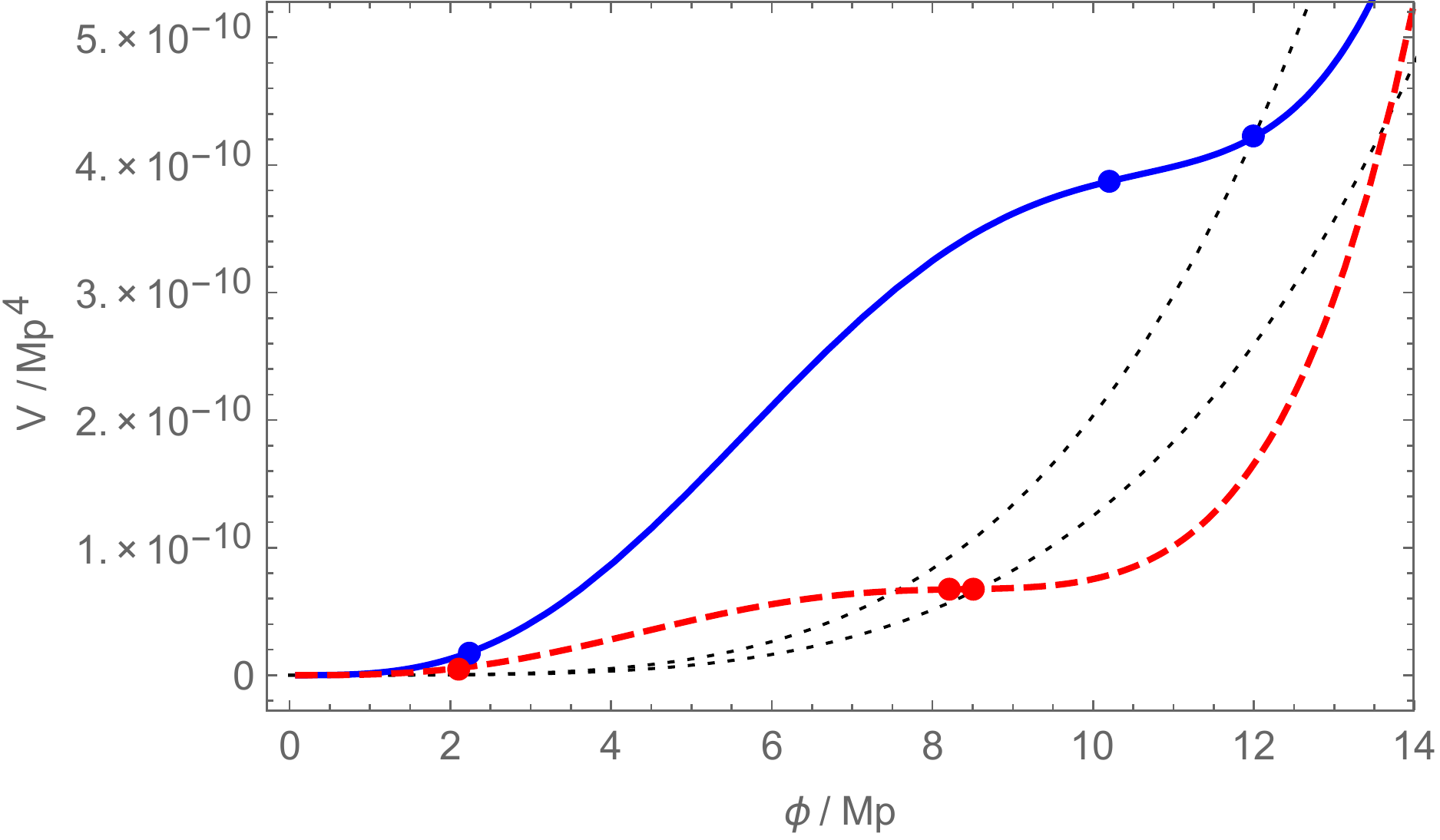}
\caption{\label{examplesquart} Examples of radiatively corrected quartic inflationary potentials. The parameters of the potentials and their predictions for inflation are listed  in Table \ref{table1}. {\it Left}: potentials in the high-$r$ branch illustrated in Figure \ref{phi4rns}. Potentials of this kind predict high $r$, high $N_e$ and low $n_s$. Still, they are currently allowed by the data. The vertical line (at $16.5 M_P$)  indicates the (approximate) common location of $\phi_i$. For comparison we show a potential with an evident false vacuum (green, long-dashed) and an unstable potential (brown, long-dashed). All these potentials have similar predictions independently of their shapes at higher field values. The dots in the exact plateau examples (continuous-blue (1) and dotted-black (3), see Table \ref{table1}) locate $\phi_0$ for those cases. {\it Right}: two examples of potentials in the low-$r$ branch, as in Figure \ref{phi4rns2}. Potentials of this type provide a comfortable fit to current data. The dots indicate the field values $\phi_0$, $\phi_i$ and $\phi_e$ in each case, from right to left. For comparison, black dotted lines represent the corresponding quartic monomials (without the radiative corrections). The potentials of both panels have $A_s(\phi_i)=2.13\times 10^{-9}$  and vanish at $\phi=0$. }
\end{figure}

We find that for $40\lesssim N_e\lesssim 70$ the field typically travels from $\phi_i$ to $\phi_e$ a distance of the order of 12--16 $M_P$ in the first branch and as low as 4.5 $M_P$ in the second branch. The plateau's location $\phi_0$ must be about  20 to  40 times $M_P$ (in the high-$r$ branch) or around 7--10 $M_P$ (in the low-$r$ branch). The effective quartic coupling at the scale $\phi_0$ must be around $10^{-13}$ (and up to $10^{-12}$ in the second branch).  For comparison, a monomial  inflation model, $V\propto\phi^N$, predicts $r=8N^2(M_P/\phi_i)^2\simeq 4 N/N_e$, which means $\phi_i\gsim 20$ $M_P$ for $N=4$. In this case, the quartic coupling is roughly of the same magnitude as for the plateau, which supports the tantalizing interpretation of the plateau (and hence of radiative corrections) as being capable of bringing standard chaotic quartic inflation back into the allowed region of parameters. Notice that a radiatively modified monomial potential that becomes unstable at large field values can also fit the data, see e.g.\ \cite{NeferSenoguz:2008nn,Okada:2014lxa,Enqvist:2013eua}. Figure \ref{examplesquart} shows several examples of successful quartic potentials that produce a spectrum of primordial perturbations compatible with the current data and enough inflation. In all of these examples $\phi_i<\phi_0$. Qualitatively, the two branches differ in the role of the plateau on the inflationary dynamics. While the plateau (or lack thereof) is actually inessential in the high-$r$ branch, it is crucial on the low-$r$ branch. This is can be seen explicitly comparing the left (high-$r$) and right (low-$r$) panels of Figure \ref{examplesquart}.

 \begin{table}[t] \centering {\small
\begin{center}
\begin{tabular}{l*{10}{c}r}
\# & $\phi_0/M_P$   &  $b_1$  & $b_2$  & $\lambda\times 10^{13}$ &  $n_s$ & $\alpha\times 10^{-4}$ & $r$ & $\phi_i/M_P$ & $\phi_e/M_P$ & $N_e$  \\
\hline
{\color{blue} 1} {\footnotesize (continuous)} & 30 & 0 & 0 & $1.37 $  & 0.961 & $-5.8$ & 0.088 & 16.94  & 1.88  & 61.7 \\
{\color{red} 2} {\footnotesize(short-dashed)} & 30 & 0.30 & 0.25 & $1.26$  &  0.961 & $ -5.8$ & 0.077 & 16.56 & 1.85 & 61.7    \\
3 {\footnotesize(dotted)} & 27 & 0 & 0 & $1.55$ &   0.960 & $ -5.5 $ & 0.074 & 16.47 & 1.85 & 61.7    \\
{\color{green} 4} {\footnotesize(dot-dashed)} & 30 & $-0.30$ & 0.20 & $1.10$  &  0.960 & $ -5.6 $ & 0.084 & 16.87 & 1.87 & 61.6    \\
{\color{amber} 5} {\footnotesize(long-dashed)} & 30 & 0 & $-0.10$ & $1.58$  &  0.960 & $ -6.3$ & 0.096 & 16.84 & 1.87 & 59.7    \\
{\color{auburn} 6} {\footnotesize(long-dashed)} & 30 & $-0.30$ & $-0.20$ & $1.04$  &  0.962 & $ -5.1$ & 0.088 & 17.61 & 1.87 & 64.7    \\
\hline
{\color{blue} 7} {\footnotesize (continuous)} & 12 & 0.25 & 0.28 & $4.88$  &  0.966 & $ -28.1$ & 0.012 & 10.21 & 1.65 & 49.5    \\
{\color{red} 8} {\footnotesize (dashed)} & 8.5 & 0.033 & 0 & $3.10$  &  0.966 & $ -29.3$ & 0.002 & 8.20 & 1.57 & 53.1    \\
\end{tabular}
\caption{Examples of radiatively corrected quartic potentials. The potentials 1--6 and 7, 8 correspond to the left and right panels of Figure \ref{examplesquart}, respectively. The primordial functions $n_s$, $r$ and $\alpha$ are evaluated at $\phi_i$. In all cases, $A_s(\phi_i)=2.13\times 10^{-9}$. The quartic coupling $\lambda$ and the deformation parameters $b_1$ and $b_2$  are given at the plateau scale $\phi_0$. The quantity $N_e$ denotes the number of e-folds from $\phi_i$ to $\phi_e$, where inflation ends. } \label{table1}
\end{center}
}
\end{table}

To end this section, it is important to highlight that, in view of the requirements imposed by current data, the preference for non-zero deformation parameters ($b_1$ and $b_2$) means that an approximate  (and hence more generic and less fine tuned)  quartic plateau performs better than an exact plateau (with $b_1=b_2=0$). In summary, generic quartic plateaus with a significant deviation from the conditions \eq{conds1} describe well the current CMB observations (and are preferred by the data).

\subsection{Quadratic plateau}

\begin{figure}[t!]
\centering
\includegraphics[scale=0.71]{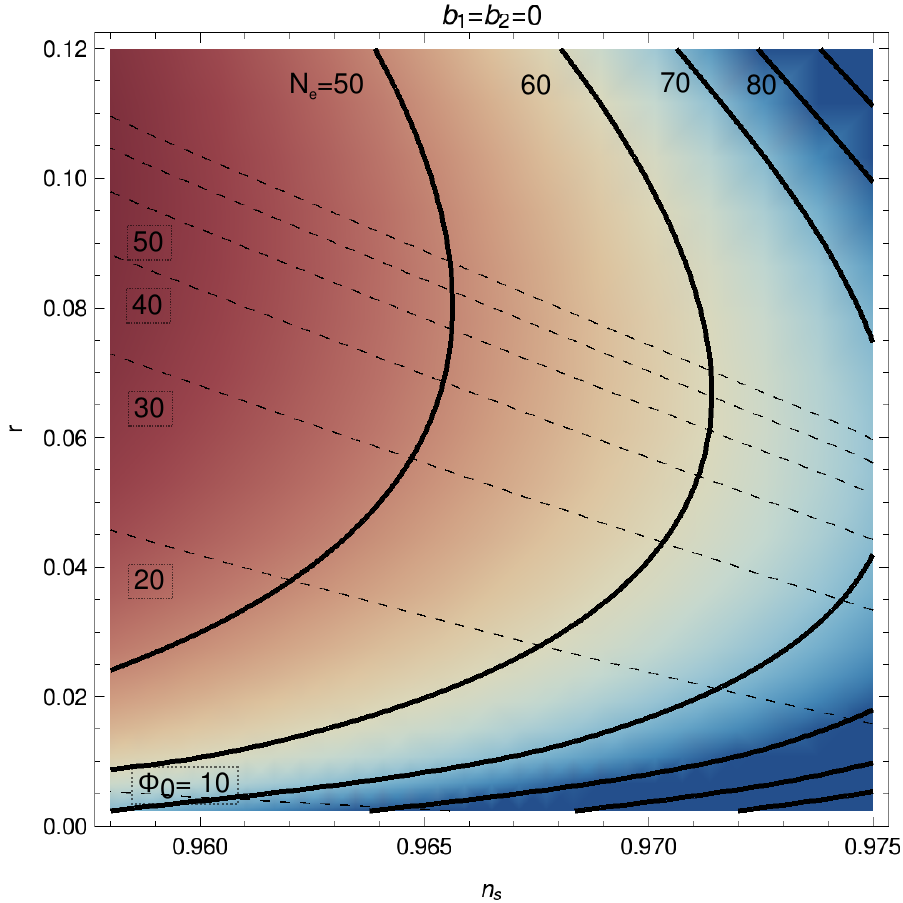}
\includegraphics[scale=0.71]{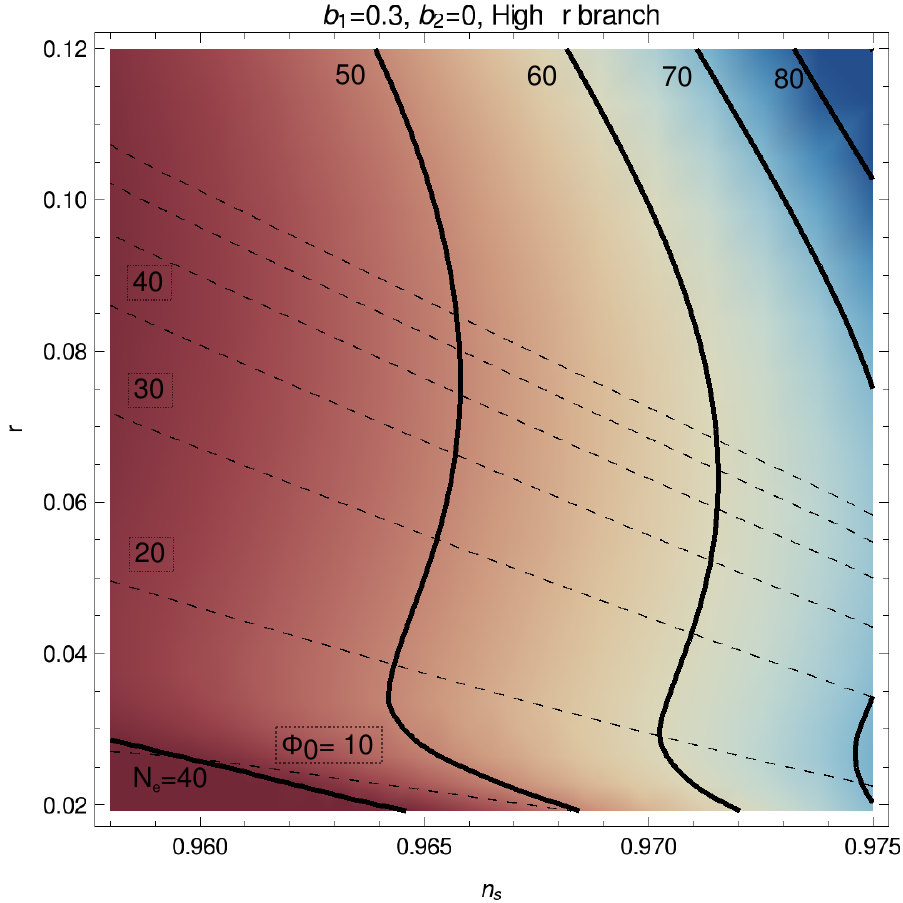}
\par 
\vspace{0.4cm}
\includegraphics[scale=0.71]{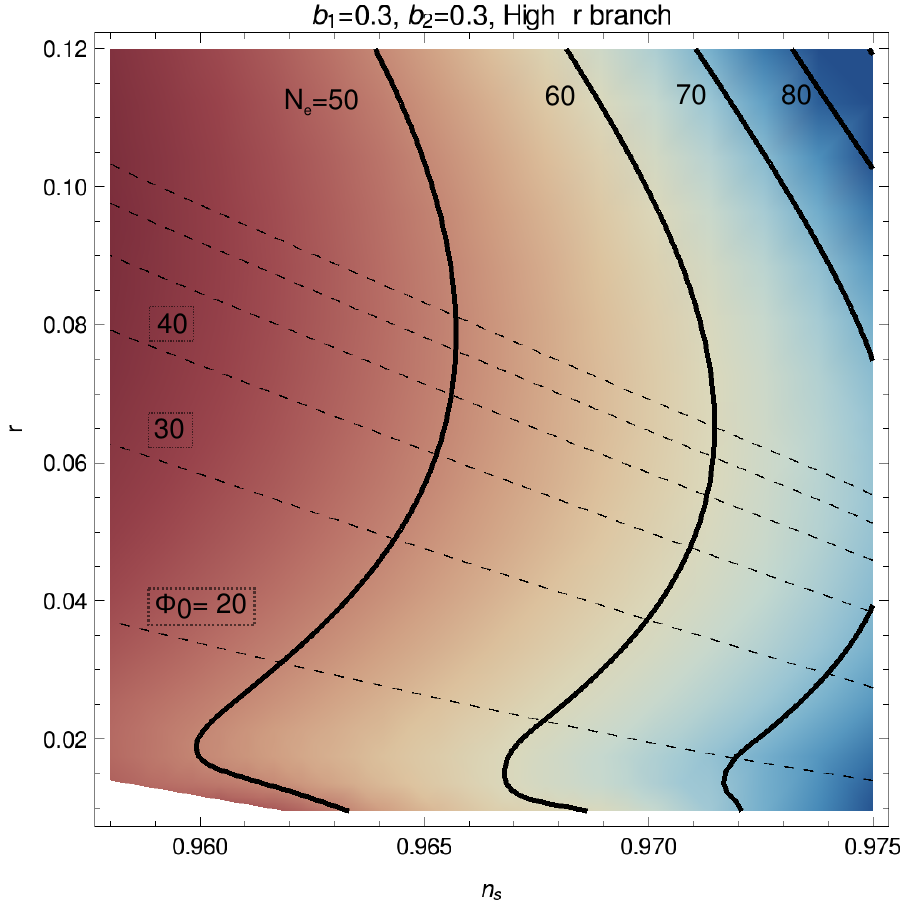}
\includegraphics[scale=0.71]{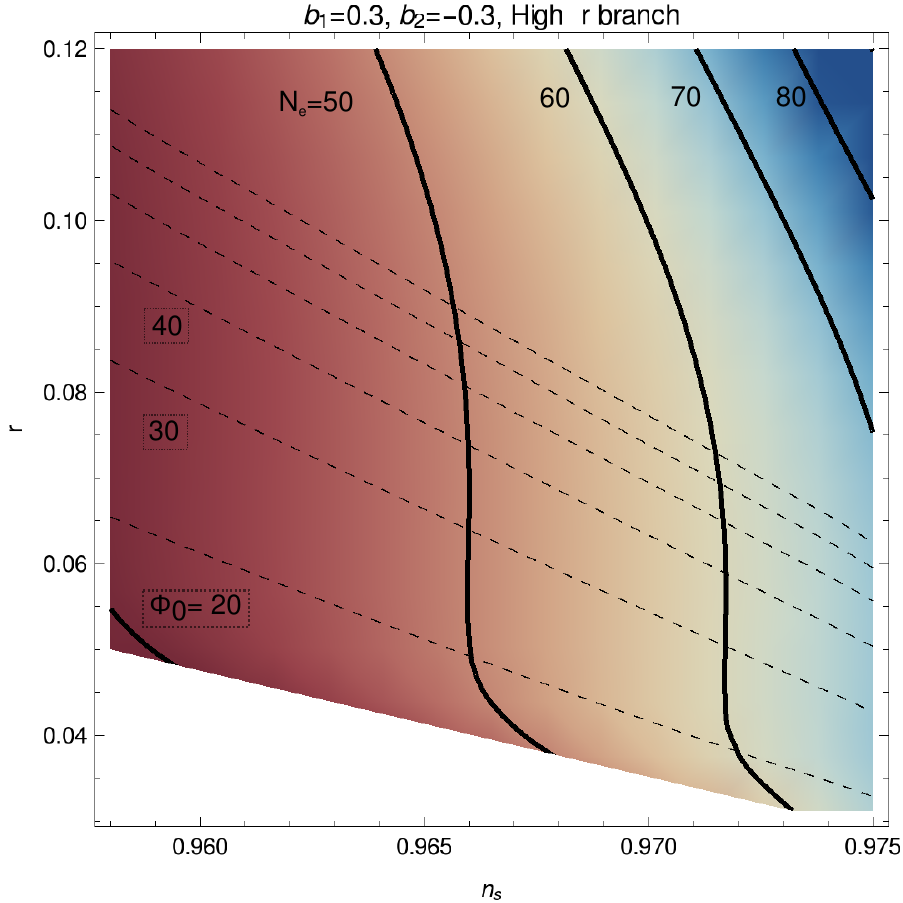}
\caption{\label{fig:quad_wide}Curves of constant $N_e$ (solid) and constant $\phi_0$ (dashed, with boxed labels) in quadratic plateau models, in terms of $r$ and $n_s$, for different choices of the deformation parameters $b_1,b_2$, and choosing the branch of solutions
giving smaller $\phi_i/\phi_0<1$.
The value of $A_s$ has been fixed to $2.142\times10^{-9}$, and the white regions are forbidden.}
  \end{figure}

\begin{figure}[t!]
\centering
\includegraphics[scale=0.8]{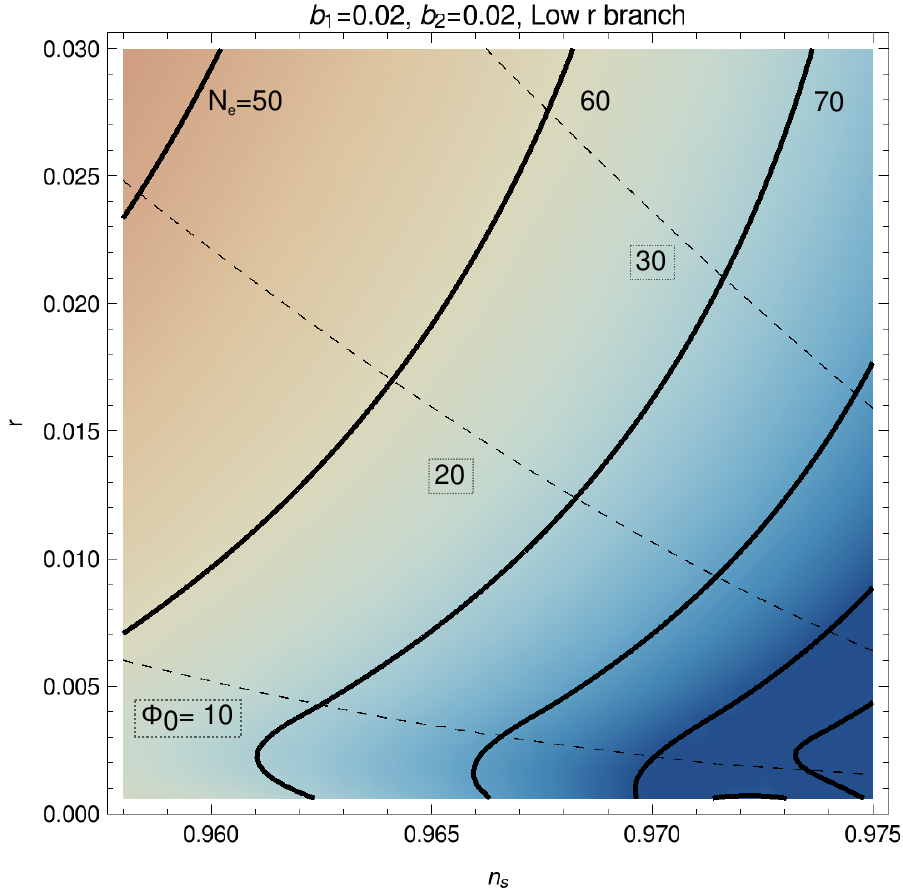}
\includegraphics[scale=0.8]{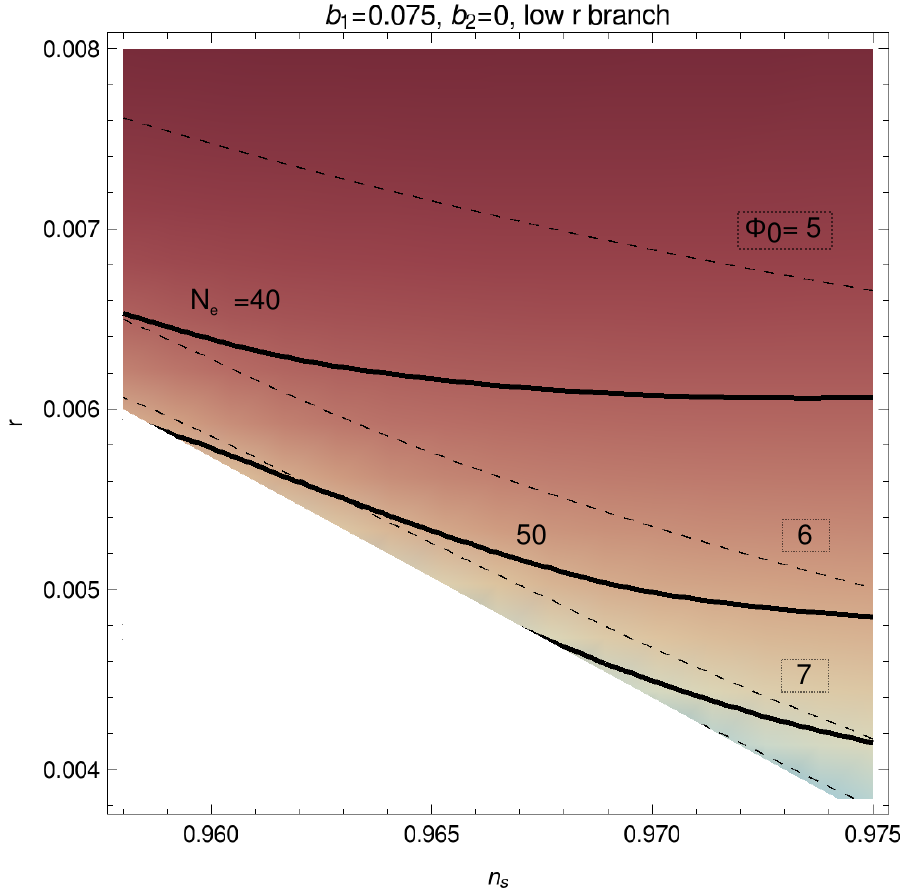}
\par 
\vspace{0.4cm}
\includegraphics[scale=0.8]{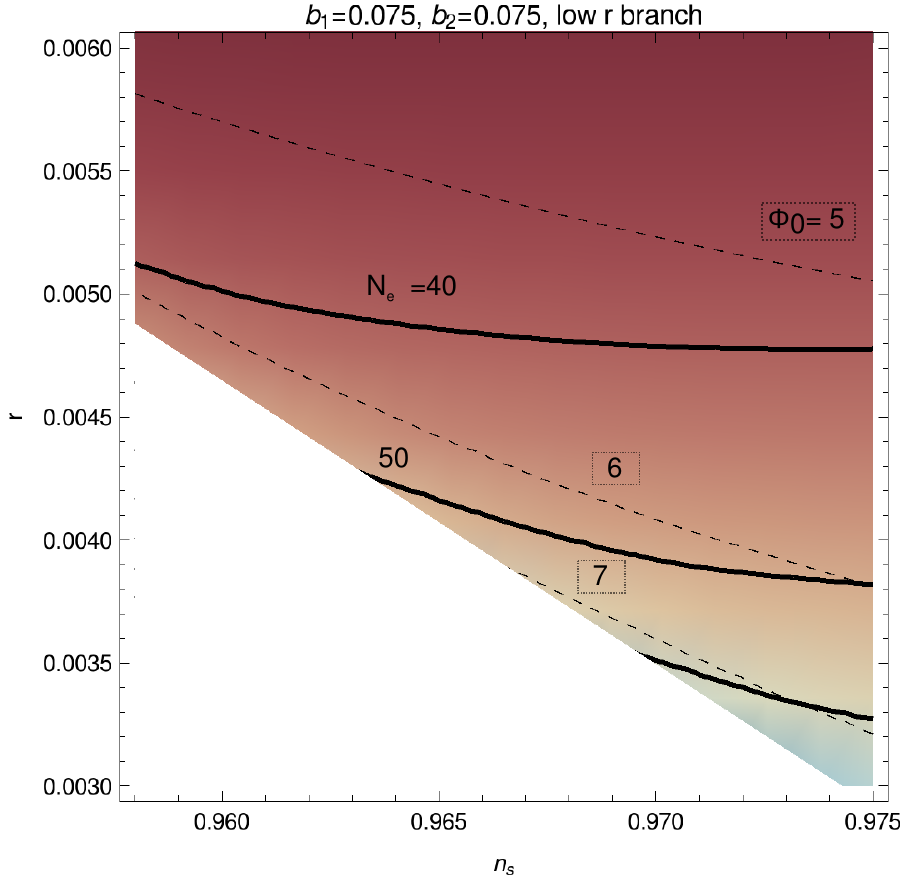}
\includegraphics[scale=0.8]{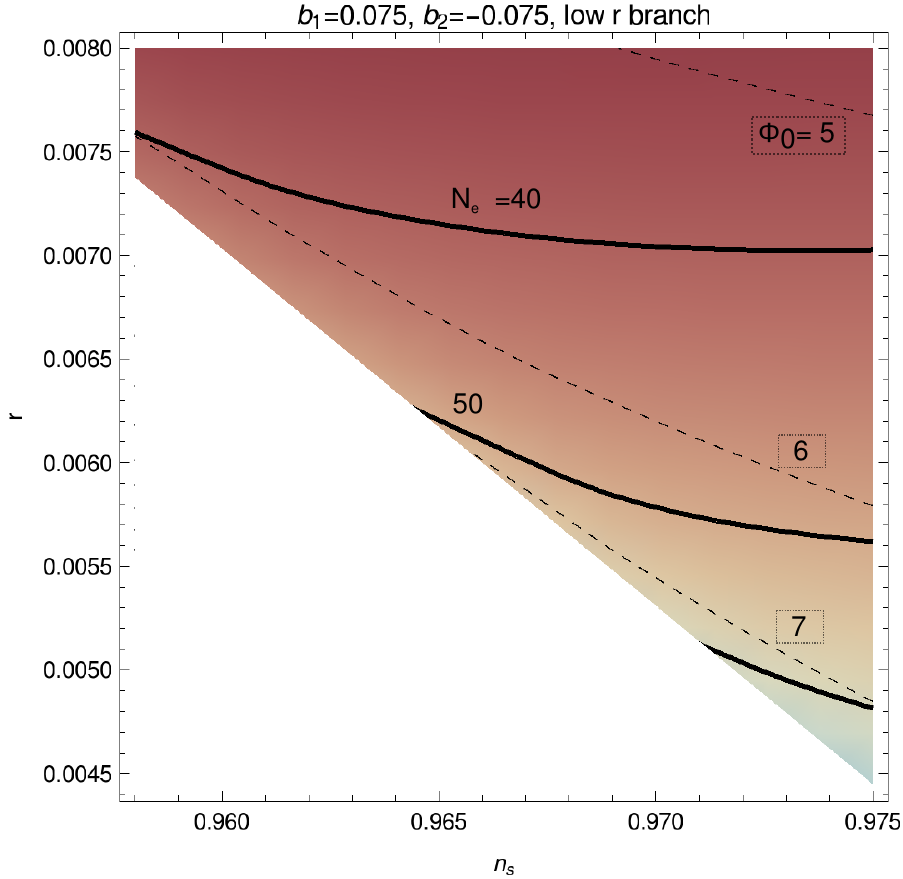}
\caption{\label{fig:quad_narrow} Curves of constant $N_e$ (solid) and constant $\phi_0$ (dashed, with boxed labels) in quadratic plateau models, in terms of $r$ and $n_s$, for different choices of the deformation parameters $b_1,b_2$, and choosing the branch of solutions
giving larger $\phi_i/\phi_0<1$.
The value of $A_s$ has been fixed to $2.142\times10^{-9}$. The white regions are forbidden}
  \end{figure}

We consider now the potential
\begin{align}
\label{quad}
  V(\phi)=\frac{m^2}{2}\left(1-\big(1-b_1\big)\log\frac{\phi^2}{\phi_0^2}+\frac{1}{2}\big(1+b_2\big)\left(\log\frac{\phi^2}{\phi_0^2}\right)^2+\cdots\right)\phi^2\,,
\end{align}
where, as in the quartic case, $b_1$ and $b_2$ parametrize the plateau's deformation. As before, we find two branches of solutions with $\phi_i<\phi_0$ for $b_1$ and $b_2$ different from zero. Again, one of the branches, which includes the $b_1=b_2=0$ case, has larger values of $r$, as shown in Fig.~\ref{fig:quad_wide}. In this case there is less tension with data than in the quartic plateaus, since a purely quadratic monomial predicts a lower $r$ than a quartic one. Successful inflation with $50<N_e<60$ is attainable with $b_1=b_2=0$. In analogy to the quartic case, increasing $b_1$ and lowering $b_2$ reduces the number of e-folds. The second branch of solutions, with larger values of $\phi_i/\phi_0$, is illustrated in Figure \ref{fig:quad_narrow}. In this branch, the values of $r$ are much lower than 0.1. The plateaus of both branches can easily reproduce the central values measured for $A_s$ and $n_s$ with $N_e$ around 50--60.

In the high-$r$ branch, the distance travelled by the inflaton for $40\lesssim N_e\lesssim 70$ is between 7--17 $M_P$ approximately, with inflation ending around $0.7M_P$, and $\phi_0$  greater than $\sim 8M_P$. In the low-$r$ branch, the typical  distance covered by $\phi$ is about 4--10 $M_P$, with inflation ending near $0.5 M_P$, and with $\phi_0>5M_P$. The values of $m^2$ needed for inflation are around $10^{-12}M_P$ in both branches. These are similar to those required in the standard $\phi^2$ monomial inflation, for which $r\simeq 8/N_e$ and $\phi_i/M_P\simeq 2 \sqrt{N_e}$, yielding $r\simeq0.13$ and $\phi\simeq 15 M_P$ with 60 e-folds of inflation. 

\begin{figure}[t!]\centering
 \includegraphics[scale=0.43]{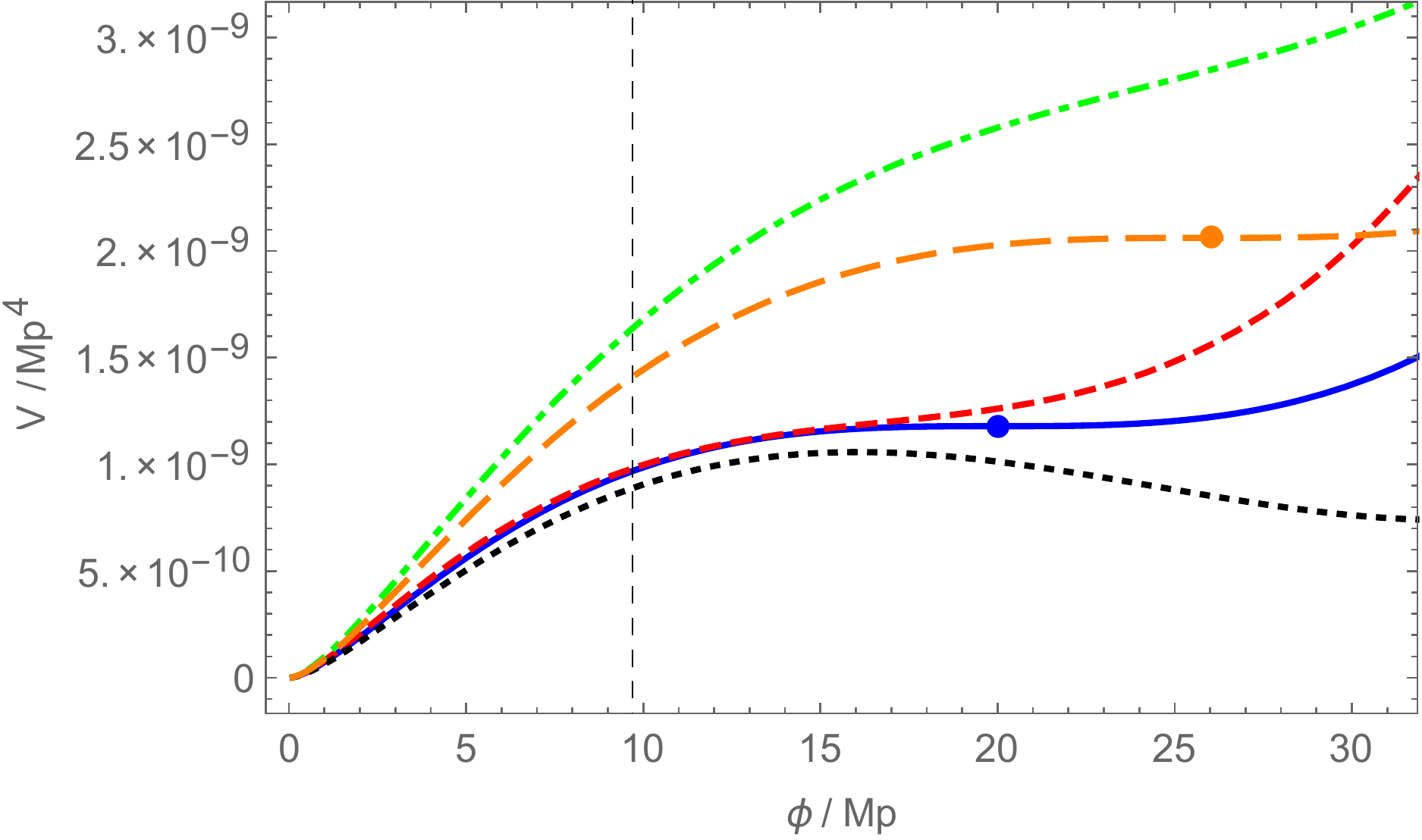}
\includegraphics[scale=0.43]{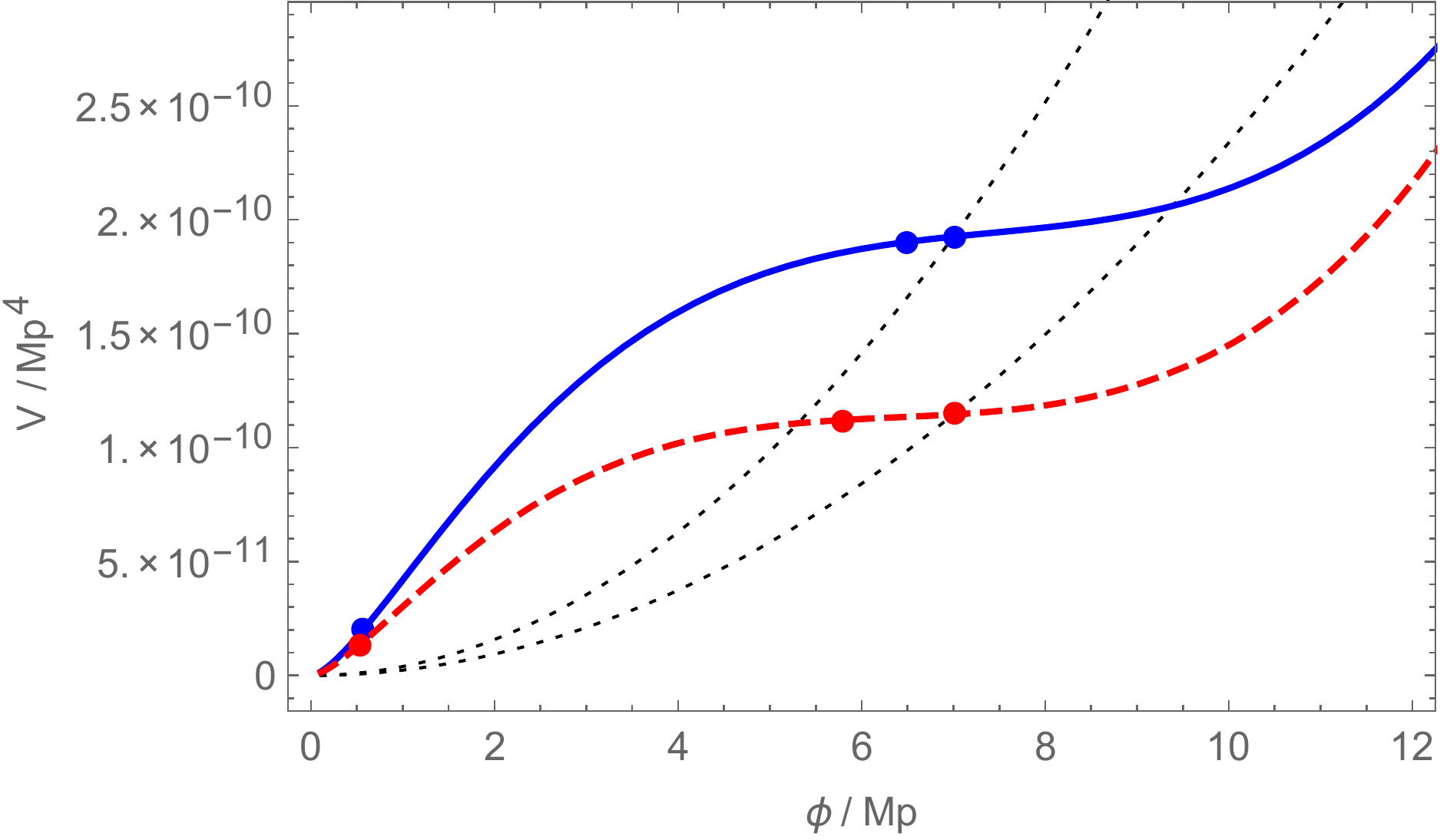}
\caption{\label{examplesquad} Examples of radiatively corrected quadratic inflation. The parameters of the potentials and their predictions for inflation are given in Table \ref{table2}. Left: potentials in the high $r$ branch illustrated in Figure \ref{examplesquart}. The vertical line (at 9.7 $M_P$) indicates the (approximate) common location of $\phi_i$. These potentials provide a comfortable fit to the data. The vertical line (at $9.7 M_P$)  indicates the (approximate) common location of $\phi_i$. The potentials have similar predictions independently of their shapes at higher field values. The dots in the exact plateau examples (continuous-blue (1) and orange long-dashed (5), see Table \ref{table2}) locate the inflection point $\phi_0$ for those cases. Right: two examples of potentials in the low $r$ branch, as in Figure \ref{phi4rns}. The dots indicate the field values $\phi_0$, $\phi_i$ and $\phi_e$ in each case, from right to left For comparison, black dotted lines represent the corresponding quartic monomials (without the radiative corrections). The potentials of both panels have $A_s(\phi_i)=2.13\times 10^{-9}$  and vanish at $\phi=0$. Both branches provide good fits to current CMB data.}
\end{figure}

As in the case of quartic potentials, the location of $\phi_i$ with respect to $\phi_0$ is qualitatively different in the two branches. While the existence of a plateau is not relevant in the high-$r$ branch (since it appears away from the region where inflaton happens), it is instead a key ingredient for inflation on the low-$r$ branch. This is can be seen explicitly comparing the left (high-$r$) and left (low-$r$) panels of Figure \ref{examplesquad}.

Radiative corrections leading to plateaus (even if they do not produce exact plateaus) help to render quadratic chaotic inflation compatible with the observations, without requiring a very fine tuning of the parameters at the two-loop level.  In Appendix \ref{reppot} we generalize the structure of quadratic (and quartic) plateaus to radiatively corrected potentials with coefficients of arbitrary sizes, highlighting the three independent parameters at order $\log^2$ in the expansion.

\begin{table}[t] \centering {\small
\begin{center}
\begin{tabular}{l*{10}{c}r}
$\#$ & $\phi_0/M_P$   &  $b_1$  & $b_2$  & $m^2/M_P^2\times 10^{12}$  & $n_s$ & $\alpha\times 10^{4}$ & $r$ & $\phi_i/M_P$ & $\phi_e/M_P$ & $N_e$  \\
\hline
{\color{blue} 1} {\footnotesize (continuous)} & 20 & 0 & 0 &  $5.89$  &  0.966 & $ -5.2$ & 0.031 & 9.69 & 0.66 & 56.8    \\
{\color{red} 2}  {\footnotesize(short-dashed)} & 20 & 0.2 & 0.1 & $6.30$  &  0.966  & $-5.8 $ & 0.031 & 9.38 & 0.65 & 55.1    \\
{\color{black} 3} {\footnotesize(dotted)}  & 22 & -0.3 & 0.2 &  $3.99$ & 0.966  & $-4.5 $ & 0.029 & 9.97 & 0.67 & 59.4 \\
{\color{green} 4} {\footnotesize(dot-dashed)} & 26 & 0.2 & -0.2 &  $8.43$ & 0.966  & $-6.6$ & 0.054 & 10.10 & 0.68 & 51.1 \\
{\color{amber} 5} {\footnotesize(long-dashed)} & 26 & 0 & 0 &  $6.10$ & 0.966  & $-6.0 $ & 0.046 & 10.09 & 0.68 & 53.1 \\
\hline
{\color{blue} 6} {\footnotesize (continuous)} & 7 & 0.075 & -0.2 & $7.86$  &  0.966 & $ -10.5$ & 0.0060 & 6.50 & 0.57 & 55.4    \\
{\color{red} 7} {\footnotesize (dashed)} & 7 & 0.075 & 0.1 & $4.68$  &  0.966 & $ -12.5$ & 0.0036 & 5.79 & 0.55 & 55.1    \\
\end{tabular}
\caption{ Examples radiatively corrected quadratic potentials. The potentials 1--5 and 6, 7 correspond to the left and right panels of Figure \ref{examplesquad}, respectively. The primordial functions $n_s$, $r$ and $\alpha$ are evaluated at $\phi_i$. In all cases, $A_s(\phi_i)=2.13\times 10^{-9}$. The mass squared $m^2$ and the deformation parameters $b_1$ and $b_2$  are given at the plateau scale $\phi_0$. The quantity $N_e$ denotes the number of e-folds from $\phi_i$ to $\phi_e$, where inflation ends. } \label{table2}
\end{center}
}
\end{table}

\section{\label{pp}Particle physics realizations}

In this section we identify particle physics models that can support quartic \eqref{potanalytic} or quadratic \eqref{potanalyticquad} plateaus. In order to do so, we look for (perturbative) solutions of the plateau conditions \eqref{conds2} or \eqref{condsquad} with specific Lagrangians. These equations  indicate that, if one strives for accuracy, the three coefficients $\lambda(\phi_0)$ (or $m^2(\phi_0)$ in the quadratic case), $c_1(\phi_0)$ and $c_2(\phi_0)$ should in principle be computed at the same loop level. Unfortunately, the one-loop approximation is insufficient because $c_2$ is zero at that order. Moreover, obtaining the two-loop effective potential in realistic models with several fields can be rather daunting. Nevertheless, for the purpose of  proving the viability of a plateau in a given particle physics model, it is sufficient to use the one-loop renormalization group (RG) improvement of the tree-level potential. This approximation comes at the cost of a lower precision in the description of the plateau, but does not undermine our goal of identifying the models that can support one. We recall that the one-loop RG-improvement of the tree-level potential, corresponds to a resummation of the potential at leading log order, i.e.\ the $N$-th $\log$ is rendered with $N$ loops, and therefore each coefficient of the resummed potential comes with a different precision.

Since the plateau conditions are two equations, they can be used to express two of the couplings of a Lagrangian (that has been shown previously to be able to support a plateau) in terms of the rest of the couplings.\footnote{If we wanted, we could solve one of the equations for the plateau scale $\phi_0$ instead of for one coupling. Then, the other equation would give us a coupling $\gamma_p$ as a function of $\phi_0$ and the rest of the couplings. As in the familiar case of the SM, in which $\gamma_p$ can be chosen as the top mass, once all couplings except $\gamma_p$ are fixed at an arbitrary scale, the plateau is achieved by tuning  $\gamma_p$, with $\phi_0$ fixed by the plateau conditions. Allowing for deformations of the plateau, i.e.\ solving the plateau conditions only approximately, alleviates the tuning.} As we will soon see, in order to obtain the first non-vanishing contributions to these two couplings at the plateau scale, we need to use the one-loop potential and the one-loop beta functions. Getting both couplings with the same level of precision requires the two-loop improvement of the one-loop effective potential, which gives each coefficient of the logarithmic expansion at next-to-leading log order \cite{Bando:1992np}, i.e.\ the $N$-th $\log$ in the effective potential is given with $N+1$ loop precision.

Once we have chosen a specific particle physics model, we may look for solutions of the plateau conditions \eqref{conds2} or \eqref{condsquad} in a perturbative expansion. For this it is convenient to write the couplings $\gamma_i$ of the Lagrangian\footnote{In practice, at the end of our computations, we will only express in this way the two couplings of the Lagrangian that we choose to solve for and the effective quartic or quadratic coupling of the potential.}  as a formal series in $\kappa=1/(16\pi^2)$, which helps to keep track of the perturbative order of the expansion:\footnote{An $N$-th loop term in a loop expansion of the potential comes with a factor $\kappa^N$ and, analogously, each contribution to a beta function coming from an $N$-loop term appears with the same power of $\kappa$.}
\begin{align}
\label{loopexpansion}
 \gamma_i=\gamma_i^{(0)}+\kappa\gamma_i^{(1)}+\kappa^2\gamma_i^{(2)}+\cdots\,.
\end{align}
In this expression  the couplings of the model are understood to be evaluated  at the scale  $\mu_0=\varepsilon\phi_0$. For simplicity, we remove $\mu_0$ from the notation, but it has to be remembered that the conditions that we will obtain on them refer only to their values at this scale. We also remind the reader that, as discussed in  Section \ref{plateaugeneral}, the effective quartic coupling and its derivatives: $\lambda(\phi_0)$, $\beta_\lambda(\phi_0)$ and $\beta^\prime_\lambda(\phi_0)$ are defined at $\phi_0$ by construction, but in the following we will denote them  simply as $\lambda$, $\beta_\lambda$ and $\beta'_\lambda$.\footnote{At any other scale, the RG equations (even at just one-loop order) generically mix the coefficients of different orders in $\kappa$. Therefore, care must be taken to avoid misinterpreting the expression \eq{loopexpansion} as an expansion of the couplings in loops.} We use the same notation with superscripts to expand the beta functions of the couplings:
\begin{align}
\beta_{\gamma_i}  =\frac{\partial \gamma_i}{\partial \log\mu}= \beta_{\gamma_i}^{(0)}+\kappa\beta_{\gamma_i}^{(1)}+\kappa^2\beta_{\gamma_i}^{(2)}+\cdots\,,
\end{align}
where each $\beta_{\gamma_i}^{(N)}$ can be written as a function of the $\gamma_i^{(M)}$. We can express the beta function $\beta_{\gamma_i}$ as a sum of its loop contributions: $\beta_{\gamma_i}=\beta_{\gamma_i}^{1l}+ \beta_{\gamma_i}^{2l}+\cdots$, where $\beta_{\gamma_i}^{nl}$ carries $n$ powers of $\kappa$. Then, expanding in Taylor series:
\begin{align}
\label{loopexpansionb} 
\beta_{\gamma_i}\left[\gamma_q^{(0)}+\kappa\gamma_q^{(1)}+\cdots\right]= \beta_{\gamma_i}\left[\gamma_q^{(0)}\right]+\kappa\sum_j \frac{\partial \beta_{\gamma_i}}{\partial \gamma_j}\Bigg{|}_{\gamma_q^{(0)}}\gamma_j^{(1)}+\cdots
\end{align}
we obtain 
\begin{align} \label{betaexpans}
\beta^{(0)}_{\gamma_i}=0\,,\quad\beta^{(1)}_{\gamma_i} = \beta^{1l}_{\gamma_i}\left[\gamma_q^{(0)}\right]\,,\quad \beta^{(2)}_{\gamma_i}= \beta^{2l}_{\gamma_i}\left[\gamma_q^{(0)}\right]+\sum_j\frac{\partial \beta_{\gamma_i}^{1l}}{\partial \gamma_j}\Bigg{|}_{\gamma_q^{(0)}}\gamma_j^{(1)}\,,\quad\ldots\,.
\end{align}
Similarly, we can expand in powers of $\kappa$ the effective quartic coupling, $\lambda$, and the effective mass squared, $m^2$, of the potentials \eq{lambdaeff} and \eq{Pq}. For concreteness, we will focus now on a quartic plateau; but analogous arguments apply to the quadratic case. We define $\lambda^{(N)}$ through the relation
\begin{align} \label{expeff1}
\lambda = \lambda^{(0)}+\kappa \lambda^{(1)}+\kappa^2 \lambda^{(2)}+\cdots\,.
\end{align}
We stress that the coupling $\lambda$  is different from the nominal quartic coupling of the Lagrangian, which we denote by $\hat\lambda$ and which we can also expand as
\begin{align} \label{expeff}
\hat\lambda = \hat\lambda^{(0)}+\kappa \hat\lambda^{(1)}+\kappa^2 \hat\lambda^{(2)}+\cdots\,.
\end{align}
The coupling $\hat\lambda$ is by definition just one among all the $\gamma_i$ introduced earlier.  At lowest order in $\kappa$, the two couplings, $\lambda$ and $\hat \lambda$, coincide, i.e.\ $\lambda^{(0)} = \hat\lambda^{(0)}$, but they differ at linear order in $\kappa$ and beyond. 

As discussed in Section \ref{plateaugeneral}, the coefficients $c_1(\phi_0)$ and $c_2(\phi_0)$ of the radiatively corrected quartic potential are related to the beta function of the effective quartic coupling and its derivative with respect to the renormalization scale $\mu$, so the plateau conditions \eqref{conds2} can be rewritten as
\begin{align}
\label{conds4}
 \beta_\lambda(\phi_0)=-4\lambda(\phi_0),\quad \beta'_\lambda(\phi_0)=-4\beta_\lambda(\phi_0).
\end{align}
These two equations can in principle be used to determine two of the couplings of any model in terms of the rest of the couplings. In what follows, we will drop again $\phi_0$ from the notation
for brevity.

Since any beta function $\beta_{\gamma_i}=\partial\gamma_i/\partial \log \mu$ is zero at order $\kappa^0$ by definition, the quantity $\beta_\lambda^{(1)}$ is just a function of  the lowest order couplings $\gamma^{(0)}_i$, whereas $\beta_\lambda^{(2)}$ depends on both $\gamma^{(0)}_i$ and $\gamma^{(1)}_i$, see \eq{betaexpans}. Analogously, $\beta'_\lambda$ is zero at order $\kappa$ (and $\kappa^0$) because $\beta'_\lambda=\partial \beta_\lambda/\partial \log \mu=\sum_i\beta_{\gamma_i}\, \partial\beta_\lambda /\partial \gamma_i$\,. Then, using $\beta_\lambda^{\prime\,(1)}=0$ in \eqref{conds4} we obtain that $ \beta^{(1)}_\lambda=0$ and thus $\lambda^{(1)}=0$. This enforces an algebraic relation among all the zeroth order couplings $\gamma_i^{(0)}$. Similarly, the first equation in \eqref{conds4} implies $\lambda^{(0)}=0$, because $\beta_\lambda^{(0)}=0$, as we have just explained. Moreover, since the tree-level quartic coupling $\hat\lambda^{(0)}$ coincides with the zeroth order effective quartic coupling $\lambda^{(0)}$, we conclude that also $\hat\lambda^{(0)}=0$. As $\hat\lambda$ and $\lambda$ differ already at order $\kappa$, their beta functions can start to be different only at order $\kappa^2$, and so $\beta_\lambda^{(1)}=\beta_{\hat\lambda}^{(1)}=0$.

The conditions \eq{conds4} imply as well that $\beta'_\lambda=16 \lambda$, that in turn leads to $ {\beta'}^{(2)}_\lambda=16\lambda^{(2)}$, which is the first non-zero contribution to the effective quartic coupling. In addition, given that ${\beta'}_\lambda$ is zero below the order $\kappa^2$, we know that ${\beta'}^{(2)}_\lambda$ depends only on the lowest order couplings  $\gamma_i^{(0)}$. Besides, the derivatives $\beta'_\lambda$ and  $\beta'_{\hat\lambda}$ differ only by terms of order $\kappa^3$ and higher, which means that ${\beta'}^{(2)}_\lambda={\beta'}^{(2)}_{\hat\lambda}$. Furthermore, ${\beta'}_{\hat\lambda}=\partial \beta_{\hat\lambda}/\partial \log \mu=\sum_i\beta_{\gamma_i}\, \partial\beta_{\hat\lambda} /\partial \gamma_i$ and so, using $ {\beta'}^{(2)}_\lambda=16\lambda^{(2)}$,  we see that  the one-loop beta functions $\beta^{(1)}_{\gamma_i}$ are sufficient to compute the first (non-zero) contribution to the effective quartic coupling at the plateau. This amounts to a one-loop RG-improvement of the tree-level potential, which captures  $c_2(\phi_0)=\beta'_\lambda(\phi_0)$ and $c_1(\phi_0)=\beta_\lambda(\phi_0)$   with leading log  precision, i.e. at orders $\kappa^2$ and $\kappa$, respectively.

The previous discussion shows that in order to establish whether a model admits a plateau it is necessary to check that  $\beta_\lambda^{(1)}=0$ should have a non-trivial solutions for which $\lambda^{(2)}={\beta'}^{(2)}_\lambda$/16 is positive. The tree-level effective potential plus the one-loop beta functions are sufficient for this purpose. However, describing the plateau in terms of the couplings of the Lagrangian, requires additional corrections to the potential or the beta functions, as discussed next. 

As we mentioned before, the equations \eq{conds4} allow to express two couplings of the model in terms of the rest of them. For convenience, we select one to be $\hat\lambda$ (or $\hat m^2$ in the quartic case) and the other can be arbitrarily chosen. In order to obtain the first non-zero contribution to the tree-level quartic coupling that is needed for a plateau, $\hat\lambda^{(1)}$, we can use the equation $\lambda^{(1)}=0$. This requires knowing the order $\kappa$ relation between $\hat\lambda$ and $\lambda$, which can be obtained from the one-loop effective potential.  The second equation in \eqref{conds4}  involves $\beta^{(2)}_\lambda$, which depends  on the zeroth and first order contributions to the couplings. This second equation allows to express the order $\kappa$ contribution to any coupling we choose in terms of the rest. Since $\hat\lambda^{(1)}$ is already fixed by $\lambda^{(1)}=0$, we can then express through that equation any other coupling at order $\kappa$. This requires knowledge of the beta functions at two-loops.

In concrete models, in order to identify the effective quartic coupling at the plateau scale, one can proceed by performing a large field expansion of the one-loop Coleman-Weinberg potential, keeping only the (logarithmically corrected) quartic terms and setting $\mu=\varepsilon\, \phi$, see Section \ref{plateaugeneral}. With this choice, the logarithmic corrections get resummed in the running of the couplings. This allows to identify immediately the effective coupling $\lambda$ from the coefficient of the surviving $\phi^4$ term.  In the quadratic case it is necessary to impose that the effective quartic and cubic terms are suppressed, and again, setting $\mu\propto\phi$ allows to identify the effective mass parameters $m^2$.  We consider quartic plateaus in \S~\ref{subquarticpp}, while quadratic plateaus will be studied in \S~\ref{subquadpp}. In all the examples that we study below, we provide the expression for $\lambda^{(2)}$ (or $m^{2(2)}$), which is the first non-vanishing contribution to the quartic (or quadratic) effective coupling at the plateau. We recall that the plateau conditions generically allow to express two of the couplings of the Lagrangian in terms of the rest, which can then be regarded as input parameters. Therefore, in the examples that follow we also give the expressions at the plateau of  both $\hat\lambda$ (or $\hat m^2$) and another coupling (that we choose conveniently in each case). 

\subsection{\label{subquarticpp}Quartic inflation}

Here we discuss the minimum ingredients for  models in which a quartic plateau \eqref{potanalytic} can be realized. As we just explained, the minimum necessary conditions for a plateau (that is exact at two-loops) are $\beta^{(1)}_\lambda=0$ and, $\beta'^{(2)}_\lambda>0$. The first of these conditions can be relaxed to $\beta^{(1)}_\lambda\simeq 0$ for approximate (i.e.\ deformed) plateaus, but the second is always necessary.  We start by considering a generic model in which the (scalar) inflaton $\phi$ couples to Weyl fermions with Yukawa couplings $\delta_a$ ($a=1,\ldots, N$) and to gauge fields with couplings $g_A$ ($A=1,\ldots,N_g$), as well as to other scalars. We will denote scalar quartic couplings of the Lagrangian --either self-interactions or mixed ones-- as  $\hat\lambda_\alpha$ ($\alpha=0,\ldots,N_s$). In particular, the coupling of the quartic self-interaction of the inflaton will be $\hat\lambda_{0}\equiv\hat\lambda$, following the notation of the previous section. Assuming that that cubic interactions can be forbidden or ignored, the one-loop beta function of the effective quartic coupling is of the general form:
\begin{align}
\label{betashape}
 \beta_{\hat\lambda}&=\sum_{a,b,c,d}^{N} \Delta^\delta_{abcd}+\sum_{A,B,C,D}^{N_g} \Delta^{g}_{ABCD}+\sum_{\alpha,\beta}^{N_s} \Delta^{\hat\lambda}_{\alpha\beta}+4\hat\lambda\gamma_\phi\,,
\end{align}
where the fermionic, gauge, and scalar quartic contributions are
\begin{equation}
\begin{aligned} \label{signs}
\Delta^\delta_{abcd}&=c_{abcd}\,\delta_a \delta_b\delta_c\delta_d\,,\quad c_{abcd}< 0,\\
\Delta^g_{ABCD}&=c_{ABCD}\,g_A g_Bg_Cg_D\,,\quad c_{ABCD}>0,\\
\Delta^{\hat\lambda}_{\alpha\beta}&=c_{\alpha\beta}\,\hat\lambda_\alpha\hat\lambda_\beta\,,\quad c_{\alpha\beta}>0.
\end{aligned}
\end{equation}
We are implicitly assuming a mass-independent subtraction scheme such as $\overline{\rm MS}$, which is the concrete scheme that we use in all the examples that follow. Then, the beta functions are given by polynomials of positive powers of the masses and couplings (having the appropriate mass dimension). This implies in particular that the beta functions of quartic couplings, which are dimensionless, can only depend on dimensionless couplings. 

In order to understand the signs of the different contributions to \eq{betashape} it is useful to notice that the freedom of redefining the fermion fields through a rotation of their phase, allows to choose the Yukawa couplings to be positive. In fact, the phases of fermion fields do not affect the beta functions of the scalar quartic couplings and masses. This is so because in a diagram with  closed  loops of Weyl fermions, in which the fermionic phases only enter through the Yukawa couplings and the fermion propagators, the rephasing of one fermion field produces anti-correlated changes between either a pair of Yukawa couplings or a Yukawa coupling and a fermion propagator, giving a net zero effect. Regarding the gauge couplings, they can always be taken positive, and this choice is radiatively stable because the beta functions of the gauge couplings are proportional to the couplings themselves. The quartic couplings are assumed to be non-negative, which ensures stability of the tree-level potential in all field directions. 

\begin{figure}
\hspace{1cm}
 \includegraphics[trim=10 10 10 10, clip, scale=0.25]{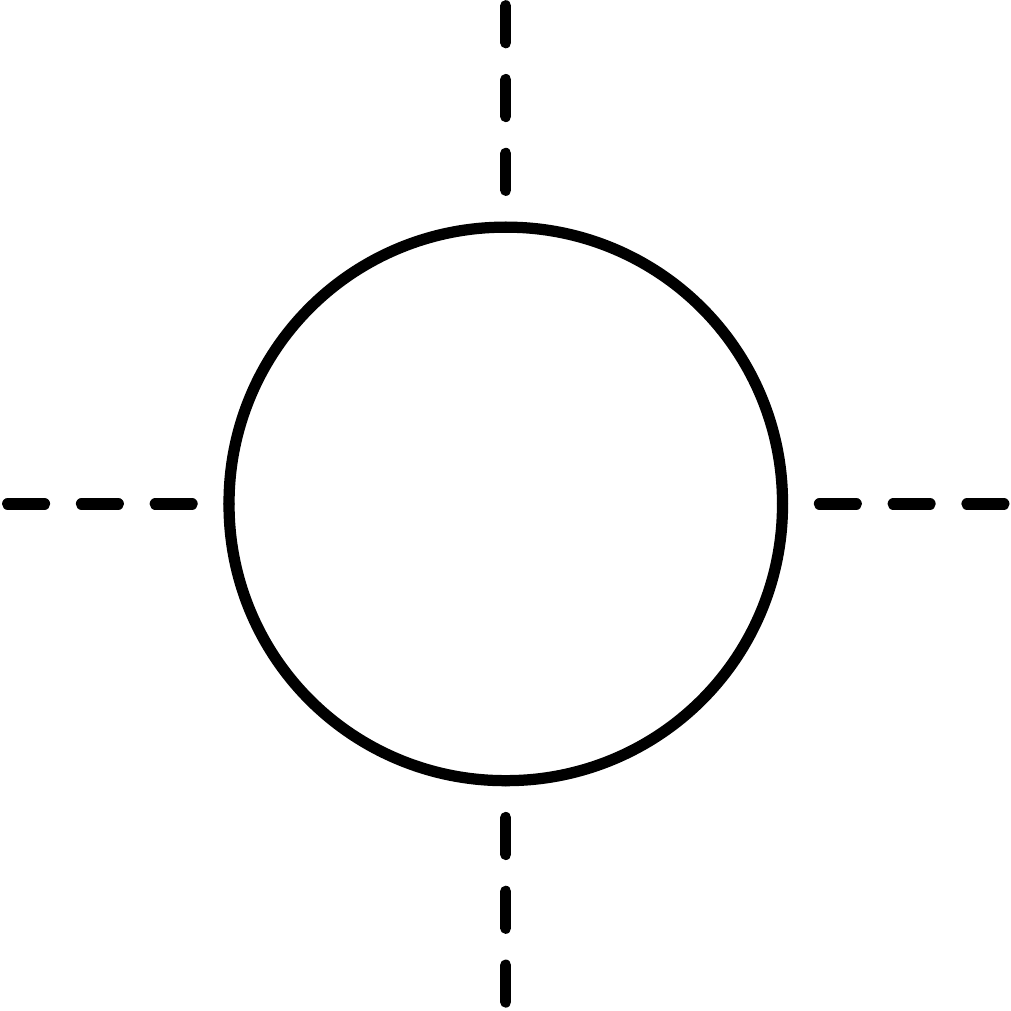}\hspace{1.5cm}
    \includegraphics[trim=10 0 10 0, clip, scale=0.25]{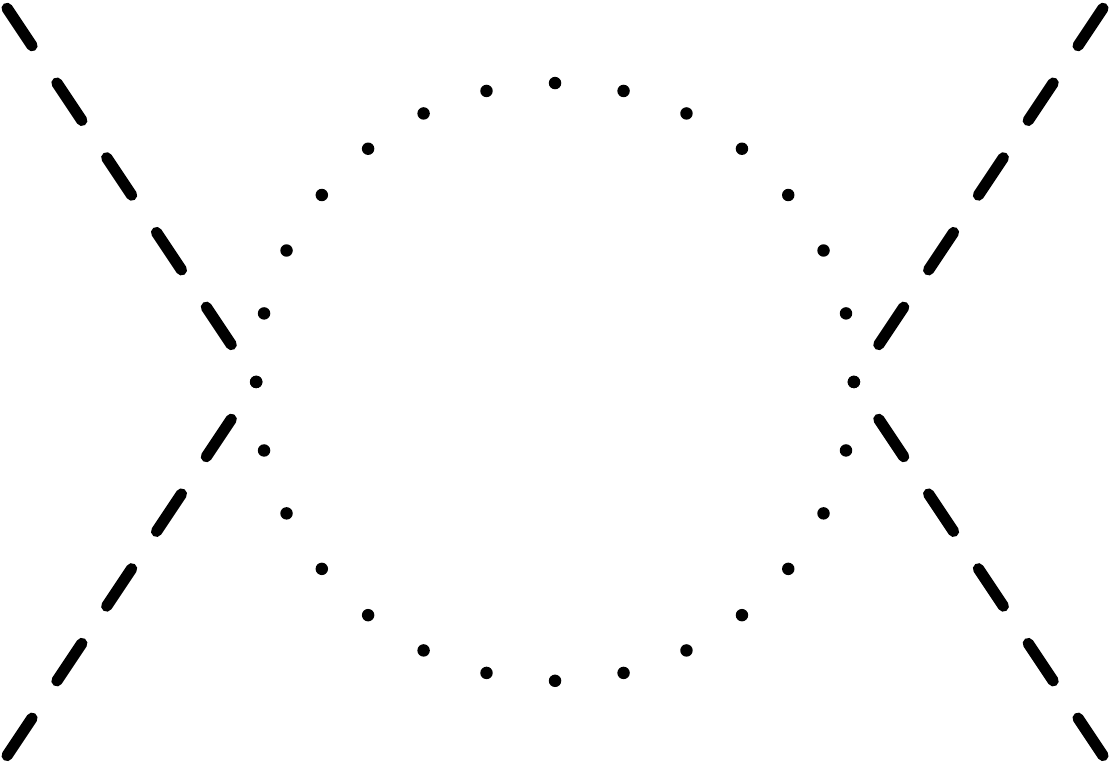}\hspace{1cm}
\includegraphics[trim=10 0 10 0, clip, scale=0.25]{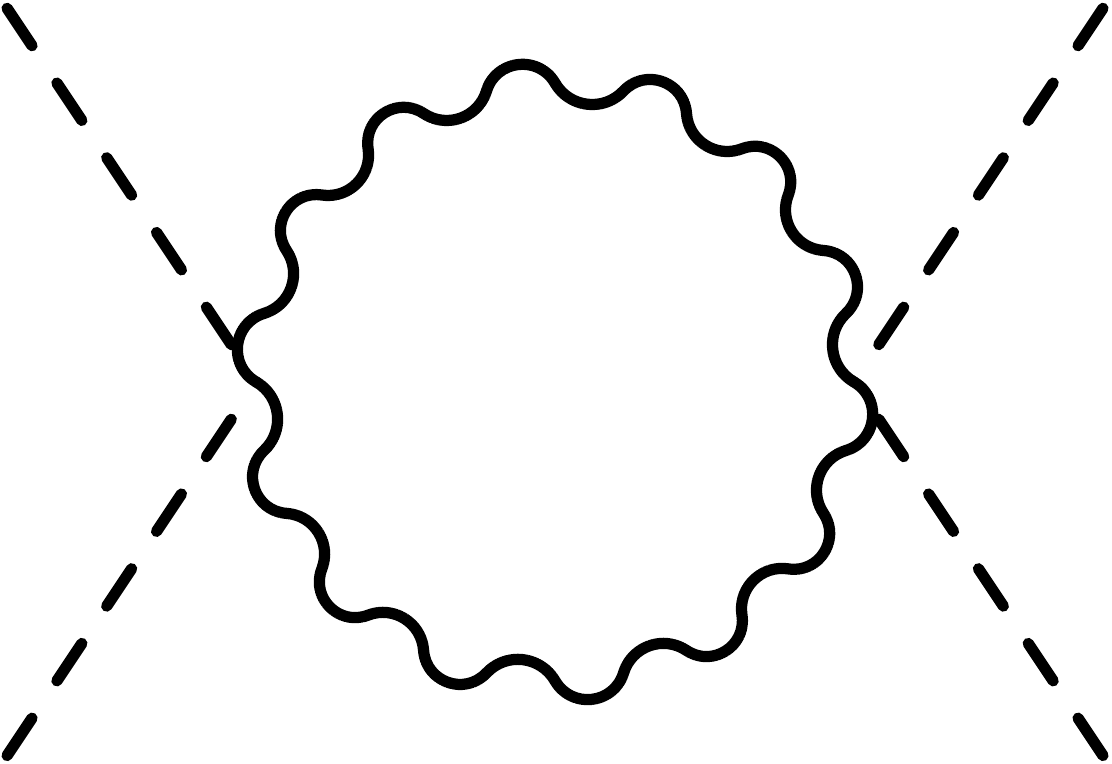}\hspace{1cm}
        \includegraphics[trim=10 0 10 0, clip, scale=0.27]{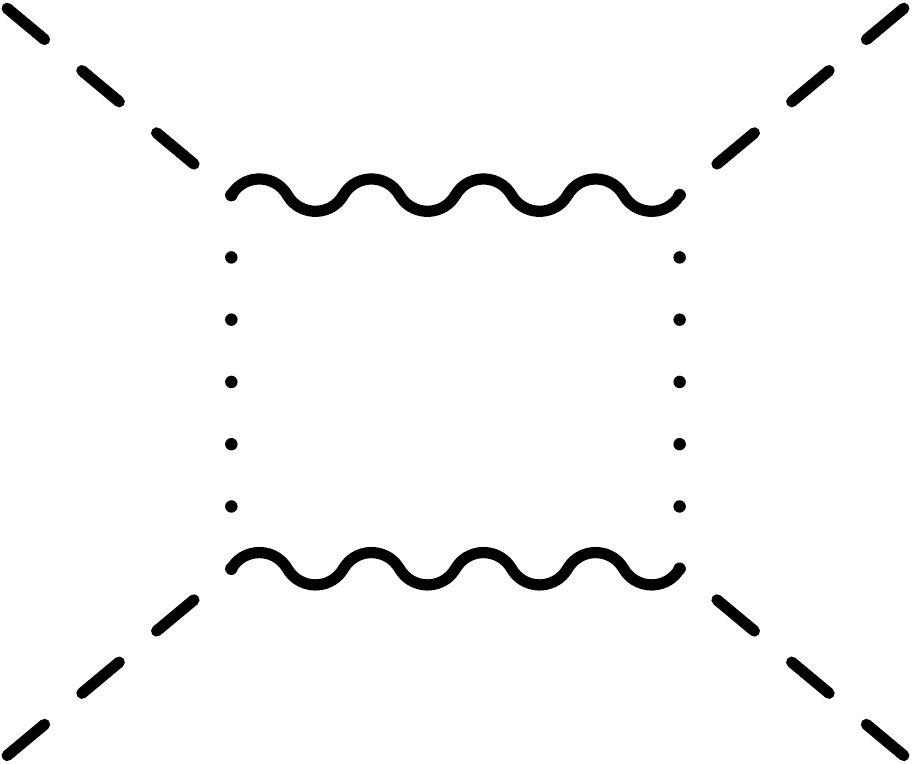}
\caption{\label{Floop} One-loop contributions to $\beta_{\hat\lambda}$. The lines represent: fermions (continuous), scalars (dotted) and gauge bosons (wiggled). Each inflaton leg is represented as a dashed line. The dotted lines inside the loops represent generic scalar fields, including the inflaton as well.}
  \end{figure}

The various types of diagrams that contribute to $ \beta_{\hat\lambda}$ at one-loop are illustrated in Figure \ref{Floop}. Given the signs of the different contributions, if the inflaton only coupled to fermions the potential would become unstable at large field values,\footnote{A model consisting of a singlet inflaton and singlet fermions  has been studied in the literature before  \cite{NeferSenoguz:2008nn,Okada:2014lxa,Enqvist:2013eua}. There, the fermions may give rise to hilltop-like inflation with an unstable potential.} whereas if it only coupled to bosons it would be monotonically increasing. In any of the two cases a plateau would be impossible. In addition, since the condition $\hat\lambda^{(0)}=0$ must be satisfied, the terms $\Delta^{\hat\lambda}_{\alpha\beta}$ that receive contributions from the inflaton's self-coupling vanish at leading order in $\kappa$. The same happens with the term that depends on the anomalous dimension $\gamma_\phi$. Therefore, we can already conclude that a plateau requires that the inflaton couples to fermions and also to gauge fields or additional scalars (different from the inflaton itself). In the second case, the couplings between the new scalars and the fermions have to be small enough, so that the quartic self-couplings of the scalars are not driven to negative values, in such a way that the contributions $\Delta^{\hat\lambda}_{\alpha\beta}$ remain non-negative.

We then have to look for solutions of the equation $\beta_{\hat\lambda}=0$, using the expression \eqref{betashape}, with all the couplings replaced by their contribution at order $\kappa^0$, and imposing in particular that $\hat\lambda^{(0)}=0$. The second  condition necessary for the plateau is  that $\beta'^{(2)}_\lambda>0$. Since ${\beta'}^{(2)}_{\hat\lambda}$ only depends on the  values of the couplings at order $\kappa^0$, the condition can already be checked with the solutions at order $\kappa^0$ obtained from the equation $\beta_{\hat\lambda}^{(1)}=0$.

Given that $\hat\lambda^{(0)}=\beta_\lambda^{(1)}=0$ at the plateau, the dominant contributions to ${\beta'_\lambda}^{(2)}$ come from three different kinds of terms:
\begin{align}
  \label{betapcont}\beta_{\delta_e}\frac{\partial \Delta^\delta_{abcd}}{\partial_{\delta_e}}\,,\quad
\beta_{g_E}\frac{\partial \Delta^{g}_{ABCD}}{\partial_{g_E}}\,,\quad
\beta_{\hat\lambda_\mu}\frac{\partial \Delta^{\hat\lambda}_{\alpha\beta}}{\partial_{\hat\lambda_\mu}}\,.
\end{align}
If the fermions are not charged under a gauge group, the first of these terms is  negative. This follows from the fact that $\Delta^\delta_{abcd}$ involves products of four Yukawas with a negative coefficient,\footnote{See equation \eqref{betashape}.} and because if the fermions are gauge singlets the beta functions $\beta_{\delta_e}$ involve cubic powers of the Yukawas with positive coefficients. However, if the fermions are charged under a gauge group, the functions $\beta_{\delta e}$ receive negative contributions involving the gauge couplings, so that the net contribution to $\beta'_\lambda$ may be positive. The second type of contributions in \eqref{betapcont} are positive if the inflaton is charged under a $U(1)$ gauge group, (otherwise they are negative for non-Abelian groups) because $\Delta^{g}_{ABCD}>0$ and $\beta_{g_E}>0$. Finally, the last class of contributions will be positive if $\beta_{\hat\lambda_\alpha}>0$. As mentioned before, the terms of this third contribution will be suppressed (or vanishing in an exact plateau) whenever $\hat\lambda_0$ (corresponding to  the inflaton's self coupling) is involved. However, additional scalars interacting with the inflaton give positive contributions provided that the beta functions of their associated quartic couplings remain positive, which will happen if they do not couple to fermions or do so weakly.\footnote{The beta functions of the quartic couplings of the additional scalars will have the same form as \eqref{betashape}, so that negative contributions will be suppressed if the couplings to fermions are small.}

From the previous discussion we conclude that  in order to achieve a plateau it is sufficient that the inflaton couples to fermions and either:
\begin{enumerate}
 \item[1)]  {\bf another scalar with weak fermionic couplings}, or
 \item[2)]  {\bf a $U(1)$ gauge group}.
\end{enumerate} 
In any of the two cases, a unique Weyl fermion  can be sufficient if it is a singlet or if it belongs to a real representation of the gauge group. Otherwise, anomaly cancellation demands several fermionic species.\footnote{Note that in the $U(1)$ case, a Weyl fermion in a real representation can be written in terms of multiple fermions with different charges.} We will now look into concrete examples of the cases 1) and 2) separately.

\subsubsection{\label{subsubsf}A singlet inflaton coupled to the Higgs and new fermions}
The models of the first kind that can develop a plateau consist of a singlet inflaton coupled to an additional scalar and one or more Weyl fermions. We will consider here the possibility of charging the fermions under a $U(1)$ gauge group, although this is not a necessary condition for achieving a plateau. The reason is that this slight complication can help to satisfy the plateau equations because the destabilizing effect of the Yukawa couplings on the running of the effective quartic coupling of the inflaton may be tamed or compensated by the gauge coupling. This is because fermions charged under a $U(1)$ give positive contributions to $\beta'_\lambda$.

Motivated by the requirement of reheating the universe after inflation and aiming to minimize the number of fields beyond those of the SM, we choose the extra scalar to be the SM Higgs and the $U(1)$ group as the SM hypercharge. We then consider $N$ pairs of Weyl fermions $\{\tilde\psi_i,\psi_i\}$, $i=1,\dots N$ with opposite hypercharges $q$ and $-q$. Assuming in addition a global $SU(N)$ symmetry for the new fermions, the Lagrangian has the terms
\begin{equation}
\begin{aligned}
\label{Lsinglet}
{\cal L}\supset&-l_1\,\phi-\frac{\hat m^2}{2}\phi^2-\frac{l_3}{3!}\phi^3-\frac{\hat\lambda}{4!}\phi^4-\frac{l_{\phi H}}{2}\phi\, H^\dagger H-\frac{\lambda_{\phi H}}{2}\phi^2 H^\dagger H-\left[M\tilde\psi\psi +y\,\phi\, \tilde\psi\psi\,+c.c.\right],
\end{aligned}
\end{equation}
where we denote collectively all the fermion pairs by $\{\tilde\psi,\psi\}$. In the expression \eq{Lsinglet} we have included all the possible renormalizble self-couplings of the inflaton ($l_1$, $l_3$, etc.) and to the Higgs. Indeed, once a Yukawa coupling between the inflaton and the fermions is present, linear and cubic terms for the singlet scalar have to be included as well. Nevertheless, it is possible to impose renormalization conditions enforcing that the physical linear and cubic terms --defined in terms of derivatives of the effective potential-- vanish at any chosen scale. In this way we can focus in practice on the case in which the quartic interactions dominate, as it is appropriate for a quartic plateau. Enforcing these conditions is by no means strictly necessary for the generation of a plateau, but we will do it here for simplicity. 

The two-loop beta functions of the couplings present in \eqref{Lsinglet} can be obtained from the expressions of Appendix \ref{app:singletbetas}.\footnote{In order to do so, it is necessary to set to zero the couplings related to the extra real scalar included in the formulae of Appendix \ref{app:singletbetas}.} From the one-loop Coleman-Weinberg potential, we can immediately obtain the effective quartic coupling of the inflaton, $\lambda(\phi_0)$ (denoted for simplicity $\lambda$ below) as explained in \S~\ref{pp}:
\begin{align}
\label{lambdaeff2}
\frac{\lambda}{Z_{\phi }^4}=\hat\lambda+\frac{3\hat\lambda^2}{2}\kappa \left(\overline\log \,\frac{\hat\lambda}{2}-\frac{3}{2}\right)+6\lambda_{\phi H}^2\,\kappa\left(\overline\log \, \frac{\lambda_{\phi H}}{2}-\frac{3}{2}\right)-24Ny^4\,\kappa\left(\overline\log \, y^2-\frac{3}{8}\right)\,.
\end{align}
Here $Z_\phi$ denotes the field renormalization factor due to the RG,  and we use the notation
\begin{align}
\overline\log \,x\equiv\log\,(x/\varepsilon^2).
\end{align}
All the couplings on the right hand side of \eqref{lambdaeff2}, as well as $Z_\phi$, are understood to be evaluated at the scale $\mu_0=\varepsilon\phi_0$. $Z_\phi$ can be set equal to 1 at the reference scale  without loss of generality, and we will do so in what follows. However, its scale dependence (given by $\gamma_\phi=-\mu/Z_\phi\, \partial Z_\phi/\partial\mu$) has to be taken into account to compute $\beta_\lambda$ from \eqref{lambdaeff2} and from the beta functions of Appendix \ref{app:singletbetas}. 

Knowing the effective quartic coupling and its beta function, we can solve the plateau conditions \eqref{conds2} using the expansion \eqref{loopexpansion}. In particular, we choose to solve them expressing $\hat\lambda$ and $\lambda_{\phi H}$ in terms of the rest of the couplings, obtaining: 
\begin{align} \label{singletsol}
\hat\lambda= -12 \kappa\, N y^4\log N+\cdots\,,\quad\lambda_{\phi H}= 2\sqrt{N}y^2+\cdots\,.
\end{align}
With these expressions, the condition $\lambda\1=0$ is indeed satisfied. The effective quartic coupling at the plateau, $\lambda=\kappa^2\lambda^{(2)}+\cdots$, is
\begin{align}
\lambda= 3\kappa^2\,\lambda_{\phi H}^2\left(\frac{9}{20}\left(8q^2 -1\right)g_1^2- \frac{9g_2^2}{4}+3\left(\lambda_H+y_t^2-{y}^2\right) +2\lambda_{\phi H}\right)+\cdots\,,
\end{align}
where $\lambda_H$ is the Higgs quartic coupling, with the convention 
\begin{align} \label{VH}
V(H)=m^2_H (H^\dagger H)+\frac{\lambda_H}{2}(H^\dagger H)^2
\end{align} 
for the Higgs potential. As we anticipated, a $U(1)$ charge for the fermions helps to make $\lambda>0$ and thus realize a plateau. For instance, setting $N=1$ and replacing the SM couplings with their values at $0.1\,M_P$, we get
$\lambda\simeq4 {y}^4 \kappa^2\left(4\, q^2+3\, {y}^2-1\right)$. If the charge is zero, $\lambda$ is positive for $y\gtrsim 1/\sqrt{3}$. Instead, for $q\neq 0$ smaller values of the Yukawa coupling are allowed and there is an extra lever to get the right order of magnitude for $\lambda\sim10^{-13}$, as it is needed for viable inflation. With $q=1$ one may for example have $y\sim0.004$. In the case $q\neq 0$ the inflaton-dependent mass stays  below the Planck mass $m_P$ during inflation.

We conclude that a singlet real scalar $\phi$ coupled to the Higgs and to generic Weyl fermions charged under $U(1)_Y$, can provide successful radiative plateau inflation. The same holds true if the $U(1)$ gauge group corresponds to a hidden sector, in which case the non-SM fermions might play the role of WIMP dark matter. In that case, matching the right thermal relic abundance requires their mass scale (for $\phi=0$) to be around the TeV scale.\footnote{Since the fermion mass scale $M$ in the case of a hidden $U(1)$ satisfies  $M\sim 1$ TeV $\ll$ $M_P$, it is consistent with the large-field approximation (and the choice $\mu =\phi$) that yields \eq{lambdaeff}.}

We also remark that these results show that a real scalar coupled to a single Weyl fermion and to an another (generic) scalar is a sufficient minimal set-up that can produce a successful quartic inflationary plateau through radiative corrections.

\subsubsection{\label{subsubgf}An inflaton charged under a $U(1)$ gauge group and coupled to fermions}

In the second type of simple models that allow a quartic plateau, the prospective inflaton is part of a complex scalar charged under a $U(1)$ gauge group and is coupled to fermions. In this case, the gauge invariance of the Yukawa interactions of the inflaton demands that some of these fermions (and possibly all of them) must be charged under the same $U(1)$ as the inflaton. 
\vskip0.5cm
{{\bf The SM case (and why it cannot support plateau inflation)}}\vskip0.5cm
The SM Higgs is charged under the hypercharge group, $U(1)_Y$, and is coupled to three generations of fermions; most strongly to the top quark. Therefore, the SM possesses the ingredients to produce a plateau. As it is well known, a plateau can indeed appear in the effective potential of the SM at high scales $\sim 10^{18}$ GeV, provided that the mass of the top quark is appropriately tuned with respect to that of the Higgs.\footnote{For a Higgs mass $m_h\simeq 125$ GeV  \cite{Aad:2015zhl}, the value of $m_t$ needed for a plateau is in the range $m_t\simeq 171.0$--$171.5$ GeV \cite{Degrassi:2012ry,Bednyakov:2015sca}, with a theoretical uncertainty around $\pm 0.5$ GeV \cite{Bednyakov:2015sca}. Some of the most recent experimental results allow this possibility. The value $m_t=171.5\pm0.5$ GeV is roughly within 0.5$\sigma$--2$\sigma$ of the CMS result $m_t= 172.38\pm0.10 (stat.)\pm0.65 (syst.) {\rm\,\,GeV}$ \cite{CMS:2014hta}, and within 1$\sigma$--2$\sigma$ of the ATLAS result $m_t= 172.99\pm0.48(stat.)\pm0.78(syst.) {\rm\,\,GeV}$\cite{Aad:2015nba}.} Here we show explicitly how the SM can satisfy the plateau conditions, and why it fails to provide successful inflation, as already pointed out, e.g.\ in \cite{Isidori:2007vm}. 

We write the tree-level Higgs potential as in \eq{VH}, where the Higgs $SU(2)$ doublet, $H$, contains the (would-be inflaton) real part of the neutral component, $h$, as well as the Goldstone modes, i.e.\ $H=(\tilde H,v+h+i G^0)/\sqrt{2}$. In the $\overline{\rm MS}$ scheme, the effective quartic coupling $\lambda$ of $h$ can be written (for large $h$)  as in the expression \eq{expeff1}, where the tree-level contribution is $\lambda^{(0l)}=3Z_H^4\lambda_H$, and at the one-loop level
\begin{align}
\frac{\lambda^{(1l)}}{Z_H^4} \simeq\frac{9\lambda_H ^2}{2}\left[\overline\log\,\frac{27\lambda^4_H}{16}-6\right]-\frac{9y_t^4}{2} \left[\overline\log\, \frac{y_t^2}{2}-\frac{3}{2}\right]+\frac{9g_2^4}{4} \left[\overline\log\,\frac{g_2^2}{4}-\frac{5}{6}\right]+ \frac{9\tilde g^4}{8}\left[\overline\log\, \frac{\tilde g^2}{4} -\frac{5}{6}\right],
\end{align}
with $\tilde g^2\equiv 3g_1^2/5 +g_2^2$\,.\footnote{Throughout this paper, we use the GUT normalization of the SM hypercharge coupling.} In this expression we are neglecting all fermion contributions except the one of the top quark. 
We choose the  top Yukawa coupling, $y_t$, and Higgs quartic coupling, $\lambda_H$, to be the couplings that we determine from the plateau conditons,\footnote{Since the SM gauge couplings are much better constrained by experimental measurements than the top Yukawa coupling, this is a convenient choice.} expanding them in $\kappa$. The first non-zero terms of the $\kappa$ expansions of these couplings are:
\begin{equation}
\begin{aligned}\label{yt0}
 y_t =\frac{1}{2}\left(\tilde g^4+2 g_2^4\right)^{1/4}\,,\quad
\lambda_H  =-\frac{3}{4}\kappa\left[g^4_2\, \,\log\, \frac{g^2_2}{4} + \frac{{\tilde g}^4}{2} \,\log\,\, \frac{{\tilde g}^2}{4}  -8 y_t^4 \left( \log\,\frac{y_t^2}{2}  -\frac{2}{3}   \right)   \right]\,.
\end{aligned}
\end{equation} 
As before, the couplings in these expressions are implicitly evaluated at the scale $\mu_0=\varepsilon \phi_0$.   

Using the two-loop RG improvement of the one-loop effective potential and assuming a Higgs mass $m_h = 125.09$ GeV \cite{Aad:2015zhl}, we find that the  plateau requires $m_t=171.75$ GeV and is located at $h_0=1.8\times 10^{18}$ GeV, with $y_t(h_0)=0.3815$, see \cite{Ballesteros:2015iua}. The simple formulae above can reproduce this result to within $\sim 1\%$ accuracy. In Appendix \ref{precision}  we provide the $O(\kappa)$ corrections, which reduce the difference to just $\sim 0.01\%$.

At the plateau, $\lambda^{(0)}$ and $\lambda^{(1)}$ vanish and the first non-zero contribution to $\lambda$ comes from
\begin{align}
\label{SMlambda}
\lambda\2=72\, y_t^4\, g_3^2-\frac{81}{2} y_t^6+\frac{\tilde g^4}{64}\left(297\, \tilde g^2-462g_2^2\right)+\frac{g^4_2}{32}\left(117\, \tilde g^2-84\,  g^2_2\right)\,.
\end{align}

As we have seen in Section \ref{phenodes}, a successful quartic inflationary plateau requires $\phi_0\sim 10\, M_P$ and
\begin{align} \label{orders}
\lambda(\mu_0)\sim|\beta_\lambda(\mu_0)|\sim\beta'_\lambda(\mu_0)\sim10^{-13}\,,
\end{align} 
where the renormalization scale $\mu_0$ is proportional to the plateau location $\phi_0$ by some small ($\ll 1$) positive constant  whose precise value can be chosen according to the induced inflaton-dependent masses. Substituting in \eq{SMlambda} the RG values of the couplings, choosing $\mu_0 =0.1 M_P$, leads to $\lambda(\mu_0)\sim 10^{-5}$, which is many orders of magnitude above the value of $10^{-13}$ yielding successful inflation. This confirms the well-known result that the plateau scenario in the SM fails to provide adequate inflation \cite{Isidori:2007vm}.

Using \eqref{expNLO}, the relation \eq{orders} corresponds to a value of the quartic coupling at the scale of the top mass of the order of $\sim10^{-12}$, which is much smaller than the value demanded by the measurements of the actual Higgs mass. Given this, and the sizable contributions of the top Yukawa and gauge couplings to the corresponding beta function and its derivatives in the SM, we can conjecture that a possible solution that would allow the Higgs to be the inflaton in a plateau (without requiring a coupling to the scalar curvature as in \cite{Bezrukov:2007ep}) could come from a new high-energy threshold. Beyond this threshold there would be new contributions to the beta functions that would nearly cancel those of the top Yukawa and gauge couplings, leading to the relation \eq{orders}. Since such cancellation should affect not only $\beta_{\lambda}$ but also $\beta'_\lambda$, this threshold would likely entail an enlarged approximate symmetry,\footnote{This could be associated to a fixed point of the RG flow.} possibly linking the values of gauge and Yukawa couplings. We do not attempt to construct such a model here.
\vskip0.5cm
{\bf \label{subcomplexinfhidden} Inflation from a complex scalar charged under a hidden $U(1)$}
\vskip0.3cm
We can consider models in which the inflaton comes from a complex scalar other that the Higgs and for which the plateau arises thanks to the interplay between Yukawa and gauge couplings. The SM results already suggest that it should be fairly easy to find viable examples. For instance, setting $g_2=g_3=0$ in the SM solution \eqref{SMlambda} we get $\lambda\2\sim {g_1}^6$. Then, the required value of the effective coupling at the plateau, $\lambda\sim10^{-13}$, can be obtained with ${g_1}\sim 0.05$. Now, we will focus on the question of whether a viable inflationary plateau can arise in models whose field content is much simpler than that of the SM. We can consider, for instance, a complex scalar, $\phi$, coupled to vector-like fermions, $\psi$ and $\tilde\psi_i$, with charges $\pm q$ under a new $U(1)$ gauge group with gauge coupling $g$:
\begin{align}
\label{Lcomplex}
{\cal L}\supset-m^2\phi^\dagger\phi -\frac{\hat\lambda}{2}(\phi^\dagger\phi)^2-\frac{\lambda_{\phi H}}{2}\phi^\dagger\phi\, H^\dagger H-\left[y\,\phi\, \tilde\psi\tilde\psi+\tilde y\,\phi^\dagger\, \psi\psi+c.c.\right].
\end{align}
Gauge invariance demands the inflaton to have charge $2q$ under the $U(1)$ group. The effective quartic coupling in the direction of the real part of $\phi$ is given by $\lambda^{(0l)}=3 Z_\phi^4 \hat\lambda$ and
\begin{align}\nn
  \frac{\lambda^{(1l)}}{Z_{\phi }^4} & =\frac{3\hat{\lambda }^2}{2} \left(\overline\log\, \frac{\hat{\lambda }}{2}+9\,\overline\log \,\frac{3 \hat{\lambda }}{2}-15\right)+18 \bar g^4 \left(  \overline\log\, \bar g^2-\frac{5}{6}\right)+6\lambda_{\phi H}^2 \left(\overline\log \,\frac{\lambda_{\phi H}}{2}-\frac{3}{2}\right)\\
 &-3 y^4 \left(\overline\log\, \frac{y^2}{2}-\frac{3}{2}\right)-3\tilde{y}^4 \left(\overline\log\, \frac{\tilde{y}^2}{2}-\frac{3}{2}\right)\,,
\end{align}
where $\bar g^2 \equiv 4 q^2 g^2$\,. Choosing $\hat\lambda$ and $y$ as the couplings to be solved from the plateau conditions, the final result for $\lambda^{(2)}$ ends up being
\begin{align}\nn
\lambda^{(2)} & =9\lambda_{\phi H}^2 \left(\lambda_H+y_t^2+\frac{2}{3}\lambda_{\phi H}- \frac{\bar{g}^2}{2}-\frac{\Delta}{3}- \frac{3}{4}g_2^2
 -\frac{3}{20}g_1^2\right)+\frac{9}{2}\bar{g}^4\left(\frac{13}{3}\bar{g}^2-3\Delta-\tilde y^2\right)\\ &+\frac{3}{2}\tilde{y}^4\left(\Delta  -\tilde{y}^2\right)\,,
 \end{align}
 with $\Delta=\left(6\bar g^4-\tilde{y}^4+2 \lambda_{\phi H}^2\right)^{1/2}$. Interestingly, if the Higgs portal coupling, $\lambda_{\phi H}$, vanishes, it can be checked that $\lambda^{(2)}$ cannot be positive. This is because the first type of contributions in \eqref{betapcont} are negative despite the positive contributions from the gauge coupling, once the relation of the latter to the Yukawa couplings is fixed by imposing $\beta^{(1)}_\lambda=0$. In consequence, viable plateaus stabilized solely by the inflaton's gauge coupling are not possible in this model.

\begin{table}[t]
\centering
\begin{tabular}{c|c|c|c}
 &$U(1)$ & $SU(N)$ & $SU(N)$\\
 \hline
$\psi$ & $q$& $\square$ & 1\\
$\tilde\psi$ & -$q$ & 1& $\overline\square$ \\
$\chi$& ${\tilde q}$ & 1 & $\square$ \\
$\tilde\chi$ &-${\tilde q}$ & $\overline\square$ & 1\\    
$\phi$ & $q$-${\tilde q}$& 1 & 1
\end{tabular}            
\caption{ \label{tab:charges} Charges of the fields  for the Lagrangian \eq{Lcomplex2}. If the $U(1)$ is Hidden, the Higgs is uncharged under this group. Another possibility consists in identifying the $U(1)$ with $U(1)_Y$ of the SM. In any of the two cases the Higgs is not charged under any of the $SU(N)$ groups.}
\end{table}

There are however examples which are only slightly more complicated and circumvent this difficulty by mimicking the SM case, in which there are several species of fermions with different $U(1)$ charges. In particular, adding $N$ pairs of  fermions $\{\chi,\tilde\chi\}$ with charges $\pm {\tilde q}$ is enough for this purpose. To simplify the Lagrangian, we impose an ${ SU(N)}\times { SU(N)}$ symmetry that forbids tree-level masses for the fermions but allows Yukawa couplings with the inflaton:
\begin{align}
\label{Lcomplex2}
{\cal L}\supset-m^2\phi^\dagger\phi -\frac{\hat\lambda}{2}(\phi^\dagger\phi)^2-\frac{\lambda_{\phi H}}{2}\phi^\dagger\phi H^\dagger H-\left[y\,\phi \tilde\psi\chi+\tilde y\phi^\dagger \psi\tilde\chi+c.c.\right].
\end{align}
The one-loop part of the effective quartic coupling in this case is
\begin{align}
\nn \frac{\lambda^{(1l)}}{Z_{\phi }^4}&=\frac{3}{2}\hat\lambda^2 \left(\overline\log\, \frac{\hat\lambda}{2}+9\, \overline\log\, \frac{3 \hat\lambda}{2}-15\right)+18{g}^4 q_s^4 \left(\overline\log\, \left({g}^2 q_s^2\right)-\frac{5}{6}\right)+6\lambda_{\phi H}^2 \left(\overline\log\, \frac{\lambda_{\phi H}}{2}-\frac{3}{2}\right)\\
 &-6N y^4 \left(\overline\log\, \frac{y^2}{2}-\frac{3}{2}\right)-6N \tilde{y}^4 \left(\overline\log\, \frac{\tilde{y}^2}{2}-\frac{3}{2}\right),
\end{align}
where we have denoted the $U(1)$ charge of the inflaton as $q_s\equiv q-\tilde q$. To show that indeed $\lambda_{\phi H}$ is not needed to realize the plateau in this case, we will set it to zero. For simplicity, we also set $\tilde y=0$, but we emphasize that turning on these couplings at any scale does not impede the appearance of the plateau.  We proceed as we did for the SM and solve for $\hat\lambda$ and the Yukawa coupling $y$ in a expansion in the parameter $\kappa$, obtaining
\begin{align}\label{plateaucomplexsol} 
 y^4=3\frac{{g}^4 q_s^4}{N}+\cdots\,,\quad \hat\lambda=-\kappa\, N\,y^4 \left(\log \frac{4N}{3}+\frac{4}{3}\right)+\cdots\,.
\end{align}
It can be checked that $\lambda^{(0)}=\lambda^{(1)}=0$, as required, and
\begin{align}
\label{lambdaU1}
\lambda\2&= {g}^2y^4 \left(4 N^2 \left(\tilde{q}^2+q^2\right)+2 N \left(5 \tilde{q}^2-q \tilde{q}+5 q^2\right)-3 \sqrt{3N}( N+1)q_s^2 \right).
\end{align}
In contrast to the singlet inflaton case of Section \ref{subsubsf}, the solution for $\lambda\2$ does not depend on any SM couplings. This is because the stabilizing role of the Higgs portal coupling in Section \ref{subsubsf} is now played by the hidden $U(1)$. The expression \eq{lambdaU1} shows that  $\lambda^{(2)}>0$ can be obtained with a small ${g}$, even in the simplest case in which we set $N=1$, $q=1$ and $\tilde q=0$. There, $\lambda\sim 10^{-12}$ can be achieved with $g\sim 0.025$. This again yields inflaton-dependent masses below $m_P$ during inflation. 

Another possibility is to identify the $U(1)$ gauge group of \eq{Lcomplex2} with the SM hypercharge group. In the limit of $\lambda_{\phi H}=\tilde y=0$, the effective quartic coupling of the inflaton is:\footnote{The coupling $\lambda^{(2)}$ is not just a straightforward substitution of $g$ for $g_1$ in \eqref{lambdaU1}, given the different beta functions.}
\begin{align}
 \lambda\2&=\frac{162 }{125}g_1^6q_s^4 \left((5+2N) \left(\tilde{q}^2+q^2\right)-q \tilde{q}+\frac{41}{4}-\frac{3}{2} \sqrt{3N}\left(1+N^{-1}\right)q^2_s\right).
\end{align}
If we choose $N=1$ and $\tilde q=0$ and we set the value of the SM $g_1$ coupling at $\mu=0.1 M_P$, i.e.\ $g_1\sim 0.6$, we obtain $\lambda\sim 10^{-13}$ for weakly charged fields with $q\sim 0.01$.

To summarize, a plateau stabilized by a $U(1)$ requires that at least two pairs of (Weyl) fermion species are charged under the $U(1)$. These charges can be integer if the $U(1)$ is Hidden from the SM. If the $U(1)$ is instead the one of the SM hypercharge, the absolute values of charges of the fermions have to be much smaller than 1.

\subsection{\label{subquadpp}Quadratic inflation}

\begin{figure}
\hspace{2cm}
 \includegraphics[trim=10 0 10 0, clip, scale=0.28]{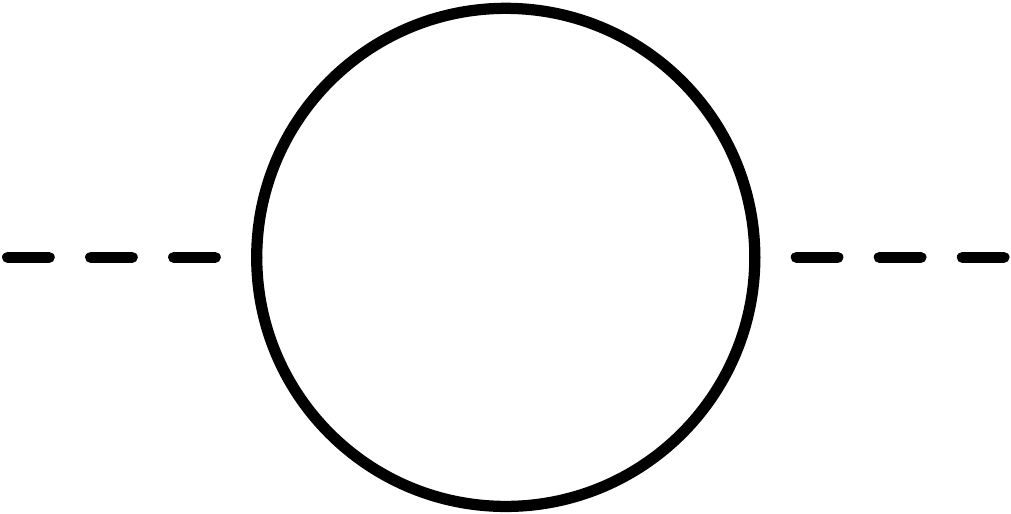}\hspace{2cm}
    \includegraphics[trim=10 0 10 0, clip, scale=0.28]{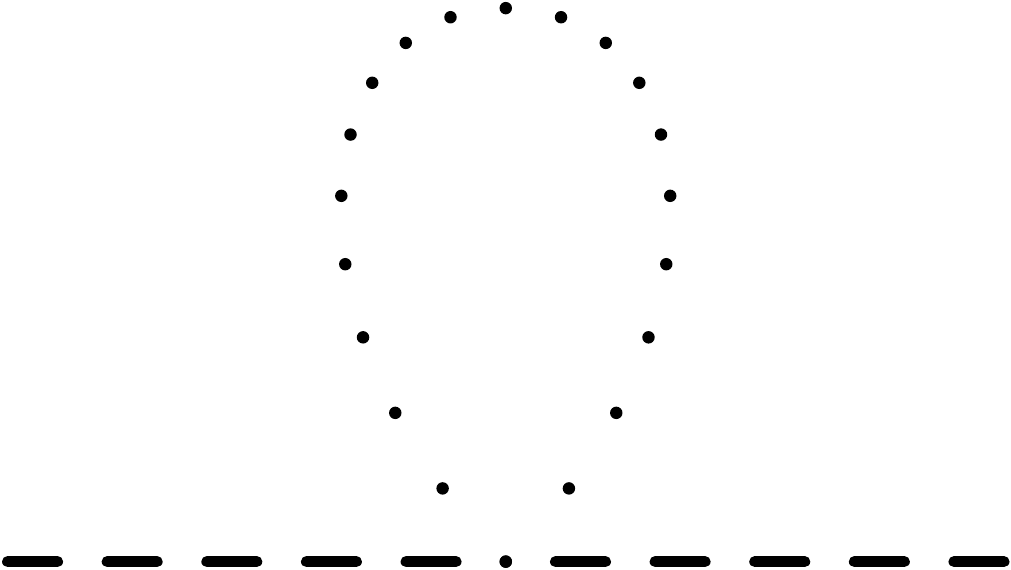}\hspace{2cm}
\includegraphics[trim=10 0 10 0, clip, scale=0.28]{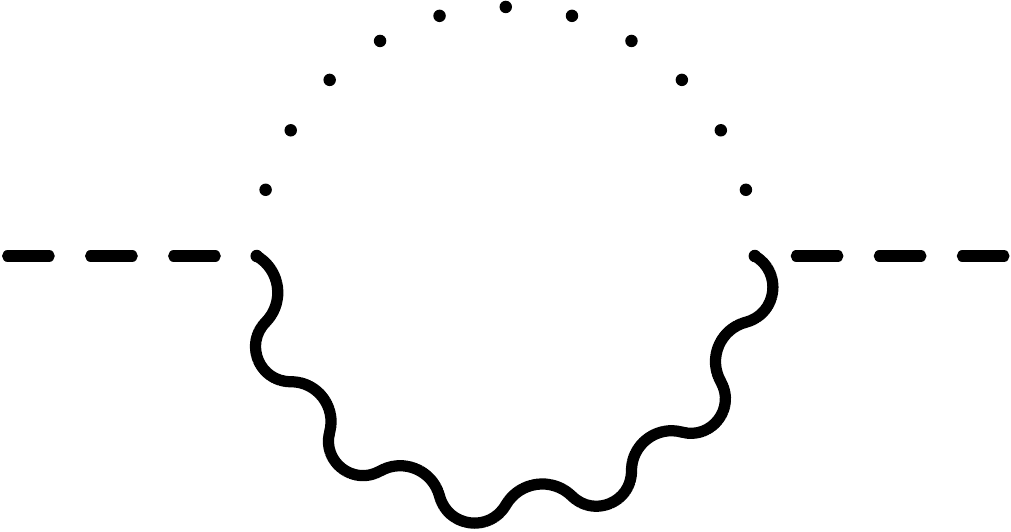}\hspace{2cm}
\caption{\label{Floop2} One-loop contributions to $\beta_{m^2}$. The lines represent: fermions (continuous), scalars (dotted) and gauge bosons (wiggled). Each inflaton leg is represented as a dashed line. The dotted lines inside the loops represent generic scalar fields, including the inflaton as well.}
  \end{figure}

In the quadratic case, assuming negligible effective quartic and cubic interactions for the range of field values relevant for inflation, the plateau conditions of \eqref{condsquad} imply  
\begin{align} \label{cq1}
  m^{2\,(0)}=m^{2\,(1)}=\beta_{m^2}^{(1)}=0,\\
  4m^{2\,(2)}={\beta'}_{m^2}^{(2)}=-2{\beta}_{m^2}^{(2)},
\end{align}
where we are using a notation that parallels that of the quartic case. Arguing also in an analogous way, the minimal requirements for a quadratic plateau are given by $\beta^{(1)}_{m^2}=0$ and $\beta'^{(2)}_{m^2}>0$ (with both conditions at the plateau scale). At linear  order in $\kappa$, the beta functions of the effective mass squared and the corresponding nominal coupling of the Lagrangian, $\hat m^2$, are related by $\beta^{(1)}_{m^2}=\beta^{(1)}_{\hat m^2}$. It is thus useful to start the analysis from the general form of the one-loop beta function of $\hat m^2$, as we did in the quartic case for $\hat\lambda$. Here we also consider a model in which the scalar $\phi$ interacts with Weyl fermions through Yukawa couplings $\delta_a$ ($a=1,\ldots, N$) and with gauge fields with couplings $g_A$ ($A=1,\ldots,N_g$), as well as to $(N_s-1)$ extra scalars. We now have to take into account the mass parameters of these scalars and fermions, $\hat m^2_{i}$ and $ m_{I}$, respectively. For negligible cubic terms and no mass mixings, the generic form of the beta function of the mass parameter of $\phi$ is 
\begin{align}
\label{eq:betam2}
\beta_{\hat m^2}&=\sum_{j,\alpha}^{N_s}Q^{\hat\lambda}_{j\alpha}+\sum_{I, J, a,b}^NQ^{\delta}_{IJ ab}+2\gamma_\phi \hat m^2 ,
\end{align}
where
\begin{equation}
\label{eq:betam22}
\begin{aligned}
Q^{\hat\lambda}_{j\alpha}&=c_{j\alpha}\,\hat\lambda_\alpha\, \hat m^2_j\,,\quad c_{j\alpha} >0,\\
Q^\delta_{IJ ab}&=c_{I J a b}\, \delta_a\, \delta_b\, m_I\, m_J\,,\quad c_{IJab} <0.
\end{aligned}
\end{equation}
The relevant diagrams are illustrated in Figure \ref{Floop2}. As in the quartic case, we are implicitly assuming a mass-independent subtraction scheme such as $\overline{\rm MS}$ (which is indeed the scheme we use in the formulae below). However, now the gauge couplings only enter through $\gamma_\phi$. The fermionic contributions to $\beta_{m^2}$ are independent of the phases of the fermion fields. The fermion masses can be chosen to be positive, so that the contributions $Q^\delta_{IJ ab}$ in \eqref{eq:betam2} and \eqref{eq:betam22} are negative.  

The conditions $\beta_{\hat m^2}^{(1)}=\beta_{m^2}^{(1)}=0$ and $m^{2(0)}=0$ of \eq{cq1} can in principle be satisfied if the model contains massive scalars aside from the inflaton (interacting with it through portal couplings), as well as fermions with tree-level mass terms. An example is the Lagrangian \eqref{Lsinglet}, that we studied in Section \ref{subsubsf}. There, the field $H$ was identified with the Higgs, but as we will shortly see, this identification cannot work in the present case. 

The effective mass squared up to the one-loop level is
\begin{align} \label{fmass}
\frac{m^2}{Z^2_\phi}=\hat m^2+\frac{\kappa}{2}\,\hat\lambda{{\hat m^2}} \left( \overline\log\, \frac{\hat\lambda}{2}-1\right)+ 2\kappa\lambda_{\phi H}{m^2_H} \left( \overline\log\, \frac{\lambda_{\phi H}}{2}-1\right)-4\kappa N y^2{M}^2 \left(3\,\overline\log\, y^2-1\right).
\end{align}
The plateau conditions can be used to express $\hat m^2$ and $\lambda_{\varphi H}$ in terms of the rest of the parameters. The tree-level quartic coupling  $\hat\lambda$ at the scale $\mu_0=\varepsilon \phi_0$ can be fixed at order $\kappa$ by demanding that the effective quartic interaction $\lambda$ of \eqref{lambdaeff} vanishes at $\phi=\phi_0$, guaranteeing that the potential is indeed dominated by the quadratic terms. Similarly, we also require that the effective cubic couplings are zero. The first non-vanishing contribution to the effective mass squared at the plateau is
\begin{align} \label{nmass}
\frac{m^{2\,(2)}}{M^2} =\frac{6}{5} N y^2 \left(9 g_1^2 \left(8 q^2-1\right)-45 g_2^2+60\left(\lambda_H +y_t^2\right)\right)-24 N (N+3) y^4+\left(12N y^2\right)^2\frac{ M^2 }{m^2{}_H}\,,
\end{align}
where, for the sake of simplicity, the only fermionic contribution from the SM that we have included is that of the top quark.  Since in the SM the parameter $m^2_H$ is negative, and its absolute value is much smaller than the scale of $10^{13}$ GeV required for $m^2$  for inflation,  it is not possible to get a successful plateau. For sufficiently high values of $M$, the last term of \eq{nmass} dominates, dragging $m^{2\,(2)}$ to negative values. We could hope to solve the problem with an extremely small value of the Yukawa coupling $y$ but that leads to a very small ratio $m^{2}/{M^2}$ and in practice there is no viable solution. A simple possible way out is to rely on a scalar field different than the Higgs, allowed to have a positive mass term. Therefore we enlarge the model of \S~\ref{subsubsf} with an additional real scalar $\varphi$, charged under a discrete $Z_2$ symmetry, having in addition to \eq{Lsinglet} the following interactions:
\begin{align}\label{Lsinglet2}
{\cal L}\supset&-\frac{\lambda_\varphi}{4!}\varphi^4-\frac{\lambda_{\varphi H}}{2}\varphi^2 H^\dagger H
-\frac{\lambda_{\phi \varphi}}{4}\phi^2 \varphi^2\,.
\end{align}
As before, provided that the quartic  and  cubic couplings of the inflaton $\phi$ are negligible at a certain scale, we can look for solutions to the quadratic plateau equations. For simplicity we also assume that $\lambda_{\phi H}$, $l_{\phi H}$ and $\lambda_{\varphi H}$ are negligible, but we note that if they take non-zero values they can aid to form the plateau. The effective mass squared in this enlarged model is then of the form \eq{fmass} with the replacement $ 2\lambda_{\phi H}\, m^2_H(\overline\log\, \lambda_{\phi H}/2-1) \rightarrow 1/2\,\lambda_{\phi \varphi} m^2_H(\overline\log\, \lambda_{\phi\varphi}/2-1)$\,. Solving for $\hat m^2$ and $\lambda_{\phi\varphi}$ we obtain ${\hat m}^{2(0)}=0$ and $\lambda_{\phi\varphi}\0=24 M^2 N y^2/m^2{}_{\varphi }$\,, whereas
\begin{align}\label{solm2}
{\hat m}^{2(1)}=\frac{\lambda_{\phi\varphi}\0}{2}\, m^2{}_{\varphi }\left(\log \frac{2y^2}{\lambda_{\phi\varphi}\0}+\frac{2}{3}\right)\,.
\end{align}
The dominant contribution to the effective mass parameter of the inflaton at the plateau can be written as
\begin{align}
\label{quadeff}
\frac{m^{2\,(2)}}{M^2}= 12 N y^2 \left(\frac{36}{5} g_1^2 q^2+\lambda _{\varphi }\right)-24N\left(N+3\right) y^4+576N^2 y^4 \frac{M^2 }{m^2{}_{\varphi }}.
\end{align}
In this case one can easily get positive values of ${m}^{2(2)}$ for positive $m^2{}_{\varphi }$ and non-zero $q$. 

The quadratic plateau will be viable as long as the effective quartic interaction can be consistently ignored for the values of the fields relevant for inflation. Imposing that the quartic coupling has to be negligible at the plateau, the beta function of the quartic coupling can be estimated to be $3\kappa (\lambda^2_{\phi\varphi}-16Ny^4)=48\kappa N(36N M^4/m^4_\varphi-1) y^4$. In this way, the quartic interaction of the inflaton is approximately
\begin{align}
\label{quarticest}
 \frac{1}{4!}\hat\lambda(\phi)\phi^4\sim\frac{1}{4!}\left(\lambda(\phi_0)+\beta_\lambda\log\frac{\phi}{\phi_0}\right)\phi^4\sim 2\kappa y^4 N\left(36N\frac{M^4}{m^4_\varphi}-1\right)\log\frac{\phi}{\phi_0}\phi^4.
\end{align}
Note that the sign of the quartic interaction is positive for $\phi>\phi_0$, so that the potential stays stable for larger values of the fields.  One can get constraints for the parameters demanding for example that the quartic interactions are suppressed by a factor of 10 with respect to the quadratic interactions in the range of $\phi$ relevant for inflation, $M_P\lesssim\phi\lesssim10M_P$. As an illustration we fix $\lambda_\varphi=0.1$, $N=q=1$, and substitute $g_1$ by the SM value at the scale of 0.1$M_P$, $g_1(0.1M_P)=0.594$. Enforcing $\kappa^2m^{2(2)}\sim10^{26}$ GeV as required for inflation fixes a relation between $y$, $M$ and $M^2/m^2_\varphi$, so that  the condition on the suppression of the quartic interaction can be recast as a lower bound on $M$ for a fixed value of $M^2/m^2_\varphi$. Once $m^2$ is fixed, increasing $M$ in \eqref{quadeff} forces a decrease in $y$, so that the quartic interaction of \eqref{quarticest} will be more suppressed. Choosing $M^2\sim m^2_\varphi$ yields then $M\gtrsim5\times10^{17}$ GeV.  This in turn yields $y\sim 5\times 10^{-4}$, which sources inflaton-dependent masses below $m_P$ for the field values relevant for inflation. We have checked that similar bounds apply for the mass scales $M$ or $(m^2_\varphi)^{1/2}$ for $0.1<M^2/m^2_\varphi<10$. Therefore, the quadratic plateau is viable when the inflaton couples to a massive scalar, other than the Higgs,  and to fermions. Both types of fields must have large masses to make the quadratic approximation to the potential consistent.

\section{Conclusions}

The latest CMB data favours flat inflationary potentials and constrains the tensor-to-scalar ratio of primordial fluctuations, $r$, to be smaller than $\sim 0.11$.\footnote{See however the note added after the acknowledgements.} We have shown how inflationary plateaus with these features can arise from radiative corrections to simple quadratic or quartic potentials. 

In the absence of quantum corrections, monomial chaotic inflation is strongly disfavoured by the data. However, loop corrections from couplings to other fields can change significantly the predictions of such simple models, allowing them to become compatible with the observations. Such couplings are  generically required for reheating the universe after inflation.

We have provided a concise, model-independent, analytic description of the effects of radiative corrections, and we have analyzed their effects on inflation. So far, the studies of loop corrections in monomial chaotic inflation have mainly focused on fermionic loops \cite{NeferSenoguz:2008nn,Enqvist:2013eua,Okada:2014lxa,Ahmed:2014cma}, which  can destabilize the inflaton's potential at large field values. We have considered a more general approach, including gauge bosons and extra scalars, with the aim of studying the possibilities leading to a stable potential with an exact or approximate plateau, as the CMB data points to. 

We find that the exact (two-loop) quartic plateau is in tension with Planck data,  while the analogous quartic case gives predictions compatible with the observations. Allowing for deviations from an exact plateau, and hence alleviating the tuning of the underlying models, the potentials we consider can easily reproduce the primordial cosmological parameters, while at the same time generate an adequate number of e-folds. Our results can thus be interpreted as bringing life back to chaotic quadratic and quartic inflation after the last burial sentence from Planck. 

In both cases --quadratic and quartic-- the plateaus are potentials with three parameters, that can be easily fitted to the data.\footnote{See appendix \ref{reppot} for these parameters.} These potentials are distributed on the parameter space on two distinct regions, which can be distinguished from each other by their comparatively larger or smaller values of  $r$. The viable potentials in the high-$r$ branch contain not only (exact and approximate) plateaus, but also a wider range of behaviours, including unstable potentials. In the branch with lower values of $r$, the plateau shape is needed for successful inflation and $r$ is typically of the order of $10^{-3}$, within the reach of upcoming experiments.

We have discussed how these plateaus can arise from two-loop radiative corrections, determining the precision required to describe the values of the parameters of the underlying model. We have shown that a one-loop improvement of the tree-level potential is sufficient to determine whether a plateau is possible in any given particle physics model. If the plateau does indeed exist, the one-loop beta functions, in combination with the one-loop effective potential, allow to calculate analytically the first non vanishing contributions to the parameters of the Lagrangian at the chosen renormalization scale. 

We have constructed specific models with the minimal particle content that is necessary to support a viable inflationary plateau. In the quartic case, a plateau requires the inflaton to couple to fermions, and to either an additional scalar or to a $U(1)$ gauge field.  In the case of a coupling to an extra scalar, this field is required to couple weakly to all the fermions. The Higgs field could play the role of this scalar and we have shown that charging the new fermions under the $U(1)$ of hypercharge helps to realize the plateau, although this is not required. If the new fermions are charged under a new $U(1)$, instead of hypercharge, they could play the role of WIMP dark matter. 

In the case in which the inflaton does not couple to another scalar and the plateau is instead made possible thanks to $U(1)$ gauge interactions, the fermions must appear in representations of the $U(1)$ group with different absolute values of their charges. The Standard Model falls into this category of models, although it is certainly not the most minimal one and fails to provide adequate inflation as it predicts too large an effective quartic coupling at the scale of the plateau. We have provided explicit constructions of minimal $U(1)$ plateau quartic models involving four Weyl fermions coupling to the inflaton. If the $U(1)$ is identified with SM hypercharge, the new fermion species have to have very small charges, whereas integer charges are possible in the case of a hidden $U(1)$. 

We have also described a basic structure for models leading to quadratic plateaus. These models involve again an additional scalar and fermions. In contrast to the quartic case, the Higgs cannot be the scalar coupling to the inflaton, because of its low mass in comparison to that of the inflaton. In order for the quadratic plateau to provide adequate inflation, while ensuring that the quadratic approximation to the potential remains valid, the fermions and the scalar coupling to the inflaton have to be very heavy, with masses around $10^{17}$ GeV, although this value can be reduced depending on the smallness of the Yukawa couplings. 

We have thus presented a way of rescuing monomial chaotic inflation which is motivated by the need of a connection between the inflaton and the SM for reheating the universe. The models of inflation that we have described require an unnatural hierarchy between the effective coupling, its beta function and the first derivative of the latter at the plateau scale. For instance, in the quartic case this is: $\lambda\sim\beta_\lambda\sim\beta_\lambda^\prime$. In this respect, it has to be noted that there is also an (implicit) tuning in most monomial chaotic models of inflation. Assuming renormalizability of the potential, the least unnatural among these is $\phi^4$, since this is the contribution that would dominate at large field values, but standard $\phi^4$ inflation is clearly ruled out by the data. It must also be recalled that an unnatural hierarchy like the one needed for a plateau might actually be realized to some extent in the SM. The latest measurements of the Higgs and top quark masses suggest that the SM potential, if extrapolated to very high energies, is likely to be metastable. This implies a similar hierarchy between the effective quartic coupling of the Higgs and its beta function at the instability scale of the SM. In fact, a small variation of the top mass with respect to the currently preferred value can lead to a plateau in the SM. Given the excellent fit that approximate plateaus give to the CMB data, it would be interesting to develop concrete models of particle physics that can produce such a hierarchy in a way that is (at least) technically natural. If future measurements\footnote{See the note added before the appendices.} confirm that plateau potentials are to be preferred, this will become an important challenge for inflationary model building. 

\section*{Acknowledgements}
We thank A.\ Casas, A.\ Notari and K.\ Schmidt-Hoberg for valuable discussions. The work of G.B.\ is funded by the European Union's Horizon 2020 research and innovation programme under the Marie Sk\l{}odowska-Curie grant agreement number 656794. G.B.\ acknowledges support of the Spanish MINECO Centro de Excelencia Severo Ochoa Programme under grant SEV-2012-0249. G.B.\ thanks the DESY theory group for hospitality and the German Science Foundation (DFG)  for funding through the  Collaborative Research Center (SFB) 676 ``Particles, Strings and the  Early Universe''. G.B.\ thanks the CERN Theory Division for hospitality. C.T.\ acknowledges support of the Spanish Government through grant FPA2011-24568 (MICINN). 

\subsubsection*{Note added}While this paper was under review, a preprint \cite{Array:2015xqh} from the BICEP2/Keck collaboration appeared pointing out that their updated upper bound on the amount of primordial gravity waves from inflation (including Planck data) is $r<0.07$ at $0.05$ Mpc$^{-1}$. This value, which is smaller than the (previous) upper bound $r\lesssim 0.11$, adds an even stronger motivation for our work, since it excludes pure $\phi^2$ and $\phi^4$ inflationary models with very high significance.

\appendix

\section{Reparametrizing the radiative potentials}\label{reppot}
To perform fits to CMB and other cosmological data, it is convenient to simplify the potentials as much as the models themselves allow. Although the examples in the tables of Section \ref{phenodes} are expressed in terms of four parameters: $\lambda$ (or $m^2/M_P^2$), $b_1$, $b_2$ and $\phi_0/M_P$, each potential depends only on three numbers. Here we write the potentials \eq{potanalyticdef} and \eq{quad} in terms of the normalized field $\Phi =\phi/M_P$, as series expansions, highlighting the independent parameters up to $\log^2$ terms. The quartic potential \eq{potanalyticdef}, can be written as 
\begin{align} \label{pn1}
V=M_P^4\left[x_0+ x_1 \log \Phi^2+ x_2 \left(\log\Phi^2\right)^2+\cdots\right]\Phi^4\,.
\end{align}
The three dimensionless parameters $x_0$, $x_1$ and $x_2$ are related to the deformations ($b_1$ and $b_2$), the effective quartic coupling, $\lambda$, and the plateau location, $\Phi_0=\phi_0/M_P$\,, as follows: 
\begin{equation}
\begin{aligned}
 x_0 =  \frac{\lambda}{24}\left[1 + 2 \left(1 - b_1\right) \log \Phi_0^2  + 2 \left(1 + b_2\right)\left(\log \Phi_0^2\right)^2\right]\,,\quad\quad\\
 x_1 = -\frac{\lambda}{12} \left(1 - b_1 + 2 \left(1 + b_2\right) \log \Phi_0^2 \right)\,,\quad
  x_2 =\frac{\lambda}{12} \left (1+b_2\right)\,.
\end{aligned}
\end{equation}
Similarly, the quadratic deformed plateu \eq{quad} can be written as
\begin{align} \label{pn2}
V=M_P^4\left[x_0+ x_1 \log \Phi^2+  x_2 \left(\log\Phi^2\right)^2+\cdots\right]\Phi^2\,,
\end{align}
where now
\begin{equation}
\begin{aligned}
 x_0 =  \frac{1}{2} \left(1 + (1 - b_1) \log \Phi_0^2 + \frac{1}{2}(1 + b_2) \left(\log \Phi_0^2\right)^2\right)\frac{m^2}{M_P^2}\,,\quad \quad\\
 x_1 = -\frac{1}{2}\left(1 - b_1 + (1 + b_2) \log \Phi_0^2 \right)\frac{m^2}{M_P^2}\,,\quad
  x_2 =\frac{1}{4}\left(1 + b_2\right)\frac{m^2}{M_P^2}\,.
\end{aligned}
\end{equation}
In both cases, assuming Planckian field values, the hierarchy $x_i\ll x_{i-1}$,  $\forall\, i\geq 1$ would normally be satisfied in a natural expansion. Plateaus that are exact (at order $\log^2$)  correspond in both cases to the condition $b_1=b_2=0$, which violates such a hierarchy. For plateaus, it is then implicitly assumed that the hierarchy is recovered from $i=3$ onwards, in such a way that higher order logarithmic terms can be neglected. The potentials \eq{pn1} and \eq{pn2} are three-parameter models (with the truncation at order $\log^2$) that, to the best of our knowledge, have not been considered in the literature so far. With arbitrary coefficients $x_i$ they generalize the plateaus studied in this paper.

\section{Beta functions} \label{2loop}

In this appendix we  provide the two-loop beta functions in the $\overline{\rm MS}$ scheme for the concrete particle physics models admitting inflationary plateaus studied in this paper. In the case of couplings that are part of the SM,  we give the modifications  of their beta functions (in the extended models) with respect to their SM counterparts. We have obtained the beta functions and anomalous dimensions following \cite{Machacek:1983tz,Machacek:1983fi,Machacek:1984zw,Luo:2002ti}. The two-loop beta functions in the SM can  also  be found in \cite{Luo:2002ey}. We use the GUT normalization for the hypercharge coupling in the SM. We remark that the one-loop beta functions are enough to derive all the results of Section \ref{pp}. The two-loop beta functions are provided here because they are required for the complementary results of Appendix \ref{precision}.

\subsection{Real singlet scalars with extra fermions}\label{app:singletbetas}

\allowdisplaybreaks[1]

Here we consider a model obtained by extending the SM with a real singlet $\phi$ (the inflaton), a second real singlet $\varphi$  and $N$ pairs of Weyl fermions $\{\psi_i, \tilde\psi_i\}$ with hypercharges $\pm q$. The multiplets $\psi$ and $\tilde\psi$ transform under a global $SU(N)$ symmetry in the fundamental and antifundamental representations, respectively. The relevant interactions of these fields are given in  \eq{Lsinglet} and \eq{Lsinglet2}. The simplified version of the model in which the singlet $\varphi$ is decoupled allows to realize a quartic plateau for the inflaton, as shown in \S~\ref{subsubsf}, while the full model can  also develop a quadratic plateau for $\phi$, provided that there is a coupling between $\phi$ and the Higgs, as described in \S~\ref{subquadpp}. Using the notation $\kappa=1/(16\pi^2)$, the two-loop beta functions in the $\overline{\rm MS}$ scheme  as well as the anomalous dimension of the scalar field $\phi$ are:
\begin{flalign*}
\beta_y&=\kappa\left[(2 N+3) y^3-\frac{18}{5}g_1^2 q^2 y\right]+\kappa^2\,\left[
y \left(\frac{\hat\lambda^2}{12}+\lambda_{\phi H}^2+\frac{\lambda_{\phi \varphi}^2}{4}\right)-2 \hat\lambda y^3-\left(12 N+\frac{9}{4}\right) y^5\right.\\ \nn
&+g_1^2 \left(6 N+\frac{72}{5}\right) q^2 y^3+y \left\{g_1^4 \left(\frac{12 N}{5}-\frac{27}{25}\right) q^4+\frac{633}{50} g_1^4 q^2\right\}\Bigg]\,,
\end{flalign*}
\begin{flalign*}
\label{betassinglet} \nn
\beta_{\hat\lambda}&=\kappa\left[3 \hat\lambda^2+8 N \hat\lambda y^2+12 \lambda_{\phi H}^2+3 \lambda_{\phi \varphi}^2-48 N y^4\right]+\kappa^2\,\left[\hat\lambda \left(28 N y^4-20 \lambda_{\phi H}^2\right)-\frac{17 \hat\lambda^3}{3}\right.\\
 &\left.-12 N \hat\lambda^2 y^2+\lambda_{\phi H}^2 \left(\frac{72 g_1^2}{5}-72 y_b^2+72 g_2^2-72 y_t^2-24 y_{\tau }^2\right)-48 \lambda_{\phi H}^3+384 N y^6-12 \lambda _{\phi \varphi }^3\right.\\
 &+g_1^2 q^2 \left(24 N y^2 \hat\lambda-\frac{576 N y^4}{5}\right)-5 \hat{\lambda } \lambda _{\phi \varphi }^2\Bigg],
\end{flalign*}
\begin{flalign*}
 \beta_{\lambda_{\phi H}}&=\kappa\left[\hat\lambda \lambda_{\phi H}+\lambda_{\phi H} \left(6 y_b^2-\frac{9 g_2^2}{2}-\frac{9 g_1^2}{10}+4 N y^2+6 y_t^2+2 y_{\tau }^2\right)+\lambda_{\phi H}(4 \lambda_{\phi H}+6 \lambda_H )\right.\\
 &+\lambda_{\varphi H} \lambda _{\phi \varphi }\bigg]+\kappa^2\,\Bigg[\lambda_{\phi H} \left\{g_1^2 \left(\frac{5 y_b^2}{4}+\frac{9 g_2^2}{8}\frac{17 y_t^2}{4}+\frac{15 y_{\tau }^2}{4}\right)-\hat\lambda \left(6 \lambda_{\phi H}^2+4 N y^2 \lambda_{\phi H}\right)\right.\\
 &-\frac{5}{6} \hat\lambda^2 \lambda_{\phi H}+g_2^2 \left(\frac{45 y_b^2}{4}+\frac{45 y_t^2}{4}+\frac{15 y_{\tau }^2}{4}\right)+g_3^2 \left(40 y_b^2+40 y_t^2\right)-21 y_b^2 y_t^2-\frac{27 y_b^4}{2}\\
 &\left.+\frac{1671 g_1^4}{400}-\frac{145 g_2^4}{16}-2 N y^4-\frac{27 y_t^4}{2}-\frac{9 y_{\tau }^4}{2}\right\}+\lambda_{\phi H}^2 \left(\frac{3 g_1^2}{5}-12 y_b^2+3 g_2^2-8 N y^2\right.\\
 &-12 y_t^2-4 y_{\tau }^2\bigg)+\lambda_H  \left\{\lambda_{\phi H} \left(\frac{36 g_1^2}{5}-36 y_b^2+36 g_2^2-36 y_t^2-12 y_{\tau }^2\right)-36 \lambda_{\phi H}^2\right\}\\
 &-15 \lambda_H ^2 \lambda_{\phi H}-\frac{21}{2} \lambda_{\phi H}^3+q^2 \left\{g_1^4 \left(\frac{3 N \lambda_{\phi H}}{5}-\frac{216 N y^2}{25}\right)+12 g_1^2 N y^2 \lambda_{\phi H}\right\}-2 \lambda_{\varphi H}^2 \lambda _{\phi \varphi }\\
 &\left.-2 \lambda_{\varphi H} \lambda _{\phi \varphi }^2-\lambda_{\phi H} \left(4 \lambda_{\varphi H} \lambda _{\phi \varphi }+\frac{\lambda_{\varphi H}^2}{2}+\frac{\lambda _{\phi \varphi }^2}{2}\right)\right],
\end{flalign*}
 \begin{flalign*}\phantom{\hskip0.5cm}
 \beta_{\lambda_{\varphi H}}&=\kappa\left[\lambda_{\varphi H} \left(6 y_b^2-\frac{9}{10}  g_1^2-\frac{9 g_2^2}{2}+6 \lambda _H+\lambda _{\varphi }+6 y_t^2+2 y_{\tau }^2\right)+4 \lambda_{\varphi H}^2+\lambda_{\phi H} \lambda _{\phi \varphi }\right]\\
 &+\kappa^2\,\left[\lambda_{\varphi H} \left\{\lambda _H \left(\frac{36 g_1^2}{5}-36 y_b^2+36 g_2^2-36 y_t^2-12 y_{\tau }^2\right)+g_1^2 \left(\frac{5 y_b^2}{4}+\frac{9 g_2^2}{8}+\frac{17 y_t^2}{4}\right.\right.\right.\\
 &\left.+\frac{15 y_{\tau }^2}{4}\right)+g_2^2 \left(\frac{45 y_b^2}{4}+\frac{45 y_t^2}{4}+\frac{15 y_{\tau }^2}{4}\right)+g_3^2 \left(40 y_b^2+40 y_t^2\right)-21 y_b^2 y_t^2-\frac{27 y_b^4}{2}+g_1^4 \left(\frac{3 N q^2}{5}\right.\\
 &\left.\left.+\frac{1671}{400}\right)-\frac{145 g_2^4}{16}-15 \lambda _H^2-\frac{5 \lambda _{\varphi }^2}{6}-4 \lambda_{\phi H} \lambda _{\phi \varphi }-\frac{\lambda_{\phi H}^2}{2}-\frac{\lambda _{\phi \varphi }^2}{2}-\frac{27 y_t^4}{2}-\frac{9 y_{\tau }^4}{2}\right\}+\lambda_{\varphi H}^2 \bigg(\frac{3 g_1^2}{5}\\
 &\left.-12 y_b^2+3 g_2^2-36 \lambda _H-6 \lambda _{\varphi }-12 y_t^2-4 y_{\tau }^2\bigg)-\frac{21}{2}  \lambda_{\varphi H}^3-2 \lambda_{\phi H}^2 \lambda _{\phi \varphi }-\lambda_{\phi H} \left(2 \lambda _{\phi \varphi }^2+4 N y^2 \lambda _{\phi \varphi }\right)\right],
 \end{flalign*}
 \begin{flalign*}\phantom{\hskip0.5cm}
 \beta_{\lambda_{\varphi}}&=\kappa\left[3 \lambda _{\varphi }^2+12 \lambda_{\varphi H}^2+3 \lambda _{\phi \varphi }^2\right]+\kappa^2\,\left[\lambda_{\varphi H}^2 \left(\frac{72 g_1^2}{5}-72 y_b^2+72 g_2^2-72 y_t^2-24 y_{\tau }^2\right)\right.&\\
 &\left.-\lambda _{\varphi } \left(20 \lambda_{\varphi H}^2+5 \lambda _{\phi \varphi }^2\right)-\frac{17}{3} \lambda _{\varphi }^3-48 \lambda_{\varphi H}^3-12 \lambda _{\phi \varphi }^3-12 N y^2 \lambda _{\phi \varphi }^2\right],&
 \end{flalign*}
 \begin{flalign*}\phantom{\hskip0.5cm}
\nn
 \beta_{\lambda_{\phi\varphi}}&=\kappa\left[4 \lambda_{\varphi H} \lambda_{\phi H}+4 \lambda _{\phi \varphi }^2+\lambda _{\phi \varphi } \left(\lambda _{\varphi }+\hat{\lambda }+4 N y^2\right)q\right]+\kappa^2\,\left[\lambda_{\varphi H} \bigg\{\lambda_{\phi H} \left(\frac{24 g_1^2}{5}-24 y_b^2\right.\right.&\\
 &+24 g_2^2-24 y_t^2-8 y_{\tau }^2\bigg)-8 \lambda_{\phi H}^2\bigg\}+\lambda _{\phi \varphi } \left(12 g_1^2 N q^2 y^2-\frac{5 \lambda _{\varphi }^2}{6}-2 \lambda_{\varphi H}^2-16 \lambda_{\varphi H} \lambda_{\phi H}-2 \lambda_{\phi H}^2\right.&\\
 &\left.\left.-\frac{5 \hat{\lambda }^2}{6}\!-2 N y^4\!-4 \hat{\lambda } N y^2\right)\!-8 \lambda_{\varphi H}^2 \lambda_{\phi H}-9 \lambda _{\phi \varphi }^3-\lambda _{\phi \varphi }^2 \left(6 \lambda _{\varphi }+6 \hat{\lambda }+8 N y^2\right)\right],&
 \end{flalign*}
 \begin{flalign*}\phantom{\hskip0.5cm}
 \gamma_{\phi}&=\frac{1}{8\pi^2}Ny^2+\kappa^2\,\left[\frac{\hat\lambda^2}{12}+\lambda_{\phi H}^2-5 N y^4+6 g_1^2 N q^2 y^2+\frac{\lambda _{\phi \varphi }^2}{4}\right],&
 \end{flalign*}
 \begin{flalign*}\phantom{\hskip0.5cm}
 \beta_M&=\kappa\,M \left[2 N y^2+3 y^2-\frac{18}{5} g_1^2 q^2\right]+\kappa^2\,\left[l_3  \left(\frac{y \hat\lambda}{12}-2 y^3\right)+ y \lambda_{\phi H}l_{\phi H}+M \bigg\{g_1^2 q^2 \bigg(6 N y^2\right.&\\
&+\frac{72 y^2}{5}\bigg)+g_1^4 \left\{\left(\frac{12 N}{5}-\frac{27}{25}\right) q^4+\frac{633 q^2}{50}\right\}-12 N y^4-\frac{9 y^4}{4}\bigg\}\Bigg],&
 \end{flalign*}
 \begin{flalign*}\phantom{\hskip0.5cm}
\beta_{l_{\phi H}}&=\kappa\left[l_{\phi H} \left(6 y_b^2-\frac{1}{10} 9 g_1^2-\frac{9 g_2^2}{2}+4 \lambda_{\phi H}+2 N y^2+6 y_t^2+2 y_{\tau }^2\right)+6 l_{\phi H}\lambda_H+l_3  \lambda_{\phi H}\right]&\\
 &+\kappa^2\,\left[\frac{l_{\phi H} \hat\lambda^2}{12}\!-\hat\lambda \left(2 l_{\phi H} \lambda_{\phi H}\!+l_3  \lambda_{\phi H}\right)\!+l_{\phi H}\lambda_H \left(\frac{36 g_1^2}{5}\!-36 y_b^2\!+36 g_2^2-36 \lambda_{\phi H}-36 y_t^2\right.\right.&\\
 &-12 y_{\tau }^2\bigg)+l_{\phi H} \Bigg\{\lambda_{\phi H} \left(\frac{3 g_1^2}{5}-12 y_b^2+3 g_2^2-12 y_t^2-4 y_{\tau }^2\right)+g_1^2 \left(\frac{5 y_b^2}{4}+\frac{9 g_2^2}{8}+\frac{17 y_t^2}{4}+\frac{15 y_{\tau }^2}{4}\right)&
 \end{flalign*}
 \begin{flalign*}\phantom{\beta_{l_{\phi H}}=}
 &+g_2^2 \left(\frac{45 y_b^2}{4}+\frac{45 y_t^2}{4}+\frac{15 y_{\tau }^2}{4}\right)+g_3^2 \left(40 y_b^2+40 y_t^2\right)-21 y_b^2 y_t^2-\frac{27 y_b^4}{2}+\frac{1671 g_1^4}{400}-\frac{145 g_2^4}{16}&\\
 &-\frac{23 \lambda_{\phi H}^2}{2}-8 N \lambda_{\phi H} y^2-5 N y^4-\frac{27 y_t^4}{2}-\frac{9 y_{\tau }^4}{2}\Bigg\}-15 l_{\phi H}\lambda_H^2+8 M N \lambda_{\phi H} y^3-l_3  \left(4 \lambda_{\phi H}^2\right.&\\
 &\left.\left.+4 N \lambda_{\phi H} y^2\right)+q^2 \left\{g_1^4 \left(\frac{3 l_{\phi H} N}{5}-\frac{207 M N y}{25}\right)+6 l_{\phi H} g_1^2 N y^2\right\}+l_{\phi H} \left(\frac{\lambda _{\phi \varphi }^2}{4}-2 \lambda_{\varphi H} \lambda _{\phi \varphi }\right.\right.&\\
 &\left.\left.-\frac{\lambda_{\varphi H}^2}{2}\right)\right],&
 \end{flalign*}
 \begin{flalign*}\phantom{\hskip0.5cm}
  \beta_{l_3}&=\kappa\left[l_3  \left(3 \hat\lambda+6 N y^2\right)+12 l_{\phi H} \lambda_{\phi H}-48 M N y^3\right]+\kappa^2\,\Bigg[l_{\phi H} \left\{\lambda_{\phi H} \left(\frac{72 g_1^2}{5}-12 \hat\lambda-72 y_b^2\right.\right.&\\
 &+72 g_2^2-72 y_t^2-24 y_{\tau }^2\bigg)-48 \lambda_{\phi H}^2\bigg\}+24 M N \hat\lambda y^3+l_3  \left(9 N y^4-\frac{23 \hat\lambda^2}{4}-12 N \hat\lambda y^2-9 \lambda_{\phi H}^2\right)&\\
 &+384 M N y^5+g_1^2 q^2 \left(18 N l_3  y^2-\frac{576}{5} M N y^3\right)-\frac{9}{4} l_3  \lambda _{\phi \varphi }^2\Bigg],&
\end{flalign*}
\begin{flalign*}\phantom{\hskip0.5cm}
 \beta_{\hat m^2}&=\kappa\left[{\hat m^2} \left(\hat\lambda+4 N y^2\right)+4 l_{\phi H}^2+4 m^2_H \lambda_{\phi H}-24 M^2 N y^2+l_3 ^2+\lambda _{\phi \varphi } m^2{}_{\varphi }\right]+\kappa^2\,\bigg[l_{\phi H}^2 \bigg(24 g_2^2&
 \\
 &\left.-2 \hat\lambda-24 y_b^2+\frac{24 g_1^2}{5}-20 \lambda_{\phi H}-24 y_t^2-8 y_{\tau }^2\right)-{\hat m^2} \left(\frac{5 \hat\lambda^2}{6}+4 N \hat\lambda y^2+2 \lambda_{\phi H}^2+2 N y^4\right)\\
 &+4 M^2 N \hat\lambda y^2-l_3 ^2 \left(\frac{5 \hat\lambda}{2}+4 N y^2\right)-8 l_{\phi H} l_3  \lambda_{\phi H}+m^2_H \left\{\lambda_{\phi H} \left(\frac{24 g_1^2}{5}-24 y_b^2+24 g_2^2-24 y_t^2\right.\right.&\\
 &\left.-8 y_{\tau }^2\bigg)-8 \lambda_{\phi H}^2\bigg\}+192 M^2 N y^4+16 M N l_3  y^3+g_1^2 q^2 \left(12 N y^2 {\hat m^2}-\frac{288}{5} M^2 N y^2\right)\right.&\\
 &\left.-2 \lambda _{\phi \varphi }^2 m^2{}_{\varphi }-\frac{1}{2} \hat{m}^2 \lambda _{\phi \varphi }^2\right],&
 \end{flalign*}
\begin{flalign*}\phantom{\hskip0.5cm}
 \beta_{l_1}&=\kappa\left(4 l_{\phi H} m^2_H+l_3  {\hat m^2}-8 M^3 N y+2 N l_1 y^2\right),&
 \end{flalign*}
 
\noindent The changes in the beta functions of the couplings shared with the SM are as follows:
 
 \begin{flalign*}\phantom{\hskip3cm}
  \beta_{g_1}&=\beta_{g_1}^{SM}+\frac{1}{20\pi^2} g_1^3 N q^2+\kappa^2\,\left[\frac{36}{25} g_1^5 N q^4-\frac{6}{5} g_1^3 N q^2 y^2\right],&
  \end{flalign*}
\begin{flalign*}\phantom{\hskip3cm}
  \beta_{g_2}&=\beta_{g_2}^{SM},&
  \end{flalign*}
\begin{flalign*}\phantom{\hskip3cm}
  \beta_{g_3}&=\beta_{g_3}^{SM},&
  \end{flalign*}
\begin{flalign*}\phantom{\hskip3cm}
  \beta_{y_t}&=\beta_{y_t}^{SM}+\kappa^2\,\left[\frac{29}{75} g_1^4 N q^2 y_t+\frac{1}{4} \lambda_{\phi H}^2 y_t+\frac{1}{4} \lambda_{\varphi H}^2 y_t\right],&
  \end{flalign*}
\begin{flalign*}\phantom{\hskip3cm}
  \beta_{y_b}&=\beta_{y_b}^{SM}+\kappa^2\,\left[\frac{1}{4} y_b \lambda_{\phi H}^2-\frac{1}{75} g_1^4 N q^2 y_b+\frac{1}{4} y_b \lambda_{\varphi H}^2\right],&
  \end{flalign*}
\begin{flalign*}\phantom{\hskip3cm}
  \beta_{y_\tau}&=\beta_{y_\tau}^{SM}+\kappa^2\,\left[\frac{33}{25} g_1^4 N q^2 y_{\tau }+\frac{1}{4} \lambda_{\phi H}^2 y_{\tau }+\frac{1}{4} \lambda_{\varphi H}^2 y_{\tau }\right],&
  \end{flalign*}
\begin{flalign*}\phantom{\hskip0.5cm}
  \beta_{\lambda_H}&=\beta^{SM}_{\lambda_H}+\kappa\left(\lambda^2_{\phi H}+\lambda^2_{\varphi H}\right)+\kappa^2\,\left[\lambda_H  \left(\frac{6}{5} g_1^4 N q^2-5 \lambda_{\phi H}^2\right)-\frac{72}{125}  g_1^6 N q^2-\frac{24}{25} g_2^2 g_1^4 N q^2\right.&\\
  &\left.-4 \lambda_{\phi H}^3-4 N y^2\lambda_{\phi H}^2-5 \lambda _H \lambda_{\varphi H}^2-4 \lambda_{\varphi H}^3\right],&
  \end{flalign*}
\begin{flalign*}\phantom{\hskip0.5cm}
  \beta_{m^2_H}&=\beta^{SM}_{m^2_H}+\kappa\left[2 l_{\phi H}^2+\lambda_{\phi H} {\hat m^2}+\lambda_{\varphi H} m^2{}_{\varphi }\right]+\kappa^2\,\left[l_{\phi H}^2 \left(\frac{3 g_1^2}{10}-6 y_b^2+\frac{3 g_2^2}{2}-6 \lambda_{\phi H}-18 \lambda_H \right.\right.&\\
  &\left.-4 N y^2-6 y_t^2-2 y_{\tau }^2\right)-2 l_{\phi H} l_3  \lambda_{\phi H}+m^2_H \left(\frac{3}{5} g_1^4 N q^2-\frac{\lambda_{\phi H}^2}{2}\right)+M^2 \left(4 N y^2 \lambda_{\phi H}-\frac{108}{25} g_1^4 N q^2\right)&\\
  &\left.-\frac{l_3 ^2 \lambda_{\phi H}}{2}-{\hat m^2} \left(2 \lambda_{\phi H}^2+4 N y^2 \lambda_{\phi H}\right)-\frac{1}{2} m^2{}_H \lambda_{\varphi H}^2-2 \lambda_{\varphi H}^2 m^2{}_{\varphi }\right].&
 \end{flalign*}

\subsection{\label{app:complexbetas}Complex scalar with extra fermions, charged under a hidden $U(1)$} \label{2loop2} 

Here we give the beta functions of the model considered in \S~\ref{subcomplexinfhidden}. This model involves a complex scalar field (leading to the inflaton) charged under a hidden $U(1)$, as well
as four types of fermions $\psi$, $\tilde\psi$, $\chi$ and $\tilde\chi$, charged under the hidden $U(1)$ with gauge coupling $g$, and a global ${ SU(N)\times SU(N)}$ symmetry as detailed in Table \ref{tab:charges}. The relevant  interactions are given in \eqref{Lcomplex}. The SM gauge couplings denoted as $g_i$ with $i=1,\dots,3$. The beta functions of the couplings beyond the SM are: 
\begin{flalign*}\phantom{\hskip0.5cm}
\beta_{g}&=\kappa\left[g^3 \left\{N \left(\frac{4 \tilde{q}^2}{3}+\frac{4 q^2}{3}\right)-\frac{2 q \tilde{q}}{3}+\frac{\tilde{q}^2}{3}+\frac{q^2}{3}\right\}\right]-\kappa^2\left[N y^2 g^3 \left(\tilde{q}^2+q^2\right)\right.&\\
&\left.+N g^3 \left(\tilde{q}^2+q^2\right) \tilde{y}^2-g^5 \left\{N \left(4 \tilde{q}^4+4 q^4\right)-16 q^3 \tilde{q}+24 q^2 \tilde{q}^2-16 q \tilde{q}^3+4 \tilde{q}^4+4 q^4\right\}\right],&
\end{flalign*}
\begin{flalign*}\phantom{\hskip0.5cm}
\beta_y&=\kappa\left[y \left\{N \tilde{y}^2-g^2 \left(3 \tilde{q}^2+3 q^2\right)\right\}+(N+1) y^3\right]+\kappa^2\,\left[y^3 \left\{\left(\frac{5 N}{2}+13\right) q^2 g^2\right.\right.&\\
&\left.+\left(\frac{5 N}{2}+13\right) g^2 \tilde{q}^2-18 q g^2 \tilde{q}-4 \hat\lambda-\frac{3 N \tilde{y}^2}{2}\right\}+y \left\{\left(\frac{4 N}{3}-\frac{14}{3}\right) q^4 g^4+\left(4 N+\frac{85}{3}\right) q^3 g^4 \tilde{q}\right.&\\
&+\tilde{y}^2 \left(\frac{5}{2} N  g^2(q^2+\tilde{q}^2)\right)+\left(\frac{8 N}{3}-\frac{151}{3}\right) q^2 g^4 \tilde{q}^2+\left(\frac{4 N}{3}-\frac{14}{3}\right) g^4 \tilde{q}^4+\left(4 N+\frac{85}{3}\right) q g^4 \tilde{q}^3&\\
&\left.\left.+\hat\lambda^2-\frac{3 N \tilde{y}^4}{2}+\lambda_{\phi H}^2\right\}+\left(\frac{7}{4}-3 N\right) y^5\right],&
\end{flalign*}
\begin{flalign*}\phantom{\hskip0.5cm}
\beta_{\tilde y}&=\kappa\left[\tilde{y} \left(N y^2-g^2 \left(3 \tilde{q}^2+3 q^2\right)\right)+(N+1) \tilde{y}^3\right]+\kappa^2\,\left[\tilde{y}^3 \left\{\left(\frac{5 N}{2}+13\right) q^2 g^2+\left(\frac{5 N}{2}\right.\right.\right.&\\
&\left.+13\bigg) g^2 \tilde{q}^2-18 q g^2 \tilde{q}-4 \hat\lambda-\frac{1}{2} 3 N y^2\right\}+\tilde{y} \left\{\left(\frac{4 N}{3}-\frac{14}{3}\right) q^4 g^4+\left(4 N+\frac{85}{3}\right) q^3 g^4 \tilde{q}\right.&\\
&+y^2 \left(\frac{5}{2} N g^2(q^2+\tilde q^2)\right)+\left(\frac{8 N}{3}-\frac{151}{3}\right) q^2 g^4 \tilde{q}^2+\left(\frac{4 N}{3}-\frac{14}{3}\right) g^4 \tilde{q}^4+\left(4 N+\frac{85}{3}\right) q g^4 \tilde{q}^3&\\
&\left.\left.+\hat\lambda^2+\lambda_{\phi H}^2-\frac{1}{2} 3 N y^4\right\}+\left(\frac{7}{4}-3 N\right) \tilde{y}^5\right],
\end{flalign*}
\begin{flalign*}\phantom{\hskip0.5cm}
\beta_{\hat\lambda}&=\kappa\left[\hat\lambda \left\{g^2 \left(24 q \tilde{q}-12 \tilde{q}^2-12 q^2\right)+4 N \tilde{y}^2+4 N y^2\right\}+g^4 \left(72 q^2 \tilde{q}^2-48 q^3 \tilde{q}-48 q \tilde{q}^3+12 \tilde{q}^4\right.\right.&\\ 
&\left.\left.+12 q^4\right)+10 \hat\lambda^2-4 N \tilde{y}^4+4 \lambda_{\phi H}^2-4 N y^4\right]+\kappa^2\,\Bigg[\hat\lambda^2 \left\{g^2 \left(56 \tilde{q}^2-112 q \tilde{q}+56 q^2\right)\right.&\\ \displaybreak[4]
&\left.-20 N \tilde{y}^2-20 N y^2\right\}+y^2 g^4 \left(144 N q^2 \tilde{q}^2-64 N q^3 \tilde{q}-64 N q \tilde{q}^3-8 N \tilde{q}^4-8 N q^4\right)&\\
&+\hat\lambda \Bigg\{y^2 g^2 \left(10 N \tilde{q}^2+10 N q^2\right)+g^2 \tilde{y}^2 \left(10 N \tilde{q}^2+10 N q^2\right)+g^4 \left\{\left(\frac{80 N}{3}+632\right) q^2 \tilde{q}^2\right.&\\
&\left.-\left(\frac{80 N}{3}+\frac{1264}{3}\right) q^3 \tilde{q}-\left(\frac{80 N}{3}+\frac{1264}{3}\right) q \tilde{q}^3+\left(\frac{40 N}{3}+\frac{316}{3}\right) \tilde{q}^4+\left(\frac{40 N}{3}+\frac{316}{3}\right) q^4\right\}&\\
&+2 N \tilde{y}^4-20 \lambda_{\phi H}^2+2 N y^4\Bigg\}+g^4 \tilde{y}^2 \left(144 N q^2 \tilde{q}^2-64 N q^3 \tilde{q}-64 N q \tilde{q}^3-8 N \tilde{q}^4-8 N q^4\right)&\\
&+g^6 \left\{\left(\frac{512 N}{3}+832\right) q^5 \tilde{q}-\left(\frac{896 N}{3}+2080\right) q^4 \tilde{q}^2+\left(\frac{1024 N}{3}+\frac{8320}{3}\right) q^3 \tilde{q}^3-\left(\frac{896 N}{3}\right.\right.&\\
&\left.+2080\Big) q^2 \tilde{q}^4+\left(\frac{512 N}{3}+832\right) q \tilde{q}^5-\left(\frac{128 N}{3}+\frac{416}{3}\right) \tilde{q}^6-\left(\frac{128 N}{3}+\frac{416}{3}\right) q^6\right\}&\\
&-16 N q y^4 g^2 \tilde{q}-16 N q g^2 \tilde{q} \tilde{y}^4-60 \hat\lambda^3+16 N \tilde{y}^6+\lambda_{\phi H}^2 \left(\frac{24 g_1^2}{5}-24 y_b^2+24 g_2^2-24 y_t^2-8 y_{\tau }^2\right)&\\
&-16 \lambda_{\phi H}^3+16 N y^6\Bigg],
\end{flalign*}
\begin{flalign*}\phantom{\hskip0.5cm}
\beta_{\lambda_{\phi H}}&=\kappa\left[\lambda_{\phi H} \left\{g^2 \left(12 q \tilde{q}-6 \tilde{q}^2-6 q^2\right)+4 \hat\lambda+2 N \tilde{y}^2+6 y_b^2-\frac{9 g_2^2}{2}-\frac{9 g_1^2}{10}+6 \lambda_H +2 N y^2+6 y_t^2\right.\right.&\\
&+2 y_{\tau }^2\bigg\}+4 \lambda_{\phi H}^2\bigg]+\kappa^2\,\left[\lambda_{\phi H}^2 \left\{g^2 \left(4 \tilde{q}^2-8 q \tilde{q}+4 q^2\right)-24 \hat\lambda-4 N \tilde{y}^2-12 y_b^2+\frac{3 g_1^2}{5}+3 g_2^2\right.\right.&\\
&-36 \lambda_H -4 N y^2-12 y_t^2-4 y_{\tau }^2\bigg\}+\lambda_{\phi H} \bigg\{\hat\lambda \left\{g^2 \left(32 \tilde{q}^2-64 q \tilde{q}+32 q^2\right)-8 N \tilde{y}^2-8 N y^2\right\}&\\
&+y^2 g^2 \left(5 N \tilde{q}^2+5 N q^2\right)+g^2 \tilde{y}^2 \left(5 N \tilde{q}^2+5 N q^2\right)+g^4 \left\{\left(\frac{40 N}{3}+172\right) q^2 \tilde{q}^2-\left(\frac{40 N}{3}\right.\right.&\\
&\left.\left.+\frac{344}{3}\right) q^3 \tilde{q}-\left(\frac{40 N}{3}+\frac{344}{3}\right) q \tilde{q}^3+\left(\frac{20 N}{3}+\frac{86}{3}\right) \tilde{q}^4+\left(\frac{20 N}{3}+\frac{86}{3}\right) q^4\right\}-10 \hat\lambda^2-3 N \tilde{y}^4&
\end{flalign*}
\begin{flalign*}\phantom{\beta_{\lambda_{\phi H}}=}
&+\lambda_H  \left(\frac{36 g_1^2}{5}-36 y_b^2+36 g_2^2-36 y_t^2-12 y_{\tau }^2\right)+g_1^2 \left(\frac{5 y_b^2}{4}+\frac{9 g_2^2}{8}+\frac{17 y_t^2}{4}+\frac{15 y_{\tau }^2}{4}\right)+g_2^2 \left(\frac{45 y_b^2}{4}\right.&\\
&\left.+\frac{45 y_t^2}{4}+\frac{15 y_{\tau }^2}{4}\right)+g_3^2 \left(40 y_b^2+40 y_t^2\right)-21 y_b^2 y_t^2-\frac{27 y_b^4}{2}+\frac{1671 g_1^4}{400}-\frac{145 g_2^4}{16}-15 \lambda_H ^2-3 N y^4&\\
&\left.-\frac{27 y_t^4}{2}-\frac{9 y_{\tau }^4}{2}\bigg\}-11 \lambda_{\phi H}^3\right],
\end{flalign*}
\begin{flalign*}\phantom{\hskip0.5cm}
\gamma_\phi&=\kappa\left[g^2 \left(6 q \tilde{q}-3 \tilde{q}^2-3 q^2\right)+N \tilde{y}^2+N y^2\right]+\kappa^2\,\left[y^2 g^2 \left(\frac{5 N \tilde{q}^2}{2}+\frac{5 N q^2}{2}\right)+g^2 \tilde{y}^2 \left(\frac{5 N \tilde{q}^2}{2}\right.\right.&\\
&\left.+\frac{5 N q^2}{2}\right)+g^4 \left\{\left(\frac{20 N}{3}+20\right) q^2 \tilde{q}^2-\left(\frac{20 N}{3}+\frac{40}{3}\right) q^3 \tilde{q}-\left(\frac{20 N}{3}+\frac{40}{3}\right) q \tilde{q}^3+\left(\frac{10 N}{3}\right.\right.&\\
&\left.\left.\left.+\frac{10}{3}\right) \tilde{q}^4+\left(\frac{10 N}{3}+\frac{10}{3}\right) q^4\right\}+\hat\lambda^2-\frac{3 N \tilde{y}^4}{2}+\lambda_{\phi H}^2-\frac{1}{2} 3 N y^4\right],
\end{flalign*}
\begin{flalign*}\phantom{\hskip0.5cm}
\beta_{{\hat m^2}}&=\kappa\left[{\hat m^2} \left\{g^2 \left(12 q \tilde{q}-6 \tilde{q}^2-6 q^2\right)+4 \hat\lambda+2 N \tilde{y}^2+2 N y^2\right\}+4 m^2_H \lambda_{\phi H}\right]&\\
&+\kappa^2\,\Bigg[{\hat m^2} \Bigg\{\hat\lambda \left\{g^2 \left(32 \tilde{q}^2-64 q \tilde{q}+32 q^2\right)-8 N \tilde{y}^2-8 N y^2\right\}+y^2 g^2 \left(5 N \tilde{q}^2+5 N q^2\right)&\\
&+g^2 \tilde{y}^2 \left(5 N \tilde{q}^2+5 N q^2\right)+g^4 \left\{\left(\frac{40 N}{3}+172\right) q^2 \tilde{q}^2-\left(\frac{40 N}{3}+\frac{344}{3}\right) q^3 \tilde{q}-\left(\frac{40 N}{3}\right.\right.&\\
&\left.\left.+\frac{344}{3}\right) q \tilde{q}^3+\left(\frac{20 N}{3}+\frac{86}{3}\right) \tilde{q}^4+\left(\frac{20 N}{3}+\frac{86}{3}\right) q^4\right\}-10 \hat\lambda^2-3 N \tilde{y}^4-2 \lambda_{\phi H}^2-3 N y^4\Bigg\}&\\
&+m^2_H \left\{\lambda_{\phi H} \left(\frac{24 g_1^2}{5}-24 y_b^2+24 g_2^2-24 y_t^2-8 y_{\tau }^2\right)-8 \lambda_{\phi H}^2\right\}\Bigg].
\end{flalign*}

\noindent The beta functions for the couplings shared with the SM are: 

\begin{flalign*}\phantom{\hskip5.5cm}
  \beta_{g_i}&=\beta_{g_i}^{SM}\quad i=1,2,3,&
  \end{flalign*}
\begin{flalign*}\phantom{\hskip5.5cm}
  \beta_{y_t}&=\beta_{y_t}^{SM}+\frac{1}{2(16\pi^2)^2} \lambda_{\phi H}^2 y_t,&
  \end{flalign*}
\begin{flalign*}\phantom{\hskip5.5cm}
  \beta_{y_b}&=\beta_{y_b}^{SM}+\frac{1}{2(16\pi^2)^2} y_b \lambda_{\phi H}^2,&
  \end{flalign*}
\begin{flalign*}\phantom{\hskip5.5cm}
  \beta_{y_\tau}&=\beta_{y_\tau}^{SM}+\frac{1}{2(16\pi^2)^2} \lambda_{\phi H}^2 y_{\tau },&
  \end{flalign*}
\begin{flalign*}\phantom{\hskip0.5cm}
  \beta_{\lambda_H}&=\beta^{SM}_{\lambda_H}+\frac{\lambda^2_{\phi H}}{8\pi^2}+\kappa^2\,\left[\lambda_{\phi H}^2 \left\{g^2 \left(16 \tilde{q}^2-32 q \tilde{q}+16 q^2\right)-4 N \tilde{y}^2-10 \lambda_H -4 N y^2\right\}-8 \lambda_{\phi H}^3\right],&
  \end{flalign*}
\begin{flalign*}\phantom{\hskip0.5cm}
  \beta_{m^2_H}&=\beta^{SM}_{m^2_H}+\frac{\lambda_{\phi H} {\hat m^2}}{8\pi^2}+\kappa^2\,\left[{\hat m^2} \left(\lambda_{\phi H} \left(g^2 \left(16 \tilde{q}^2-32 q \tilde{q}+16 q^2\right)-4 N \tilde{y}^2-4 N y^2\right)-4 \lambda_{\phi H}^2\right)\right.&\\
  &\left.-m^2_H \lambda_{\phi H}^2\right].&
\end{flalign*}

\section{Plateau solutions with $O(\kappa)$ precision} \label{precision}

As explained in Section \ref{pp}, the existence or not of a plateau can be  determined from the one-loop RG improvement of the tree-level potential. 
However, if the plateau exists, solving for the tree-level couplings of the model at the plateau scale, requires additional corrections. We have shown that the one-loop improvement of the one-loop potential is enough to determine the lowest non-zero contributions to the $\kappa$ expansion of the couplings of the Lagrangian at the plateau, see Sections \ref{subquarticpp} and \ref{subquadpp}. This allowed us to compute the quartic and quadratic couplings (depending on the type of plateau) at the plateau scale itself to $O(\kappa)$. In those sections we gave the extra relation among the couplings, enforced by the plateau equations, at order $\kappa^0$ (which is enough for the first non-vanishing contribution). In order to obtain both relations at the same level of precision (i.e.\ at order $\kappa$), it is  necessary to use the full two-loop potential or, instead,  the two-loop improvement of the one-loop effective potential.\footnote{The precision with which the two-loop improvement of the one-loop effective potential allows to compare the three coefficients $c_0(\phi_0)$, $c_1(\phi_0)$ and $c_2(\phi_0)$ of \eq{expNLO}  is one-loop. This is because with this approximation the coefficient of each $\log^N$ of the potential is rendered with $(N+1)$-loop precision. This means that $c_2$ is obtained at 3-loops, $c_1$ at two-loops and $c_0$ at one-loop.}  Applying the second method, with the two-loop beta functions of Appendix \ref{2loop}, we collect here the $O(\kappa)$ corrections to the formulae \eqref{singletsol}, \eqref{yt0}, \eqref{plateaucomplexsol} and the relation before \eqref{solm2}. We use the same assumptions concerning vanishing couplings as specified for each model in Section \ref{pp}.

\subsection*{Singlet inflaton coupled to the Higgs and new fermions}

The order $\kappa$ contribution to the coupling $\lambda_{\phi H}$ at the plateau scale in the model \eqref{Lsinglet}, which correcting the 
zeroth order contribution \eqref{singletsol}, is given by:
\begin{equation}
\begin{aligned}
\lambda_{\phi H}\1&=\frac{\sqrt{N}}{20}  {y} \bigg[3 {y} \Big\{g_1^2 \left(40 q^2-11\right)-55 g_2^2\Big\}+20 \left(12 \sqrt{N}-11\right) {y}^3+60 \lambda  {y}-60 (\log N-3)y y_t^2\\
&+ 2 {y} \left\{\left(9 g_1^2+45 g_2^2-60 \lambda -80 \sqrt{N} {y}^2\right)\log \left(\sqrt{N} {y}^2\right)
+12 \log {y}^2\left(5 {y}^2-6 g_1^2 q^2-5 y_t^2\right)\right\}\bigg].
 \end{aligned}
 \end{equation}

\subsection*{Standard Model}

The order $\kappa$ contribution to the value of $y_t$ at the plateau scale, which corrects \eqref{yt0}, is given by
\begin{equation}
\label{yt1}
\begin{aligned}
 y_t\1&=\,\frac{1}{960\ 3^{3/4} \sqrt{5} \Omega^{3/2}}\left[18 \Omega^2 \left\{16 \left(20 {g_3}^2+g_1^2\right)-3 \sqrt{3} \Omega\right\} \log \frac{\sqrt{3} \Omega}{40}-4800 \Omega^2 {g_3}^2\right.\\
&+20 \tilde g^2 \log \frac{\tilde g^2}{4} \left(450 \tilde g^4-45 \sqrt{3} \Omega\tilde g^2+657 g_1^4 -725 g_2^4\right)+9 \sqrt{3} \Omega \left(325 \tilde g^4 -160 g_1^4\right.\\
 &\left.-150 g_2^4\right)-200 g_2^4 \left(9 \sqrt{3} \Omega-27 g_1^2+55 g_2^2\right) \log \frac{g_2^2}{4}-2 \left(6363 g_1^6+16065 g_2^2 g_1^4\right.\\
 &\left.+16375 g_2^4 g_1^2-74875 g_2^6\right)\Bigg],
 \end{aligned}
 \end{equation}
 with $\Omega=\,(3 g_1^4+10 g_2^2 g_1^2+25 g_2^4)^{1/2}$ and $\tilde g^2=3/5g^2_1+g_2^2$. For a fixed value of the Higgs mass, we can use our analytic expressions \eqref{yt0} and \eqref{yt1} to calculate the value of $y_t$ at the plateau scale, which in turn can be used to determine the corresponding physical value
 of $m_t$ that yields a plateau. To ensure compatibility with the SM measurements (other than the value of $m_t$, viewed as a prediction from the plateau scenario), this can be done with an iterative procedure, in which boundary conditions are imposed at $\phi_0$ and the weak scale, and 
 $\phi_0$ is scanned until $\lambda_H(\phi_0)$ is compatible with the assumed value of the Higgs mass. As commented in Section \ref{subquarticpp}, a purely numerical computation using the two-loop improvement of the full one-loop effective potential,  without using the analytic expressions that we have obtained, gives the SM plateau at a scale around $h=1.8\times 10^{18}$ GeV, with $y_t(\phi_0)=0.3815$. Using \eqref{yt0} and \eqref{yt1} and substituting the SM couplings at the scale of $1.8\cdot10^{18}$ GeV (having run the couplings from the weak scale with $m_t=173.34$ GeV \cite{ATLAS:2014wva} as a first approximation), yields $y_t\0+\kappa y_t\1=0.3816$, with $y_t\0=0.3892$ and $y_t\1=-1.1969$, in a remarkable $0.03\%$ agreement  with the value obtained by purely numerical methods, despite this being a first iteration in the procedure mentioned above. This is a two-order of magnitude improvement with respect to the order $\kappa^0$ determination of the top Yukawa coupling done in Section \ref{subquarticpp}, which was to be expected since $\kappa\simeq 6\times 10^{-3}$. Indeed, the perturbative $\kappa$ expansion works well, with $|y_t\0|\gg\kappa |y_t\1|$. This shows explicitly that $\lambda\sim\beta_\lambda\sim\beta'_\lambda$ does not imply a breakdown of perturbation theory, but rather a tuning of the tree-level couplings.
 
The value of $m_t$ at the SM plateau scale is very close to the absolute stability bound obtained by demanding that the SM potential stays positive up to the Planck scale, the difference being of the order of $0.4$ GeV. The semi-analytic methods described here, based on the RG and the analytic equations  \eqref{yt0} and \eqref{yt1}, allow an efficient way for computing approximate stability bounds for $m_t$ as a function of $m_h$, including next-to-leading-log corrections to the Higgs potential.

\subsection*{Complex inflaton charged under $U(1)$ and coupled to multiple fermions}

The value of the Yukawa coupling $y$ at the plateau scale in the model of \eqref{Lcomplex}, given at lowest order by \eqref{plateaucomplexsol}, receives the following correction at order $\kappa$:
\begin{equation}
\begin{aligned}
y\1&=\,\frac{g^3 q_s}{12\ 3^{3/4} N^{3/4}} \Bigg[6 \left\{2 \sqrt{N} \left(q \tilde{q}-5 \tilde{q}^2-5 q^2\right)-4 N^{3/2} \left(\tilde{q}^2+q^2\right)+3 \sqrt{3} (N+1) q_s^2\right\} \left(\log g^2\right.\\
&\left.+2 \log q_s\right)+\tilde{q}^2 \left\{28 N^{3/2}+9 \sqrt{3} N+94 \sqrt{N}-9 \sqrt{3} \left(5+\log \frac{4}{3}\right)\right\}-9 \log N \left\{\sqrt{3} \tilde{q}^2\right.\\
&\left.-2 \left(3 \sqrt{N}+\sqrt{3}\right) q \tilde{q}+\sqrt{3} q^2\right\}+2 q \tilde{q}\, \bigg\{27 \sqrt{3} N+\sqrt{N} (54 \log 2-27 \log 3-49)\\
&\left.+9 \sqrt{3} \left(5+\log \frac{4}{3}\right)\right\}+q^2 \left\{28 N^{3/2}+9 \sqrt{3} N+94 \sqrt{N}-9 \sqrt{3} \left(5+\log \frac{4}{3}\right)\right\}\Bigg].
\end{aligned}
\end{equation}

\subsection*{Singlet inflaton coupled to a singlet scalar and new fermions}

In the model with a quadratic plateau containing the interactions \eqref{Lsinglet} and \eqref{Lsinglet2}, the $O(\kappa)$ correction to the zero order value of $\lambda_{\phi\varphi}$ written before \eqref{solm2} is
\begin{align}
\frac{\lambda_{\phi\varphi}\1}{\lambda_{\phi\varphi}\0}=y^2 \left\{\left(N+3\right)\log y^4 +\frac{4N}{3}-1\right\}-g_1^2 q^2 \left(\frac{18}{5} \log y^4+6\right)-\left(\lambda _{\varphi }+2\lambda_{\phi\varphi}\0\right) \log\frac{\lambda_{\phi\varphi}\0}{2}-\frac{\lambda _{\varphi }}{2}\,.
\end{align}

\bibliographystyle{hunsrt}  
\bibliography{bibreheat}

\begin{thebibliography}{10}

\bibitem{Starobinsky:1979ty}
Alexei~A. Starobinsky.
\newblock {Spectrum of relict gravitational radiation and the early state of
  the universe}.
\newblock {\em JETP Lett.}, 30:682--685, 1979.
\newblock [Pisma Zh. Eksp. Teor. Fiz.30,719(1979)].

\bibitem{Starobinsky:1980te}
Alexei~A. Starobinsky.
\newblock {A New Type of Isotropic Cosmological Models Without Singularity}.
\newblock {\em Phys. Lett.}, B91:99--102, 1980.

\bibitem{Guth:1980zm}
Alan~H. Guth.
\newblock {The Inflationary Universe: A Possible Solution to the Horizon and
  Flatness Problems}.
\newblock {\em Phys. Rev.}, D23:347--356, 1981.

\bibitem{Kazanas:1980tx}
D.~Kazanas.
\newblock {Dynamics of the Universe and Spontaneous Symmetry Breaking}.
\newblock {\em Astrophys. J.}, 241:L59--L63, 1980.

\bibitem{Mukhanov:1981xt}
Viatcheslav~F. Mukhanov and G.~V. Chibisov.
\newblock {Quantum Fluctuation and Nonsingular Universe. (In Russian)}.
\newblock {\em JETP Lett.}, 33:532--535, 1981.
\newblock [Pisma Zh. Eksp. Teor. Fiz.33,549(1981)].

\bibitem{Linde:1981mu}
Andrei~D. Linde.
\newblock {A New Inflationary Universe Scenario: A Possible Solution of the
  Horizon, Flatness, Homogeneity, Isotropy and Primordial Monopole Problems}.
\newblock {\em Phys. Lett.}, B108:389--393, 1982.

\bibitem{Albrecht:1982wi}
Andreas Albrecht and Paul~J. Steinhardt.
\newblock {Cosmology for Grand Unified Theories with Radiatively Induced
  Symmetry Breaking}.
\newblock {\em Phys. Rev. Lett.}, 48:1220--1223, 1982.

\bibitem{Mukhanov:1982nu}
Viatcheslav~F. Mukhanov and G.~V. Chibisov.
\newblock {The Vacuum energy and large scale structure of the universe}.
\newblock {\em Sov. Phys. JETP}, 56:258--265, 1982.
\newblock [Zh. Eksp. Teor. Fiz.83,475(1982)].

\bibitem{Hawking:1982cz}
S.~W. Hawking.
\newblock {The Development of Irregularities in a Single Bubble Inflationary
  Universe}.
\newblock {\em Phys. Lett.}, B115:295, 1982.

\bibitem{Starobinsky:1982ee}
Alexei~A. Starobinsky.
\newblock {Dynamics of Phase Transition in the New Inflationary Universe
  Scenario and Generation of Perturbations}.
\newblock {\em Phys. Lett.}, B117:175--178, 1982.

\bibitem{Guth:1982ec}
Alan~H. Guth and S.~Y. Pi.
\newblock {Fluctuations in the New Inflationary Universe}.
\newblock {\em Phys. Rev. Lett.}, 49:1110--1113, 1982.

\bibitem{Bardeen:1983qw}
James~M. Bardeen, Paul~J. Steinhardt, and Michael~S. Turner.
\newblock {Spontaneous Creation of Almost Scale - Free Density Perturbations in
  an Inflationary Universe}.
\newblock {\em Phys. Rev.}, D28:679, 1983.

\bibitem{Linde:1983gd}
Andrei~D. Linde.
\newblock {Chaotic Inflation}.
\newblock {\em Phys.Lett.}, B129:177--181, 1983.

\bibitem{Planck:2015xua}
P.A.R. Ade et~al.
\newblock {Planck 2015 results. XIII. Cosmological parameters}.
\newblock 2015, 1502.01589.

\bibitem{Ade:2015lrj}
P.~A.~R. Ade et~al.
\newblock {Planck 2015 results. XX. Constraints on inflation}.
\newblock 2015, 1502.02114.

\bibitem{Ade:2015tva}
P. A. R. Ade et~al.
\newblock {Joint Analysis of BICEP2/$Keck  Array$ and $Planck$ Data}.
\newblock {\em Phys. Rev. Lett.}, 114:101301, 2015, 1502.00612.

\bibitem{Creminelli:2015oda}
Paolo Creminelli, Diana~L\'opez Nacir, Marko Simonovi\'c, Gabriele Trevisan,
  and Matias Zaldarriaga.
\newblock {Detecting Primordial $B$-Modes after Planck}.
\newblock 2015, 1502.01983.

\bibitem{Matsumura:2013aja}
T.~Matsumura et~al.
\newblock {Mission design of LiteBIRD}.
\newblock 2013, 1311.2847.

\bibitem{Andre:2013afa}
Philippe Andre et~al.
\newblock {PRISM (Polarized Radiation Imaging and Spectroscopy Mission): A
  White Paper on the Ultimate Polarimetric Spectro-Imaging of the Microwave and
  Far-Infrared Sky}.
\newblock 2013, 1306.2259.

\bibitem{Book:2011dz}
Laura Book, Marc Kamionkowski, and Fabian Schmidt.
\newblock {Lensing of 21-cm Fluctuations by Primordial Gravitational Waves}.
\newblock {\em Phys. Rev. Lett.}, 108:211301, 2012, 1112.0567.

\bibitem{Nakayama:2013jka}
Kazunori Nakayama, Fuminobu Takahashi, and Tsutomu~T. Yanagida.
\newblock {Polynomial Chaotic Inflation in the Planck Era}.
\newblock {\em Phys. Lett.}, B725:111--114, 2013, 1303.7315.

\bibitem{Kallosh:2013tua}
Renata Kallosh, Andrei Linde, and Diederik Roest.
\newblock {Universal Attractor for Inflation at Strong Coupling}.
\newblock {\em Phys. Rev. Lett.}, 112(1):011303, 2014, 1310.3950.

\bibitem{Ashoorioon:2013eia}
Amjad Ashoorioon, Konstantinos Dimopoulos, M.~M. Sheikh-Jabbari, and Gary Shiu.
\newblock {Reconciliation of High Energy Scale Models of Inflation with
  Planck}.
\newblock {\em JCAP}, 1402:025, 2014, 1306.4914.

\bibitem{Nakayama:2014wpa}
Kazunori Nakayama, Fuminobu Takahashi, and Tsutomu~T. Yanagida.
\newblock {Polynomial Chaotic Inflation in Supergravity Revisited}.
\newblock {\em Phys. Lett.}, B737:151--155, 2014, 1407.7082.

\bibitem{Harigaya:2014fca}
Keisuke Harigaya, Masahiro Kawasaki, and Tsutomu~T. Yanagida.
\newblock {Lower bound of the tensor-to-scalar ratio $r \mathop{}_{\textstyle
  \sim}^{\textstyle >} 0.1$ in a nearly quadratic chaotic inflation model in
  supergravity}.
\newblock {\em Phys.Lett.}, B741:267--271, 2015, 1410.7163.

\bibitem{Li:2015mwa}
Tianjun Li, Zhijin Li, and Dimitri~V. Nanopoulos.
\newblock {Symmetry Breaking Indication for Supergravity Inflation in Light of
  the Planck 2015}.
\newblock {\em JCAP}, 1509(09):006, 2015, 1502.05005.

\bibitem{Achucarro:2015rfa}
Ana Achucarro, Vicente Atal, and Yvette Welling.
\newblock {On the viability of m**2 phi**2 and natural inflation}.
\newblock 2015, 1503.07486.

\bibitem{Harigaya:2015pea}
Keisuke Harigaya, Masahiro Ibe, Masahiro Kawasaki, and Tsutomu~T. Yanagida.
\newblock {Revisiting the Minimal Chaotic Inflation Model}.
\newblock 2015, 1506.05250.

\bibitem{Pallis:2015mga}
Constantinos Pallis.
\newblock {Kinetically modified nonminimal chaotic inflation}.
\newblock {\em Phys. Rev.}, D91(12):123508, 2015, 1503.05887.

\bibitem{Kannike:2015apa}
Kristjan Kannike, Gert Hütsi, Liberato Pizza, Antonio Racioppi, Martti Raidal,
  Alberto Salvio, and Alessandro Strumia.
\newblock {Dynamically Induced Planck Scale and Inflation}.
\newblock {\em JHEP}, 05:065, 2015, 1502.01334.

\bibitem{Boubekeur:2015xza}
Lotfi Boubekeur, Elena Giusarma, Olga Mena, and H\'ector Ram\'irez.
\newblock {Does Current Data Prefer a Non-minimally Coupled Inflaton?}
\newblock {\em Phys. Rev.}, D91:103004, 2015, 1502.05193.

\bibitem{Buchmuller:2015oma}
Wilfried Buchmuller, Emilian Dudas, Lucien Heurtier, Alexander Westphal,
  Clemens Wieck, and Martin~Wolfgang Winkler.
\newblock {Challenges for Large-Field Inflation and Moduli Stabilization}.
\newblock {\em JHEP}, 04:058, 2015, 1501.05812.

\bibitem{NeferSenoguz:2008nn}
Vedat~Nefer Senoguz and Qaisar Shafi.
\newblock {Chaotic inflation, radiative corrections and precision cosmology}.
\newblock {\em Phys. Lett.}, B668:6--10, 2008, 0806.2798.

\bibitem{Enqvist:2013eua}
Kari Enqvist and Mindaugas Karciauskas.
\newblock {Does Planck really rule out monomial inflation?}
\newblock {\em JCAP}, 1402:034, 2014, 1312.5944.

\bibitem{Okada:2014lxa}
Nobuchika Okada, Vedat~Nefer Şenoğuz, and Qaisar Shafi.
\newblock {The Observational Status of Simple Inflationary Models: an Update}.
\newblock 2014, 1403.6403.

\bibitem{Ahmed:2014cma}
Waqas Ahmed, Ommair Ishaque, and Mansoor~Ur Rehman.
\newblock {Quantum Smearing in Hybrid Inflation with Chaotic Potentials}.
\newblock 2014, 1501.00173.

\bibitem{Croon:2015fza}
Djuna Croon, Veronica Sanz, and Jack Setford.
\newblock {Goldstone Inflation}.
\newblock {\em JHEP}, 10:020, 2015, 1503.08097.

\bibitem{Isidori:2007vm}
Gino Isidori, Vyacheslav~S. Rychkov, Alessandro Strumia, and Nikolaos Tetradis.
\newblock {Gravitational corrections to standard model vacuum decay}.
\newblock {\em Phys.Rev.}, D77:025034, 2008, 0712.0242.

\bibitem{Fairbairn:2014nxa}
Malcolm Fairbairn, Philipp Grothaus, and Robert Hogan.
\newblock {The Problem with False Vacuum Higgs Inflation}.
\newblock {\em JCAP}, 1406:039, 2014, 1403.7483.

\bibitem{Ballesteros:2015iua}
Guillermo Ballesteros and Carlos Tamarit.
\newblock {Higgs portal valleys, stability and inflation}.
\newblock {\em JHEP}, 09:210, 2015, 1505.07476.

\bibitem{Masina:2012yd}
Isabella Masina and Alessio Notari.
\newblock {Inflation from the Higgs field false vacuum with hybrid potential}.
\newblock {\em JCAP}, 1211:031, 2012, 1204.4155.

\bibitem{Notari:2014noa}
Alessio Notari.
\newblock {Higgs Mass and Gravity Waves in Standard Model False Vacuum
  Inflation}.
\newblock {\em Phys. Rev.}, D91:063527, 2015, 1405.6943.

\bibitem{Kannike:2014mia}
Kristjan Kannike, Antonio Racioppi, and Martti Raidal.
\newblock {Embedding inflation into the Standard Model - more evidence for
  classical scale invariance}.
\newblock {\em JHEP}, 06:154, 2014, 1405.3987.

\bibitem{Kannike:2015kda}
Kristjan Kannike, Antonio Racioppi, and Martti Raidal.
\newblock {Linear inflation from quartic potential}.
\newblock 2015, 1509.05423.

\bibitem{Ballesteros:2005eg}
Guillermo Ballesteros, J.~A. Casas, and J.~R. Espinosa.
\newblock {Running spectral index as a probe of physics at high scales}.
\newblock {\em JCAP}, 0603:001, 2006, hep-ph/0601134.

\bibitem{Ballesteros:2007te}
Guillermo Ballesteros, J.~A. Casas, J.~R. Espinosa, R.~Ruiz~de Austri, and
  R.~Trotta.
\newblock {Flat Tree-level Inflationary Potentials in Light of CMB and LSS
  Data}.
\newblock {\em JCAP}, 0803:018, 2008, 0711.3436.

\bibitem{BuenoSanchez:2006xk}
J.~C. Bueno~Sanchez, Konstantinos Dimopoulos, and David~H. Lyth.
\newblock {A-term inflation and the MSSM}.
\newblock {\em JCAP}, 0701:015, 2007, hep-ph/0608299.

\bibitem{Allahverdi:2006we}
Rouzbeh Allahverdi, Kari Enqvist, Juan Garcia-Bellido, Asko Jokinen, and Anupam
  Mazumdar.
\newblock {MSSM flat direction inflation: Slow roll, stability, fine tunning
  and reheating}.
\newblock {\em JCAP}, 0706:019, 2007, hep-ph/0610134.

\bibitem{Baumann:2007np}
Daniel Baumann, Anatoly Dymarsky, Igor~R. Klebanov, Liam McAllister, and
  Paul~J. Steinhardt.
\newblock {A Delicate universe}.
\newblock {\em Phys. Rev. Lett.}, 99:141601, 2007, 0705.3837.

\bibitem{Baumann:2007ah}
Daniel Baumann, Anatoly Dymarsky, Igor~R. Klebanov, and Liam McAllister.
\newblock {Towards an Explicit Model of D-brane Inflation}.
\newblock {\em JCAP}, 0801:024, 2008, 0706.0360.

\bibitem{Itzhaki:2007nk}
Nissan Itzhaki and Ely~D. Kovetz.
\newblock {Inflection Point Inflation and Time Dependent Potentials in String
  Theory}.
\newblock {\em JHEP}, 10:054, 2007, 0708.2798.

\bibitem{Linde:2007jn}
Andrei~D. Linde and Alexander Westphal.
\newblock {Accidental Inflation in String Theory}.
\newblock {\em JCAP}, 0803:005, 2008, 0712.1610.

\bibitem{Cicoli:2008gp}
M.~Cicoli, C.~P. Burgess, and F.~Quevedo.
\newblock {Fibre Inflation: Observable Gravity Waves from IIB String
  Compactifications}.
\newblock {\em JCAP}, 0903:013, 2009, 0808.0691.

\bibitem{Badziak:2008gv}
Marcin Badziak and Marek Olechowski.
\newblock {Volume modulus inflection point inflation and the gravitino mass
  problem}.
\newblock {\em JCAP}, 0902:010, 2009, 0810.4251.

\bibitem{Chen:2009nk}
Heng-Yu Chen, Ling-Yan Hung, and Gary Shiu.
\newblock {Inflation on an Open Racetrack}.
\newblock {\em JHEP}, 03:083, 2009, 0901.0267.

\bibitem{Enqvist:2010vd}
Kari Enqvist, Anupam Mazumdar, and Philip Stephens.
\newblock {Inflection point inflation within supersymmetry}.
\newblock {\em JCAP}, 1006:020, 2010, 1004.3724.

\bibitem{Hotchkiss:2011gz}
Shaun Hotchkiss, Anupam Mazumdar, and Seshadri Nadathur.
\newblock {Observable gravitational waves from inflation with small field
  excursions}.
\newblock {\em JCAP}, 1202:008, 2012, 1110.5389.

\bibitem{Mazumdar:2011ih}
Anupam Mazumdar, Seshadri Nadathur, and Philip Stephens.
\newblock {Inflation with large supergravity corrections}.
\newblock {\em Phys. Rev.}, D85:045001, 2012, 1105.0430.

\bibitem{Cerezo:2012ub}
Rafael Cerezo and Joao~G. Rosa.
\newblock {Warm Inflection}.
\newblock {\em JHEP}, 01:024, 2013, 1210.7975.

\bibitem{Ballesteros:2014yva}
Guillermo Ballesteros and J.~Alberto Casas.
\newblock {Large tensor-to-scalar ratio and running of the scalar spectral
  index with Instep Inflation}.
\newblock {\em Phys. Rev.}, D91:043502, 2015, 1406.3342.

\bibitem{Martin:2013nzq}
J\'er\^ome Martin, Christophe Ringeval, Roberto Trotta, and Vincent Vennin.
\newblock {The Best Inflationary Models After Planck}.
\newblock {\em JCAP}, 1403:039, 2014, 1312.3529.

\bibitem{Ijjas:2013vea}
Anna Ijjas, Paul~J. Steinhardt, and Abraham Loeb.
\newblock {Inflationary paradigm in trouble after Planck2013}.
\newblock {\em Phys. Lett.}, B723:261--266, 2013, 1304.2785.

\bibitem{Guth:2013sya}
Alan~H. Guth, David~I. Kaiser, and Yasunori Nomura.
\newblock {Inflationary paradigm after Planck 2013}.
\newblock {\em Phys. Lett.}, B733:112--119, 2014, 1312.7619.

\bibitem{Martin:2015dha}
Jerome Martin.
\newblock {The Observational Status of Cosmic Inflation after Planck}.
\newblock 2015, 1502.05733.

\bibitem{Coleman:1973jx}
Sidney~R. Coleman and Erick~J. Weinberg.
\newblock {Radiative Corrections as the Origin of Spontaneous Symmetry
  Breaking}.
\newblock {\em Phys.Rev.}, D7:1888--1910, 1973.

\bibitem{Bando:1992wy}
Masako Bando, Taichiro Kugo, Nobuhiro Maekawa, and Hiroaki Nakano.
\newblock {Improving the effective potential: Multimass scale case}.
\newblock {\em Prog. Theor. Phys.}, 90:405--418, 1993, hep-ph/9210229.

\bibitem{Casas:1998cf}
J.~A. Casas, V.~Di~Clemente, and M.~Quiros.
\newblock {The Effective potential in the presence of several mass scales}.
\newblock {\em Nucl. Phys.}, B553:511--530, 1999, hep-ph/9809275.

\bibitem{Bando:1992np}
Masako Bando, Taichiro Kugo, Nobuhiro Maekawa, and Hiroaki Nakano.
\newblock {Improving the effective potential}.
\newblock {\em Phys. Lett.}, B301:83--89, 1993, hep-ph/9210228.

\bibitem{Liddle:2003as}
Andrew~R Liddle and Samuel~M Leach.
\newblock {How long before the end of inflation were observable perturbations
  produced?}
\newblock {\em Phys. Rev.}, D68:103503, 2003, astro-ph/0305263.

\bibitem{Adshead:2010mc}
Peter Adshead, Richard Easther, Jonathan Pritchard, and Abraham Loeb.
\newblock {Inflation and the Scale Dependent Spectral Index: Prospects and
  Strategies}.
\newblock {\em JCAP}, 1102:021, 2011, 1007.3748.

\bibitem{Shimabukuro:2014ava}
Hayato Shimabukuro, Kiyotomo Ichiki, Susumu Inoue, and Shuichiro Yokoyama.
\newblock {Probing small-scale cosmological fluctuations with the 21 cm forest:
  Effects of neutrino mass, running spectral index, and warm dark matter}.
\newblock {\em Phys. Rev.}, D90(8):083003, 2014, 1403.1605.

\bibitem{Martin:2014rqa}
Jerome Martin, Christophe Ringeval, and Vincent Vennin.
\newblock {How Well Can Future CMB Missions Constrain Cosmic Inflation?}
\newblock {\em JCAP}, 1410(10):038, 2014, 1407.4034.

\bibitem{Aad:2015zhl}
Georges Aad et~al.
\newblock {Combined Measurement of the Higgs Boson Mass in $pp$ Collisions at
  $\sqrt{s}=7$ and 8 TeV with the ATLAS and CMS Experiments}.
\newblock {\em Phys. Rev. Lett.}, 114:191803, 2015, 1503.07589.

\bibitem{Degrassi:2012ry}
Giuseppe Degrassi, Stefano Di~Vita, Joan Elias-Miro, Jose~R. Espinosa, Gian~F.
  Giudice, et~al.
\newblock {Higgs mass and vacuum stability in the Standard Model at NNLO}.
\newblock {\em JHEP}, 1208:098, 2012, 1205.6497.

\bibitem{Bednyakov:2015sca}
A.~V. Bednyakov, B.~A. Kniehl, A.~F. Pikelner, and O.~L. Veretin.
\newblock {Stability of the Electroweak Vacuum: Gauge Independence and Advanced
  Precision}.
\newblock 2015, 1507.08833.

\bibitem{CMS:2014hta}
CMS.
\newblock {Combination of the CMS top-quark mass measurements from Run 1 of the
  LHC}.
\newblock 2014.

\bibitem{Aad:2015nba}
Georges Aad et~al.
\newblock {Measurement of the top quark mass in the $t\bar t \to {\rm
  lepton+jets}$ and $t\bar t \to {\rm dilepton}$ channels using $\sqrt{s}=7$
  TeV ATLAS data}.
\newblock 2015, 1503.05427.

\bibitem{Bezrukov:2007ep}
Fedor~L. Bezrukov and Mikhail Shaposhnikov.
\newblock {The Standard Model Higgs boson as the inflaton}.
\newblock {\em Phys. Lett.}, B659:703--706, 2008, 0710.3755.

\bibitem{Array:2015xqh}
P.~A.~R. Ade et~al.
\newblock {BICEP2 / Keck Array VI: Improved Constraints On Cosmology and
  Foregrounds When Adding 95 GHz Data From Keck Array}.
\newblock 2015, 1510.09217.

\bibitem{Machacek:1983tz}
Marie~E. Machacek and Michael~T. Vaughn.
\newblock {Two Loop Renormalization Group Equations in a General Quantum Field
  Theory. 1. Wave Function Renormalization}.
\newblock {\em Nucl.Phys.}, B222:83, 1983.

\bibitem{Machacek:1983fi}
Marie~E. Machacek and Michael~T. Vaughn.
\newblock {Two Loop Renormalization Group Equations in a General Quantum Field
  Theory. 2. Yukawa Couplings}.
\newblock {\em Nucl.Phys.}, B236:221, 1984.

\bibitem{Machacek:1984zw}
Marie~E. Machacek and Michael~T. Vaughn.
\newblock {Two Loop Renormalization Group Equations in a General Quantum Field
  Theory. 3. Scalar Quartic Couplings}.
\newblock {\em Nucl.Phys.}, B249:70, 1985.

\bibitem{Luo:2002ti}
Ming-xing Luo, Hua-wen Wang, and Yong Xiao.
\newblock {Two loop renormalization group equations in general gauge field
  theories}.
\newblock {\em Phys.Rev.}, D67:065019, 2003, hep-ph/0211440.

\bibitem{Luo:2002ey}
Ming-xing Luo and Yong Xiao.
\newblock {Two loop renormalization group equations in the standard model}.
\newblock {\em Phys. Rev. Lett.}, 90:011601, 2003, hep-ph/0207271.

\bibitem{ATLAS:2014wva}
{First combination of Tevatron and LHC measurements of the top-quark mass}.
\newblock 2014, 1403.4427.

\end{thebibliography}

\end{document}